\documentclass[structabstract]{aa}
\usepackage{adjustbox}
\usepackage{color}
\usepackage{graphicx}
\usepackage{amsmath}% Advanced maths commands
\usepackage[table]{xcolor}

\usepackage{natbib}
\usepackage{graphicx}
\usepackage{rotating}
\usepackage{txfonts}
\usepackage{lscape}
\usepackage{booktabs}
\usepackage{dcolumn}
\usepackage{float}
\usepackage{dblfloatfix}
\usepackage{array}
\usepackage{footnote}
\usepackage{wasysym}
\makesavenoteenv{tabular}
\makesavenoteenv{table}
\usepackage{diagbox}

\usepackage{txfonts}
\usepackage[colorlinks=true,
            linkcolor=blue,
            citecolor=magenta]{hyperref}

\begin{document} 

\title{Impact of non-thermal particles on the spectral and structural properties of M87}
%Impact of the nonthermal emission properties onto the M87 jet opening angle}

\titlerunning{Impact of non-thermal emission on the M\,87 jet}
\authorrunning{Fromm et al.}
\author{Christian M. Fromm \inst{1,2,3}\fnmsep\thanks{cfromm@th.physik.uni-frankfurt.de}, Alejandro Cruz-Osorio \inst{2} \thanks{osorio@itp.uni-frankfurt.de}, Yosuke Mizuno\inst{4,2}, Antonios Nathanail 
\inst{5,2}, \\
Ziri Younsi\inst{6}, Oliver Porth\inst{7}, Hector Olivares\inst{2,8}, Jordy Davelaar\inst{9,10},
Heino Falcke\inst{8,3}, Michael Kramer\inst{3,11} \\and Luciano Rezzolla\inst{2,12,13} }
\institute{Black Hole Initiative at Harvard University, 20 Garden Street, Cambridge, MA 02138, USA\
\and Institut f\"ur Theoretische Physik, Goethe Universit\"at, Max-von-Laue-Str. 1, D-60438 Frankfurt, Germany 
\and Max-Planck-Institut f\"ur Radioastronomie, Auf dem H\"ugel 69, D-53121 Bonn, Germany\
\and Tsung-Dao Lee Institute and School of Physics and Astronomy, Shanghai Jiao Tong University, Shanghai, 200240, People's Republic of China\
\and Department of Physics, National and Kapodistrian University of Athens, Panepistimiopolis, GR 15783 Zografos, Greece \
\and Mullard Space Science Laboratory, University College London, Holmbury St.\,Mary, Dorking, Surrey RH5 6NT, UK \
\and Anton Pannekoek Institute for Astronomy, University of Amsterdam, Science Park 904, 1098 XH Amsterdam, The Netherlands\
\and Department of Astrophysics/IMAPP, Radboud University Nijmegen, P.O. Box 9010, 6500 GL Nijmegen, The Netherlands \
\and Department of Astronomy and Columbia Astrophysics Laboratory, Columbia University, 550 W 120th St, New York, NY 10027, USA \
\and Center for Computational Astrophysics, Flatiron Institute, 162 Fifth Avenue, New York, NY 10010, USA \
\and Jodrell Bank Centre for Astrophysics, University of Manchester, Machester M13 9PL, UK \
\and Frankfurt Institute for Advanced Studies, Ruth-Moufang-Strasse 1, 60438 Frankfurt, Germany \
\and School of Mathematics, Trinity College, Dublin 2, Ireland \\
\email{cfromm@th.physik.uni-frankfurt.de, osorio@itp.uni-frankfurt.de}}

\date{
Draft 1.0: \today
%Received September 15, 1996; accepted March 16, 1997
}
\abstract
  {The recent 230\,GHz observations of the Event Horizon Telescope (EHT) are able to image the innermost structure of the M\,87 and show a ring-like structure which is in agreement with thermal 
synchrotron emission generated in a torus surrounding a supermassive black hole. However, at lower frequencies M\,87 is characterised by a large-scale and edge-brightened jet with clear 
signatures of non-thermal emission. In order to bridge the gap between these scales and to provide a theoretical interpretation of these observations we perform general relativistic 
magnetohydrodynamic simulations of accretion on to black holes and jet launching.}
  {M\,87 has been the target for multiple observations across the entire electromagnetic spectrum. Among these VLBI observations provide unique details on the collimation profile of the jet down to 
several gravitational radii. In this work we aim to model the observed broad-band spectrum of M\,87 from the radio to the NIR regime and at the same time fit the jet structure as observed with 
Global mm-VLBI at 86\,GHz.}
  {We use general relativistic magnetohydrodynamics and simulate the accretion of the magnetised plasma onto Kerr-black holes in 3D. The radiative signatures of these simulations are computed 
taking different electron distribution functions into account and a detailed parameter survey is performed in order to match the observations.}
  {The results of our simulations show that magnetically arrested disks around fast spinning black holes ($a_\star\geq0.5$) together with a mixture of thermal and non-thermal particle distributions 
are able to model simultaneously the broad-band spectrum and the innermost jet structure of M\,87}
  {}

\maketitle

\section{Introduction}\label{sec:intro}

The recent observations of the elliptical galaxy M87, at 230 GHz
performed by the Event Horizon Telescope (EHT)  using the Very Long
Baseline Interferometry (VLBI) technique are consistent with
a supermassive black hole (SMBH) with mass $M_{\rm BH}=6.5^{+0.7}_{-0.7}
\times 10^{9} M_{\odot}$ at the centre of the galaxy \cite{EHT_M87_PaperI}. 
The compact radio source that was captured by the EHT revealed an asymmetric ring of
diameter $d=\, 42\, \pm\, 3\, {\mu \rm as}$, which is supposed to be
located beneath the actual base of a relativistic jet.

The M\,87 jet has been systematically monitored from radio to $\gamma$-rays
\citep[see, e.g.,][]{Reid1982,Hada2013,Kim2018a,Snios2019,MAGIC2020}. During the EHT 2017
observations, several international facilities in space and on the ground
partnered for a quasi-simultaneous multi-wavelength campaign of the M\,87
nuclei \citep{Algaba2021}. Such efforts can reveal the underlying
jet-launching and particle acceleration physics. Moreover, multi-wavelength observations can shed light to 
the theoretical expectations of a black
hole jet and a magnetized disk wind \cite{Blandford1977,
Blandford:1982di,Blandford2019,LyndenBell2006}.

More specifically, the jet collimation profile can be revealed from multi-frequency (10 -230 GHz) VLBI observations. Such observations have shown that, till hundreds of Schwarzschild radii\footnote{$R_{\rm S}=2GM/c^2$, with gravitational constant $G$, mass of the black hole $M$, and speed of light $c$}, the
M87 jet has a parabolic shape \cite{Asada2012,Doeleman2012,Hada2013}.
The shape and width of a jet can be best described by its opening angle,
and for the M87 case, the jet base has revealed a rather wide structure
\citep{Kim2018a}. The EHT observations of M\,87 show a ring-like emission structure in agreement with theoretical models of accretion onto a SMBH \citep{EHT_M87_PaperV}. The key physical ingredients are the spin of the SMBH, the flow properties and the
magnetic field, which is supposed to be responsible for jet-launching. A
theoretical model needs to capture the compact ring-like image (on horizon scales)
and fit the spectrum, but also account for the morphology and kinematics
of the observed jet \citep{Mertens2016,Walker2018}. To solve this multi-parametric
problem, works have developed semi-analytical techniques to fit the
spectrum and model the basic jet characteristics
\citep{DiMatteo2003,Broderick2009}. To draw a self-consistent picture of accretion onto a black hole,
general relativistic magnetohydrodynamic (GRMHD) simulations together with general relativistic radiative transfer (GRRT) calculations \citep[see, e.g.][]{Dexter2012} are employed. The characteristics of the M87 radio core, of a flat spectrum and an
increasing size with wavelength can be reproduced by a two-temperature
accretion flow and a hot single-temperature jet \citep{Moscibrodzka2016}.
The addition of non-thermal electrons is expected to better reproduce the
NIR flux and the flat radio spectrum, but also acquire a more extended jet 
structure at 43\,GHz and 86\,GHz \citep{Davelaar2019}.
\newline In this work we study the impact of a non-thermal emission model on the
structure and morphology of the M87 jet and its spectrum. We extended the 
study presented in \cite{Cruz2021b} and included additional models. 
Furthermore, we perform a detailed parameter survey to understand the 
impact of the various emission parameters on the resulting broad-band 
spectrum and image structures. To this end we perform long-term, 
high-resolution, 3D GRMHD simulations. The two important parameters 
of these simulations are the black hole spin, $a_{\star}=cJ/GM^2$ with angular momentum $J$, and the 
magnetic flux across the horizon, $\phi_{\rm BH}$. Depending on the magnetic 
flux the models can be divided into Standard And Normal Evolution (SANE) 
$\left(\Psi=\phi_{\rm BH}/\sqrt{\dot{M}}< 10 \rm{\,where\,\dot{M}\, is\, the\, mass\,accretion\,rate} \right)$ and Magnetically Arrested Disks  (MAD) $\left(\Psi \geq15\right)$ Notice, that our definition of the MAD parameter $\Psi$ differs by   a factor $\sqrt{4\pi}$ from the one defined in \citet{Tchekhovskoy2011}. In this work we use SANE 
and MAD models with spins $a_\star=$ -0.9375, -0.50, 0, 0.50, and 0.9375. 
The output of these simulations is
coupled with GRRT calculations, where we employ a mixture of thermal and
non-thermal electrons using the kappa distribution see \citet{Pierrard2010} for a review and
\citet{Davelaar2019} for the application to M\,87. We present a comprehensive analysis focusing on
the impact of the different parameters in the emission model on the spectrum and image structure of M\,87. More specifically, the free parameters in the emission model are: the local ratio of the proton
to electron temperature, the maximum magnetization in the jet spine and
the percentage of the magnetic energy on the kappa model. 
\newline The paper is organized as follows: In Section \ref{sec:GRMHD} we discuss
the details of the GRMHD simulations, whereas in Section \ref{sec:GRRT}
we describe the GRRT post process calculations. In Section
\ref{sec:results} we analyse the impact of the free parameters of our model on the broad-band spectrum and image structure of M\,87, followed by a summary and discussion in Section \ref{sec:discussion}. Lastly, we present our conclusions and an outlook in Section
\ref{sec:conc}.
\newline Throughout this work we use a black hole mass of 
$6.5\times10^9\,M_{\astrosun}$ for M\,87 and a distance of 16.8\,Mpc. 

\section{General Relativistic Magneto-Hydrodynamic simulations}\label{sec:GRMHD}

We use the state-of-the-art GRMHD code \textit{Black Hole Accretion Code} \texttt{BHAC} \cite{Porth2019, Olivares2020} to simulate accretion onto black holes and jet launching. \texttt{BHAC} solves the three-dimensional 
GRMHD equations written in a coordinate basis $(t, x^i)$ on a generic space-time four-metric $(g_{\mu\nu})$ with metric determinant $g$, while keeping the magnetic field divergence free. In this work Greek indices run through $\left[0,\,1,\,2,\,3\right]$ while Roman indices cover $\left[1,\,2,\,3\right]$. For the purpose of this work we choose Kerr black holes in spherical coordinates.  The equations solved in \texttt{BHAC} in geometric 
units ($GM=c=1$ and $1/\sqrt{4\pi}$ is absorbed in the magnetic field) are the conservation of mass, the local conservation of energy-momentum and covariant Maxwell equations:

\begin{equation}
    \nabla_{\mu} (\rho u^{\mu})=0\,, \quad \nabla_{\mu} T^{\mu \nu}=0\,, \quad \nabla_{\mu} {}^{*}\!F^{\mu \nu}=0\,,
\end{equation}

where $\rho$ is the rest-mass density and $u^\mu$ is the four-velocity.
The stress-energy tensor for ideal MHD is defined as:
\begin{equation}
   T^{\mu\nu} = (\rho h_{\rm tot}) u^\mu u^\nu + \left(p + \frac{1}{2} b^2\right) g^{\mu\nu} - b^\mu b^\nu,
\end{equation}
where $h_{\rm tot}=h+b^2/\rho$ is the total specific enthalpy and $p$ the fluid pressure.
The magnetic field four-vector $b^\mu$ is given by:
\begin{equation}
b^t = B^i u^\mu g_{i\mu} \quad  b^i = (B^i + b^t u^i)/u^t,
\end{equation}
where $B^i$ is the magnetic field measured by an Eulerian observer and is evolved as a primitive variable. For the closure of the system of conservation laws we assume an equation of state for an 
ideal gas, connecting the specific enthalpy, $h$, to the pressure, $p$, and density, $\rho$:
\begin{equation}
 h= 1+ \frac{\hat{\gamma} p}{\left(\hat{\gamma} -1\right)\rho},   
\end{equation}
where $\hat{\gamma}$ is the adiabatic index (see \cite{Anton06,Lora2015, Porth2017,Cruz2020} for more details of the evolution equations).
We solve the GRMHD equations on a spherical polar grid $(r,\theta,\phi)$ where the grid spacing is logarithmic in radial and linear in polar and azimuthal direction. The dimensions and resolution of 
the grid are presented in Table \ref{tab:resolution}.

\begin{table}[h!]
\caption{Dimension and resolution of the numerical grid used for the GRMHD simulations}   
\centering
\begin{tabular}{lcccc}
\hline
 Model  &  r [M] & $\theta \,[\rm rad]$ & $ \phi\,[\rm rad]$ & $ N_r,N_\theta,N_\phi$ \\
\hline
\hline
SANE & 1.18$r_{\rm EH}$\,-- 3333 &  0 -- $\pi$ & 0 -- 2$\pi$ & 512,192,192\\
MAD  & 1.18$r_{\rm EH}$\,-- 2500 &  0 -- $\pi$ & 0 -- 2$\pi$ & 384,192,192\\
\hline
\multicolumn{5}{l}{$r_{\rm EH}=1+\sqrt{1-a_{\star}^2}$ is the radius of the event horizon}\\
\multicolumn{5}{l}{in geometric units}
\end{tabular}
\label{tab:resolution}
\end{table} 

We use inflow boundary condition and outflow boundary conditions in radial coordinate, solid reflective wall along the polar boundaries \citep{Olivares2018a}, and in the azimuthal direction, we have 
employed a periodic boundary conditions to all physical quantities across the cells at $\phi=0$. To solve the GRMHD equations we have employed finite volume and high resolution shock capturing 
methods; the LLF flux formula in combination with PPM reconstruction and a two-step predictor-corrector scheme. At the same time, the magnetic fields are evolved with the staggered upwind 
constrained transport scheme \citep{Olivares2020}.
We initialise our 3D GRMHD simulations with a magnetised torus in hydrodynamic equilibrium with the central rotating Kerr BH. The torus has a constant specific angular momentum and weak 
poloidal magnetic field with a single loop is added on the top of the torus, defined by vector potential:
\begin{equation}
   A_{\phi}\propto\mathrm{max}(q-0.2,0), 
\end{equation} where 
\begin{eqnarray}
 q &=& \frac{\rho} {\rho_{\rm max}}\left(\frac{r}{r_{\rm in}}\right)^3 \sin^3 \theta \exp \left(\frac{-r}{400}\right) \quad \mathrm{for\, MAD}\\
 q &=& \frac{\rho} {\rho_{\rm max}}  \qquad \qquad \qquad \qquad  \qquad \mathrm{\,\,for\,SANE}
\end{eqnarray}
see, e.g., \cite{Fishbone76,Font02b,Rezzolla_book:2013}.

The initial conditions of the GRMHD simulations are the constant specific angular momentum, $l$, adiabatic index of the fluid $\hat{\gamma}$, plasma beta, $\beta=2p/b^2$, inner $r_{\rm in}$ radius and radius of the density maximum in the torus, 
$\rm r_{c}$, as well as the dimensionless spin parameter spin parameter $a_{\star}$.  In Table \ref{tab:TorusID} we list the used values where models names \texttt{MT.M.*} refer to MAD states and \texttt{MT.S.*} to SANE states. To trigger the magneto-rotational instability (MRI), we perturbed the equilibrium torus by a white noise to the fluid pressure so that its resulting value is $p = p (1 + X_p)$ with a random number $|X_p|<0.04$.

\begin{table}[h!]
\caption{Physical parameters of the equilibrium magnetised torus for MAD and SANE accretion models (for details see text).}
\begin{adjustbox}{max width=0.48\textwidth}
\centering
\begin{tabular}{lcccccccl}
\hline
 Model                  & $\rm r_{in} [M]$& $\rm r_{c} [M]$& $l$        & ${\hat{\gamma}}$ &$\beta$&$a_{\star}$ & $\rm acc$\\
\hline
\hline
$\texttt{MT.M.1}$&  $20$    & $40$    &  $6.92$  & $4/3$    &$10^{2}$& $-15/16$& MAD \\
$\texttt{MT.M.2}$&  $20$    & $40$    &  $6.88$  & $4/3$    &$10^{2}$& $-1/2$  & MAD \\
$\texttt{MT.M.3}$&  $20$    & $40$    &  $6.84$  & $4/3$    &$10^{2}$& $0$     & MAD \\
$\texttt{MT.M.4}$&  $20$    & $40$    &  $6.80$  & $4/3$    &$10^{2}$& $+1/2$  & MAD \\
$\texttt{MT.M.5}$&  $20$    & $40$    &  $6.76$  & $4/3$    &$10^{2}$& $+15/16$& MAD\\
\hline
$\texttt{MT.S.1}$&  $10$   & $17$    &  $5.26$  & $4/3$    &$10^{2}$& $-15/16$& SANE\\
$\texttt{MT.S.2}$&  $8$    & $15$    &  $5.00$  & $4/3$    &$10^{2}$& $-1/2$  & SANE\\
$\texttt{MT.S.3}$&  $6$    & $15$    &  $4.84$  & $4/3$    &$10^{2}$& $0$     & SANE\\
$\texttt{MT.S.4}$&  $6$    & $13$    &  $4.52$  & $4/3$    &$10^{2}$& $+1/2$  & SANE\\
$\texttt{MT.S.5}$&  $6$    & $12$    &  $5.46$  & $4/3$    &$10^{2}$& $+15/16$& SANE \\
\hline
\end{tabular}
\label{tab:TorusID}
\end{adjustbox}
\end{table} 

\begin{figure}[h!]
\centering
\includegraphics[width=0.49\textwidth]{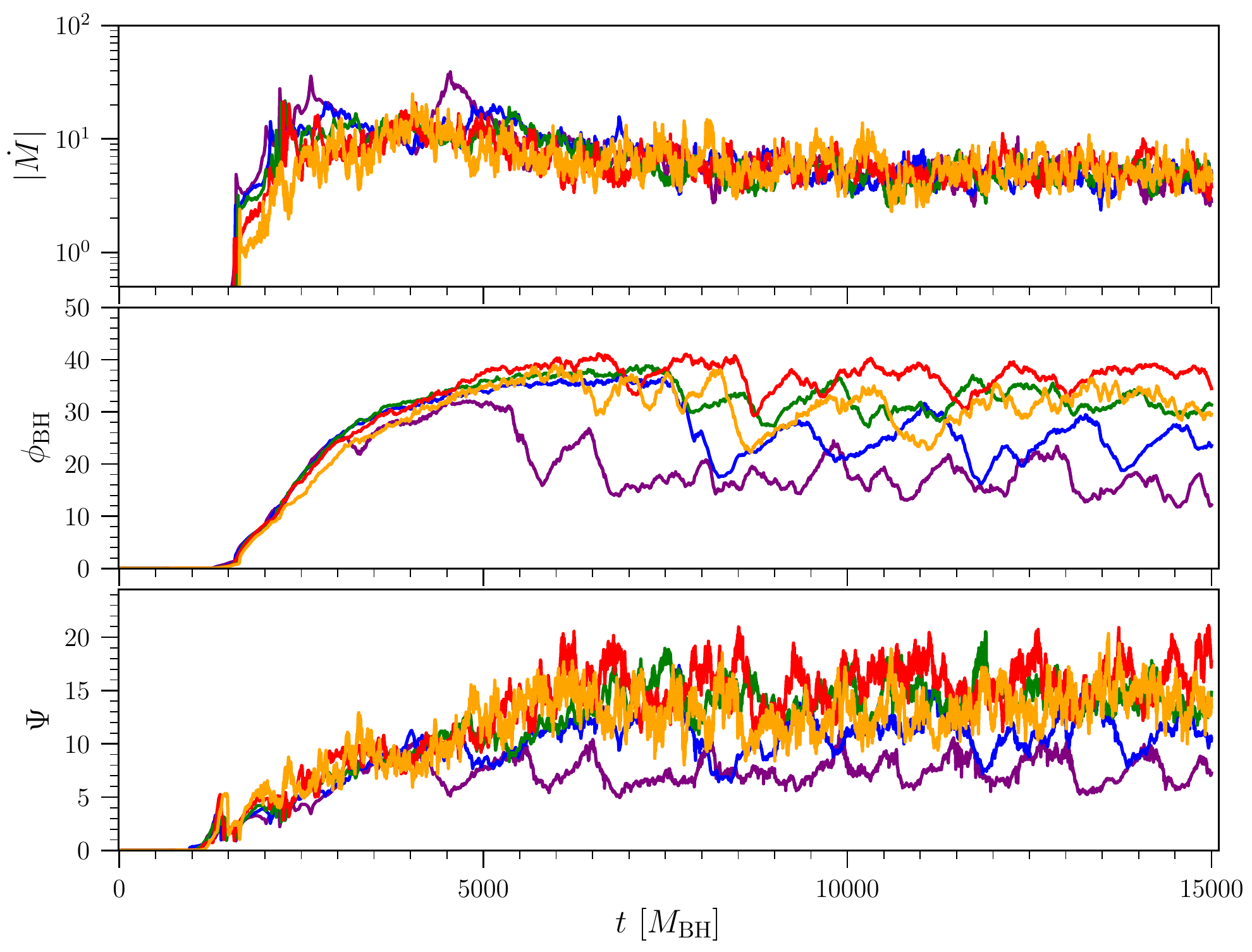}
\includegraphics[width=0.49\textwidth]{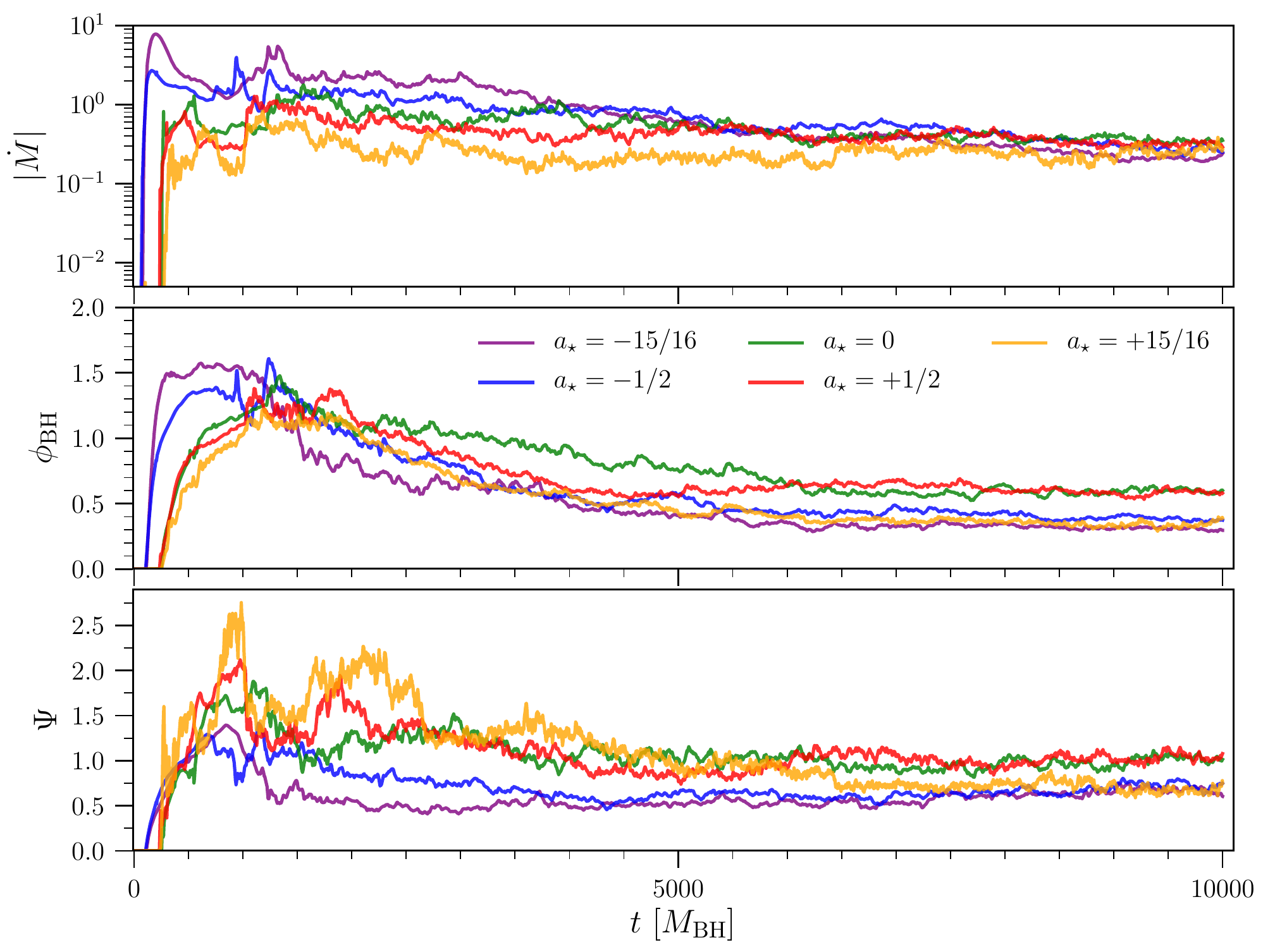}
 \caption{Mass accretion rates $\dot{M}$, magnetic flux $\phi$, and MAD flux parameter $\Psi$ in code units for black holes with different dimensionless spins. The first three panels correspond to the MAD state and 
bottom three panels to SANE simulations. Notice the different scales for the MAD and SANE models.}
\label{fig:rates}
\end{figure}

During the course of the GRMHD simulations we monitor the mass accretion rate, $\dot{M}$, and the accreted magnetic flux across the horizon, $\phi_{\rm BH}$. In Fig. \ref{fig:rates} we present 
$\dot{M}$, 
$\phi_{\rm BH}$ and the MAD flux parameter $\Psi=\phi_{\rm BH}/\sqrt{\dot{M}}$ for the MAD models (top three panels) and for the SANE models (bottom three panels). 
Since the inner edge of the MAD torus is located at a larger radii from the black hole as in the SANE configurations (see Table \ref{tab:TorusID}) the saturation times are different between the models 
($t_{\rm sat}\sim 10,000\,\rm M$ for MADs and $t_{\rm sat}\sim7,000\,\rm M$ for SANEs, see Fig. \ref{fig:rates}). Once the simulations obtain a quasi-stable mass accretion rate ($t \in [8,000M,\ 
10,000M]$ for the SANE models and $t \in [13,000M,\ 15,000M]$ for the MAD ones) we compute the average mass accretion rate, $\langle \dot{M}\rangle$, average magnetic flux across the 
horizon, $\langle \phi_{\rm BH}\rangle$, and the average MAD parameter $\langle \Psi\rangle$ (see Table \ref{tab:rates}). The average mass accretion rate approach a constant value 
$\langle\dot{M}_{\rm 
MAD}\rangle \sim 5$ and $\langle\dot{M}_{\rm SANE}\rangle \sim 0.3$ independently of the BH spin. On the other hand, the average magnetic flux and average MAD parameter 
increase with spin and exhibit a maximum for a BH spin of $a=+1/2$ for both, SANE and MAD models. All MAD models except model $\texttt{MT.M.1}$ reach the MAD state according to the 
threshold value of $\Psi \sim 10$ \citep{Tchekhovskoy2011}. Models $\texttt{MT.S.*}$ are well within the low magnetisation regime $\langle \Psi\rangle \leq 1$ and in agreement with results 
presented in \cite{Porth2019}. 

\begin{table}[h!]
\caption{Averaged accretion rates and their standard deviations for different spins $a_{\rm \star}$ and accretion models. All quantities are presented in code units. For details see text.} 
\begin{adjustbox}{max width=0.48\textwidth}
\centering
\begin{tabular}{lcccc}
\hline
 Model      & $a_{\rm \star}$& $\langle \dot{M}\rangle$& $\langle \phi_{\rm BH} \rangle$& $\langle \Psi\rangle$ \\
\hline
\hline
$\texttt{MT.M.1}$& $-15/16$& $4.7 \pm 0.96$& $15 \pm 2$& $7  \pm 1$\\
$\texttt{MT.M.2}$& $-1/2$  & $4.7 \pm 0.90$& $25 \pm 3$& $11 \pm 2$\\
$\texttt{MT.M.3}$& $0$     & $5.0 \pm 0.68$& $32 \pm 1$& $14 \pm 1$\\
$\texttt{MT.M.4}$& $+1/2$  & $5.6 \pm 1.40$& $37 \pm 1$& $16 \pm 2$\\
$\texttt{MT.M.5}$&$+15/16$ & $5.2 \pm 1.20$& $33 \pm 2$& $15 \pm 2$\\
\hline
$\texttt{MT.S.1}$& $-15/16$& $0.23 \pm 0.03$& $0.32 \pm 0.01$& $0.66 \pm 0.04$\\
$\texttt{MT.S.2}$& $-1/2$  & $0.32 \pm 0.05$& $0.39 \pm 0.02$& $0.70 \pm 0.05$\\
$\texttt{MT.S.3}$& $0$     & $0.36 \pm 0.03$& $0.60 \pm 0.02$& $0.99 \pm 0.05$\\
$\texttt{MT.S.4}$& $+1/2$  & $0.33 \pm 0.03$& $0.59 \pm 0.02$& $1.00 \pm 0.05$\\
$\texttt{MT.S.5}$&$+15/16$ & $0.24 \pm 0.04$& $0.34 \pm 0.02$& $0.71 \pm 0.06$\\
\hline
\end{tabular}
\label{tab:rates}
\end{adjustbox}
\end{table} 

\begin{figure*}[h!]
\centering
\includegraphics[width=0.7\textwidth]{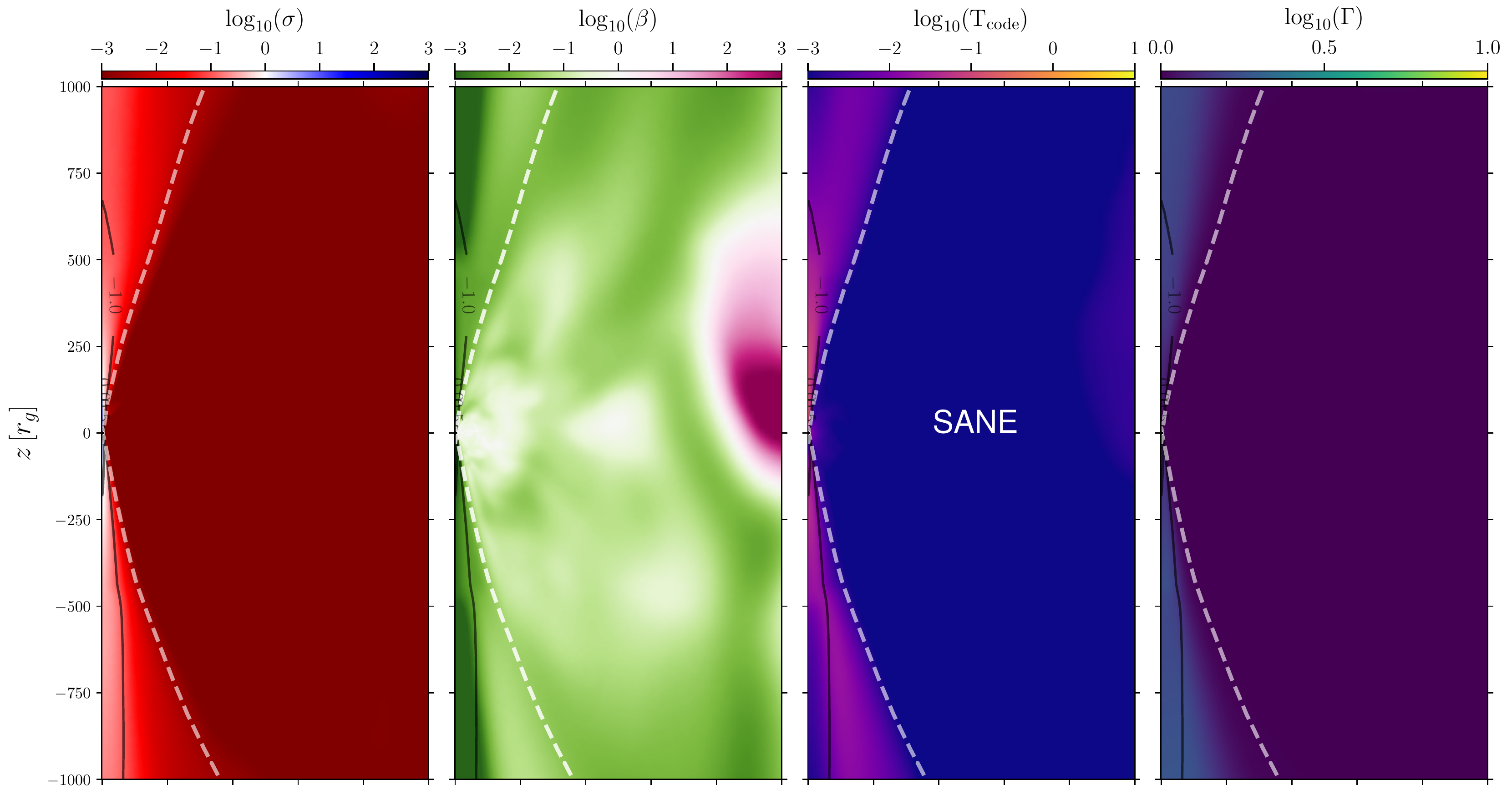}
\includegraphics[width=0.7\textwidth]{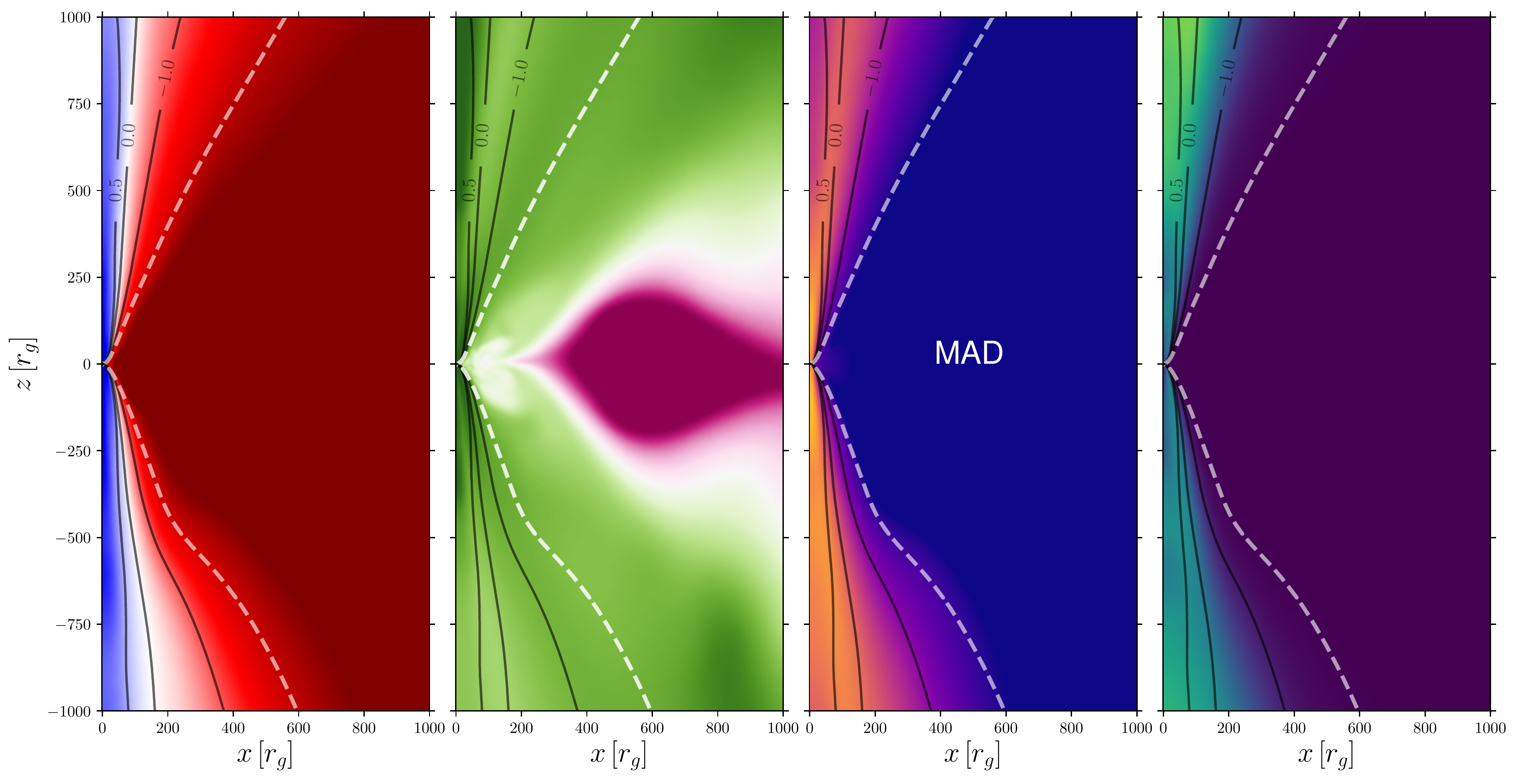}
 \caption{From left to right the panels show the time and azimuthal averaged magnetisation, $\sigma$,  plasma $\beta$, gas temperature (proton) in code units, $\rm T_{\rm code}$, and Lorentz 
factor $\Gamma$ for a BH with $a=+15/16$. All parameters are plotted in logarithmic scale.
 The top row shows the aforementioned quantities for the SANE accretion and the bottom one for the MAD case. The dashed white line represents the boundary between bounded ($\rm Be<1$) 
and unbounded ($\rm Be>1.02$) plasma defined through the Bernoulli parameter $\rm Be$, black contour lines $\log_{10} \sigma =-1.0,\,0,\,\mathrm{and}\, 0.5$ corresponds to jet sheath and jet 
spine boundaries.}
\label{fig:Morphology}
\end{figure*} 

In Figure \ref{fig:Morphology} we show the time\footnote{for SANE $t=\rm [8,000M,\ 10,000M]$ and for MAD $t=\rm [13,000M,\ 15,000M]$} and azimuthal averaged large scale morphology of 
the plasma magnetisation, $\sigma=b^2/\rho$, plasma beta, $\beta$, gas temperature, $T_{\rm code}=p/\rho$, and Lorentz factor, $\Gamma$, for a BH with spin $a=+15/16$. The top row 
corresponds to a SANE model and the bottom row to a MAD one. The morphology of our simulations can be divided into two distinct regions: the jet and the disk. We use the Bernoulli parameter $-hu_t<1.02$ to separate disk (bound plasma) from the jet region (unbound plasma). The jet region can be further 
divided into a jet spine and a jet sheath region. The former is characterised by a high magnetisation $\sigma > \sigma_{\rm cut}$\footnote{commonly a value of $\sigma_{\rm cut}=1$ \citep[see, 
e.g.,][]{EHT_M87_PaperV} is used} and low plasma beta ($\beta \ll 1$). In Fig. \ref{fig:Morphology} the solid black lines indicate $\log_{10}\sigma=-1.0\,,0,\, \mathrm{and}\,0.5$. Within the jet spine, 
the hot plasma is accelerated along the poles reaching Lorentz factors $\Gamma\sim 10$ in the MAD case and $\Gamma\sim 5$ for the SANE models. We define as jet sheath the region where 
$\sigma 
< \sigma_{\rm cut}$ while the Bernoulli parameter $\mathrm{Be} =-hu_{t}>1.02$, indicating out-flowing plasma (dashed white line in Fig. \ref{fig:Morphology}). In the jet sheath we find typical 
Lorentz factors of $\Gamma\sim 8$ for MAD models and $\Gamma\sim 5$ for the SANE ones. The regions outside of these boundaries we attribute to the disk and disk wind.
In Fig. \ref{fig:Morphology} the major differences between MAD and SANE models are clearly visible:
\begin{itemize}
    \item MAD models show larger jet opening angles than SANEs (compare the area contained within the dashed white lines and jet axis ($x=0$) in Fig. \ref{fig:Morphology})
    \item MAD models show higher magnetisation in the jet spine and sheath than SANEs (see first and fifth panel in Fig. \ref{fig:Morphology})
    \item MAD models produce hotter jets i.e. higher temperature in the jet region than SANEs (see third and seventh panel in Fig. \ref{fig:Morphology})
    \item MAD models generate faster jets than their SANE counterparts (see fourth and eight panel in Fig. \ref{fig:Morphology})
\end{itemize} 

\section{General Relativistic Radiative Transfer (GRRT) calculations}\label{sec:GRRT}
In order to compare our GRMHD models with single dish and VLBI observations of M\,87 we need to compute their radiative signatures across the electromagnetic spectrum. The covariant 
radiative transfer equation can be written as:
\begin{equation}
    \frac{d\mathcal{I}}{d\tau_\nu}=-\mathcal{I} +\frac{\eta}{\chi},
\end{equation}
where $\mathcal{I}$ is the Lorentz invariant specific intensity and is connected to the specific intensity, $I_\nu$, by: 
\begin{equation}
\mathcal{I}=\frac{I_\nu}{\nu^3}.
\end{equation}
and  Lorentz invariant emissivity, $\eta$, and absorptivity, $\chi$, are related to the emission, $j_\nu$ and absorption coefficients, $\alpha_\nu$ evaluated at frequency $\nu$ via:
\begin{equation}
\eta=j_{0,\nu}/\nu^2 \qquad \chi=\alpha_{0,\nu}\nu,
\end{equation}
where the subscript "0" indicates quantities measured in the local-rest frame of the plasma. Using the definition of the optical depth $d\tau_\nu=\int \alpha_\nu ds$, the equation for the covariant 
radiative transport can be decoupled into two differential equations:
\begin{eqnarray}
 \frac{d\tau_{\nu}}{d\lambda}= \gamma^{-1} \alpha_{0, \nu}\,, \quad \quad 
 \frac{d{\mathcal{I} }}{d\lambda}= \gamma^{-1}  \left(\frac{j_{0,\nu}}{\nu^3}\right) \exp\left(-\tau_{\nu }\right)\,, \label{eq:grrt}
\end{eqnarray}
with affine parameter $\lambda$ and energy shift between oberver's and co-moving frame $\gamma^{-1}=\nu_0/\nu=-k_\alpha u^\alpha\rvert_\lambda/k_\beta u^{\,\beta}\rvert_\infty$ where 
$k_\alpha$ is the wave vector of the photon. For more details on the radiative transfer scheme see \citet{Younsi2012}. The emission and absorption coefficients depend on the assumed emission process and 
electron distribution function \citep[see, e.g.,][]{Pandya2016}.

The GRRT equations (Eq. \ref{eq:grrt}) are solved along the geodesic trajectories in the fast-light approximation, assuming that dynamical time of the GRMHD simulations is greater than 
light-crossing 
time using the \textit{Black Hole Observations in Stationary Spacetimes} (\texttt{BHOSS}) code \cite{Younsi2020}. Within \texttt{BHOSS} the null geodesics (propagation of the 
electromagnetic radiation) are solved using a Runge-Kutta-Fehlberg (RKF45) integrator with fourth order adaptive step sizing and fifth order error control. The radiative transfer equations 
are evolved using an Eulerian method where the step size is provided from the geodesic integrator. Throughout our GRRT calculations we used a field of view of $10^3$\,M (which corresponds to 
4\,mas for the black hole mass and distance of M\,87) with a resolution of $800\times800$ pixels. The accuracy of the geodesic integration within the RKF45 method is set to 
$\Delta\epsilon=10^{-14}$ 
and an optical depth cut of $\tau_{\nu,\,\mathrm{cut}}=5$ is used to speed up the GRRT calculations. The GRMHD simulations described in the previous section only 
provide the protons temperature in the plasma. Therefore, the properties of the radiating electrons, especially the electron temperature, $T_{e}$, need to be reconstructed from plasma quantities. Here, we follow 
the work of \citet{Moscibrodzka2016} and \citet{EHT_M87_PaperV} and compute the electron temperature via the so-called R-$\beta$ model which assumes a plasma-$\beta$ depending 
temperature ratio between protons and electrons:

\begin{equation}
\Theta_{\rm e}= \frac{p m_{\rm p}/m_{\rm e} }{ \rho T_{\rm ratio}},\,  \quad T_{\rm ratio} \equiv \frac{T_{\rm p}}{T_{\rm e}}=\frac{R_{\rm low} + R_{\rm high} \beta^{2}}{1+\beta^{2}},
\label{eq:Te}
\end{equation}
where $m_{\rm p}$ and $m_{\rm e}$ are respectively the
proton and electron masses. Besides the plasma-$\beta$ which is provided from our GRMHD simulations two additional free parameters are involved in the calculation of the temperature ratio: 
$\rm 
R_{low}$ and $\rm R_{high}$. 

\begin{figure}[h!]
\centering
    \includegraphics[width=0.49\textwidth]{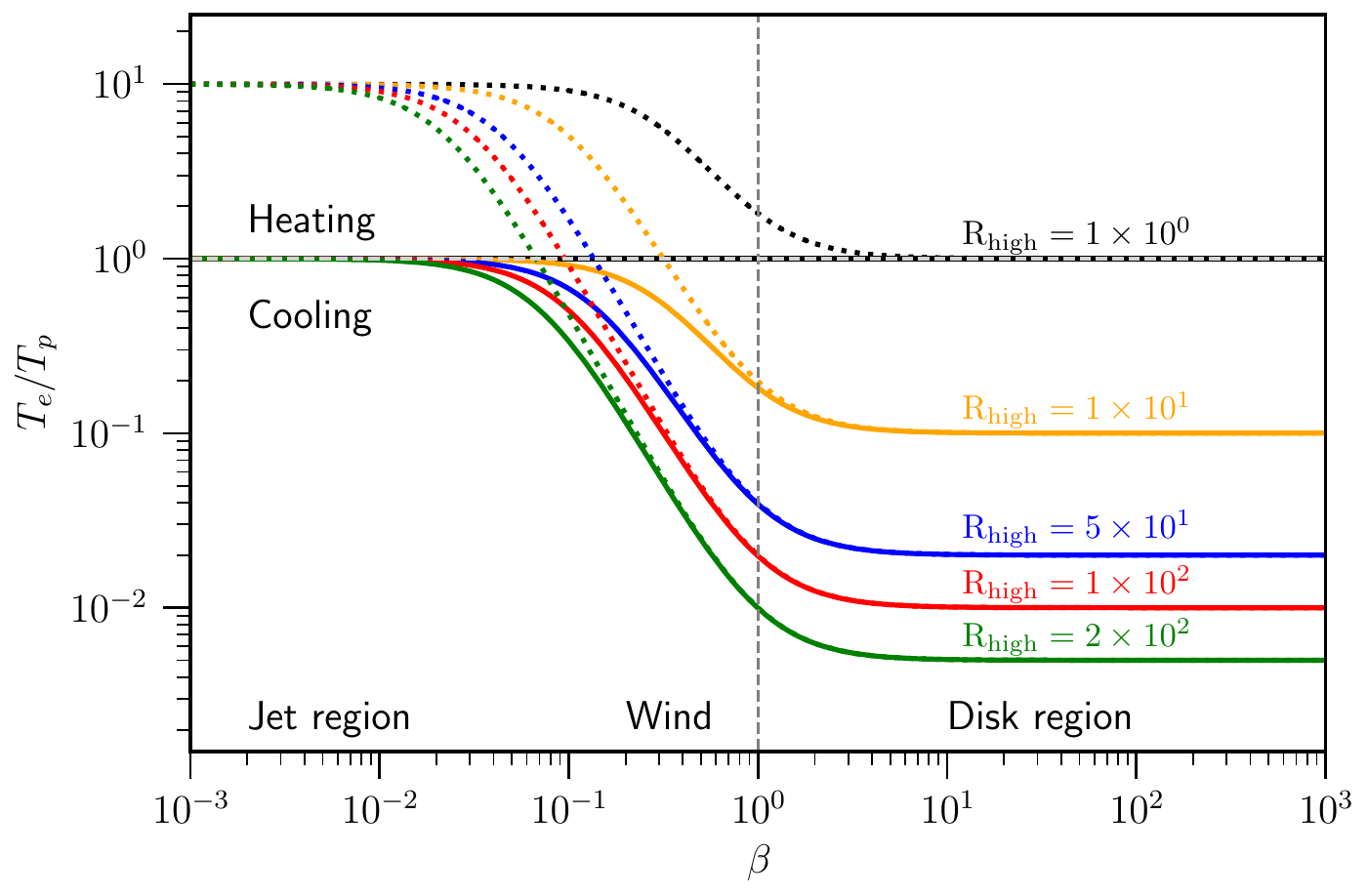}
       \caption{Electron-to-proton temperature ratio $T_{\rm e}/T_{\rm p}$ as function of plasma $\beta$ varying the $R_{\rm low}$ and $R_{\rm high}$ (definition given below in Eq. \ref{eq:Te}). The 
vertical dashed black line marks the transition between the jet sheath region ($\beta<1$) and the disk region (($\beta>1$ ). The solid lines corresponds to 
models with $R_{\rm low}=1$, and dotted lines to models with $R_{\rm low}=0.1$. The different colours indicated different $R_{\rm high}$ values.}
        \label{fig:Te}
\end{figure}

To better understand the impact of $\rm R_{low}$ and $\rm R_{high}$ on the electron temperature we evaluated Eq. \ref{eq:Te} for large range of plasma-$\beta$ ($10^{-3} - 10^3$, a typically range 
found in our GRMHD simulations) for several selected values for $\rm R_{low}$ and $\rm R_{high}$. The result of this calculation can be found in Fig. \ref{fig:Te}. Notice that we plot the 
electron-to-proton 
temperature ratio, i.e., $(T_{\rm p}/T_{\rm e})^{-1}$. The solid lines corresponds to $\rm R_{low}=1.0$ and the dotted lines to $\rm R_{low}=0.1$. The color of the lines indicate different 
$\rm 
R_{high}$ values from black ($\rm R_{high}=1$) to green ($\rm R_{high}=160$). The vertical dashed line marks the transition between jet region ($\beta< 1$) and disk regions ($\beta\geq1$) 
( see also second and sixth panel in Fig. \ref{fig:Morphology}). The variation of the electron-to-proton temperature ration with plasma-$\beta$ resembles the shape of a sigmoid-curve, i.e., 
saturating for low plasma-$\beta$ to the value of $1/R_{\rm low}$ and for large plasma-$\beta$ to the one given by $1/R_{\rm high}$. In other words, increasing the $R_{\rm high}$ value while 
keeping the $R_{\rm low}$ value fixed reduces the electron temperature in the disk regions while not altering the electron temperature in the jet. 
%Therefore, models with large $R_{\rm  high}$ are  commonly referred as "jet dominated" temperature models whereas low $R_{\rm  high}$ are labelled as "disk dominated" temperature models. 

Once the electron temperature, $\Theta_e$, is computed we need to assign an emission process and the electron distribution function (eDF).  These three parameters together with the plasma 
properties (e.g., the magnetic field) define the emission, $j_{0,\nu}$, and absorption, $\alpha_{0,\nu}$, coefficients required for radiative transfer (see. Eq. \ref{eq:grrt}). As mentioned earlier we 
assume synchrotron emission i.e., the gyration of relativistic electrons around the magnetic field line as the main radiative process responsible for the generation of the observed emission in M\,87 
from the radio to NIR range \citep[see, e.g.,][]{Yuan2014}. Additional radiation processes such as Bremsstrahlung can provide additional seed photons important during inverse Compton scattering and thus be responsible for 
the high energy emission in M\,87 \citep[see, e.g.,][]{Yarza2020}. However, given their contributions to the frequency regime under investigation in this paper 
$(10^9\,\mathrm{Hz}\leq10^{16}\,\mathrm{Hz})$ 
these processes can be neglected. 

The two eDFs we explore in this work are the Maxwell-J\"uttner distribution function for thermal electrons and the kappa distribution function for non-thermal electrons. The application of the former 
is well justified by the modelling of the horizon scale emission of M\,87 which is in good agreement with a thermal eDF \citep{EHT_M87_PaperV}. The use of the kappa eDF is motivated from the 
distribution of the spectral index, $\alpha$, computed from VLBI observations of M\,87 at 22\,GHz, 43\,GHz and 86\,GHz which exhibit values between $-1\leq \alpha \leq 0$ within a distance of 
$r<4\,\mathrm{mas}$ from the core \citep[see, Fig. 15 in][]{Algaba2021}. These values indicate a power-law electron distribution with $2\leq p \leq 4$. In addition, the work of \citet{Davelaar2019} 
showed that the M\,87 spectrum between 
$(10^9\,\mathrm{Hz}\leq10^{16}\,\mathrm{Hz})$ consisting of a flat radio spectrum and a steep NIR part can be well reproduced by a kappa eDF. In the following paragraph we provide details on the 
two eDFs and how we connect them to our GRMHD simulations and refer to the work of \citet{Xiao2006,Pandya2016, Davelaar2019} for further details. The Maxwell-J\"uttner distribution function is 
given by:

\begin{equation}
\frac{dn_{\rm e}}{d\gamma_{\rm e}} = \frac{n_{\rm e}}{4 \pi \Theta_{\rm e}} \frac{\gamma_{\rm e} \sqrt{\gamma_{\rm e}^2 - 1}}{K_2\left(1/\Theta_{\rm e}\right)} \exp \left(- \frac{\gamma_{\rm e}}{\Theta_{\rm e}}\right) 
\label{eq:mjedf}, 
\end{equation}
where $n_{\rm e}$ is the electron number density, $\gamma_{e}$ is the electron Lorentz factor and $K_{2}$ is the Bessel functions of second kind. The second eDF is the kappa distribution, which 
smoothly connects a thermal distribution for small $\gamma_e$ with a power law tail for larger $\gamma_e$. The kappa distribution can be written as:

\begin{equation}
\frac{dn_{\rm e}}{d\gamma_{\rm e}} = \frac{N}{4 \pi} \gamma_{\rm e} \sqrt{\gamma_{\rm e}^2 - 1} \left(1 + \frac{\gamma_{\rm e}-1}{\kappa w}\right)^{-(\kappa+1)},
\label{eq:kappaedf}
\end{equation}

where $N$ is a normalisation factor \citep[see][for details]{Pandya2016}. The power-law exponent of a non-thermal particle distribution $dn_{\rm e}/d\gamma_{\rm e}\propto \gamma_{\rm e}^{-s}$ is 
related to the $\kappa$ value by $s=\kappa -1$. The energy of the kappa distribution is set by its width, $w$. We follow the work of \citet{Davelaar2019} and attribute besides the thermal energy 
also a fraction $\varepsilon$ of the magnetic energy to the the width of the kappa distribution:
\begin{equation}
   w:= \frac{ \kappa -3 }{\kappa} \Theta_{\rm e} + \frac{\varepsilon}{2}\left[1+\tanh(r-r_{\rm inj})\right]\, \frac{ \kappa -3 }{6 \kappa} \frac{m_{\rm p}}{m_{\rm e}} \sigma \label{eq:w} 
\end{equation}
In the equation above $r_{\rm inj}$ corresponds injection radius, i.e., the distance from where we start injecting electrons with magnetic energy contribution \citep[see also][]{Davelaar2019}.
Throughout this work we set $r_{\rm inj}=10\,M$ which is in agreement with the jet stagnation surface $\left(u^r=0\right)$ typically located between $\rm 5\,M$ and $10\,M$ 
\citep{Nakamura2018}.
In this work we assume magnetic reconnection as main particle acceleration mechanism. Since this acceleration mechanism acts on scales which we cannot resolve in our global GRMHD 
simulations we follow again \citet{Davelaar2019} and employ a particle-in-cell simulation based sub-grid model from \citet{Ball2018a}. More precisely, we use the magnetisation and plasma-$\beta$ 
dependent power-law slope of the electron distribution and connect this to the $\kappa$ value:
\begin{equation}
    \kappa :=2.8 +0.7\sigma^{-1/2} + 3.7\sigma^{-0.19}\tanh{(23.4\sigma^{0.26} \beta)}\label{eq:kappa} 
\end{equation}
Given the width, $w$, of the kappa distribution (see Eq.\ref{eq:w}) together with the numerical approximations for the emission, $j_{\nu}$, and absorption coefficients, $\alpha_{\nu}$, 
\citep[see][]{Pandya2016} 
the values for $\kappa$ are restricted to the interval $3< \kappa\leq 8$.

\begin{figure}[h!]
\centering
    \includegraphics[width=0.45\textwidth]{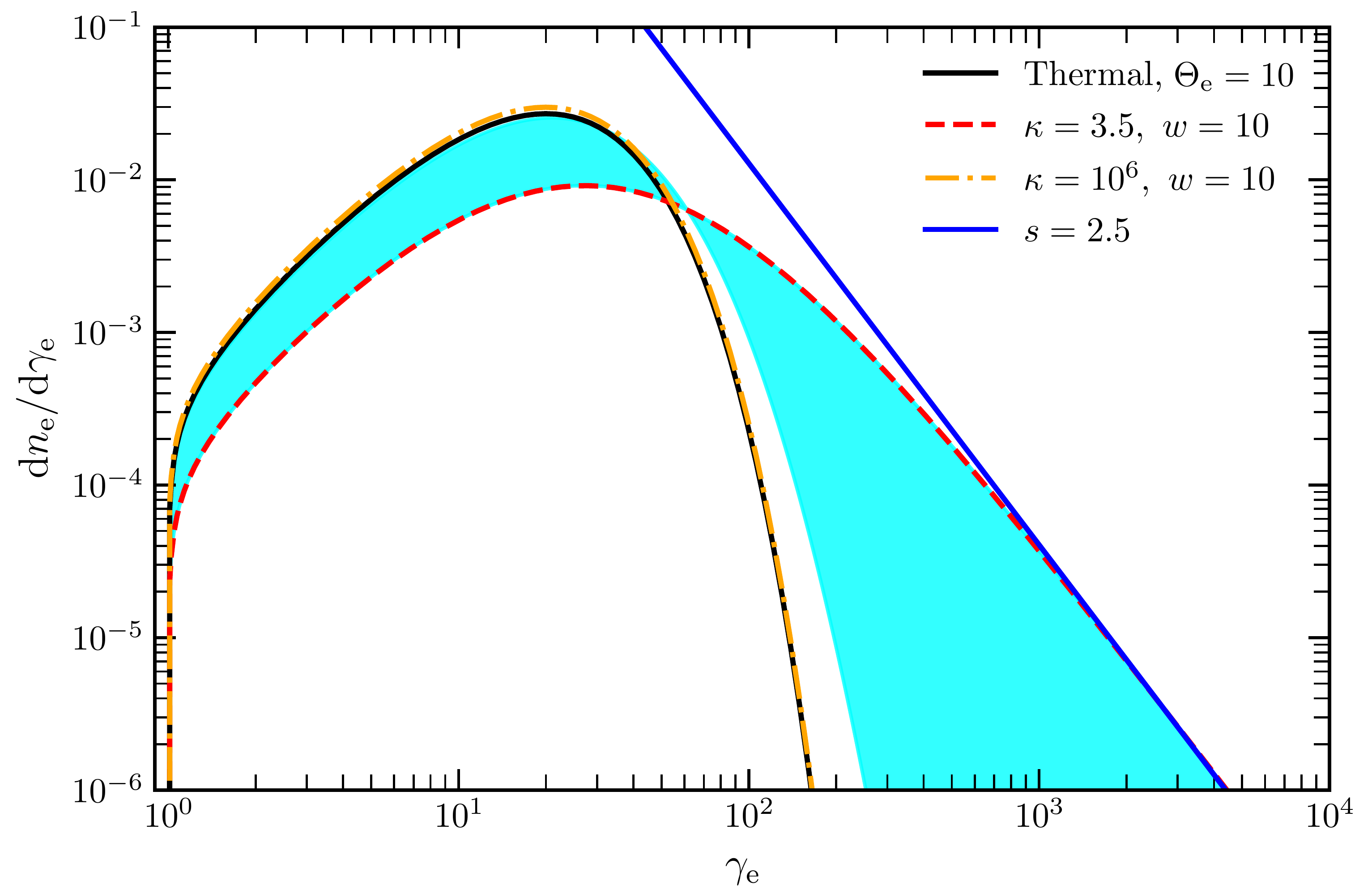}
       \caption{Comparision of eDFs: Maxwell-J\"uttner electron distribution (black), power-law electron distribution (blue) with $s=2.5$, $\kappa$ electron distribution for $\kappa=3.5$ and $w=10$ 
(red) and for $\kappa=10^{6}$ and $w=10$ (orange). For more details see text.}
    \label{fig:EF}
\end{figure}

In Figure \ref{fig:EF} we compare the Maxwell-J\"uttner and the kappa electron distribution functions. The black solid lines corresponds to a Maxwell-J\"uttner eDF with an electron temperature of 
$\Theta_e=10$. As expected from Eq. \ref{eq:mjedf} the Maxwell-J\"uttner eDF decreases exponentially for large electron Lorentz factors, $\gamma_e$. The kappa eDF for $\kappa=3.5$ and a width of 
$w=10$ is plotted as dashed red curve. The power-law tail of the kappa eDF is clearly visible for $\gamma_e>10^3$ and is well matched by a power-law with exponent $s=2.5$ shown as solid blue 
curve. Keeping the width of the kappa eDF fixed while increasing the $\kappa$ value, the kappa eDF approximates a Maxwell-J\"uttner eDF. This behaviour is clearly illustrated by the dashed 
orange curve corresponding to a kappa eDF with $w=10$ and $\kappa=10^6$. Using Eq. \ref{eq:kappa} together with typical values for the magnetisation, $\sigma$, and plasma-$\beta$ in the jet 
sheath from our GRMHD simulations (see Fig. \ref{fig:Morphology}) we find $3\leq \kappa \leq 7$. The corresponding kappa eDFs using a fixed width of w=10 are shown as cyan region in Fig. 
\ref{fig:EF}.

\begin{figure*}[h!]
\centering
     \includegraphics[width=0.65\textwidth]{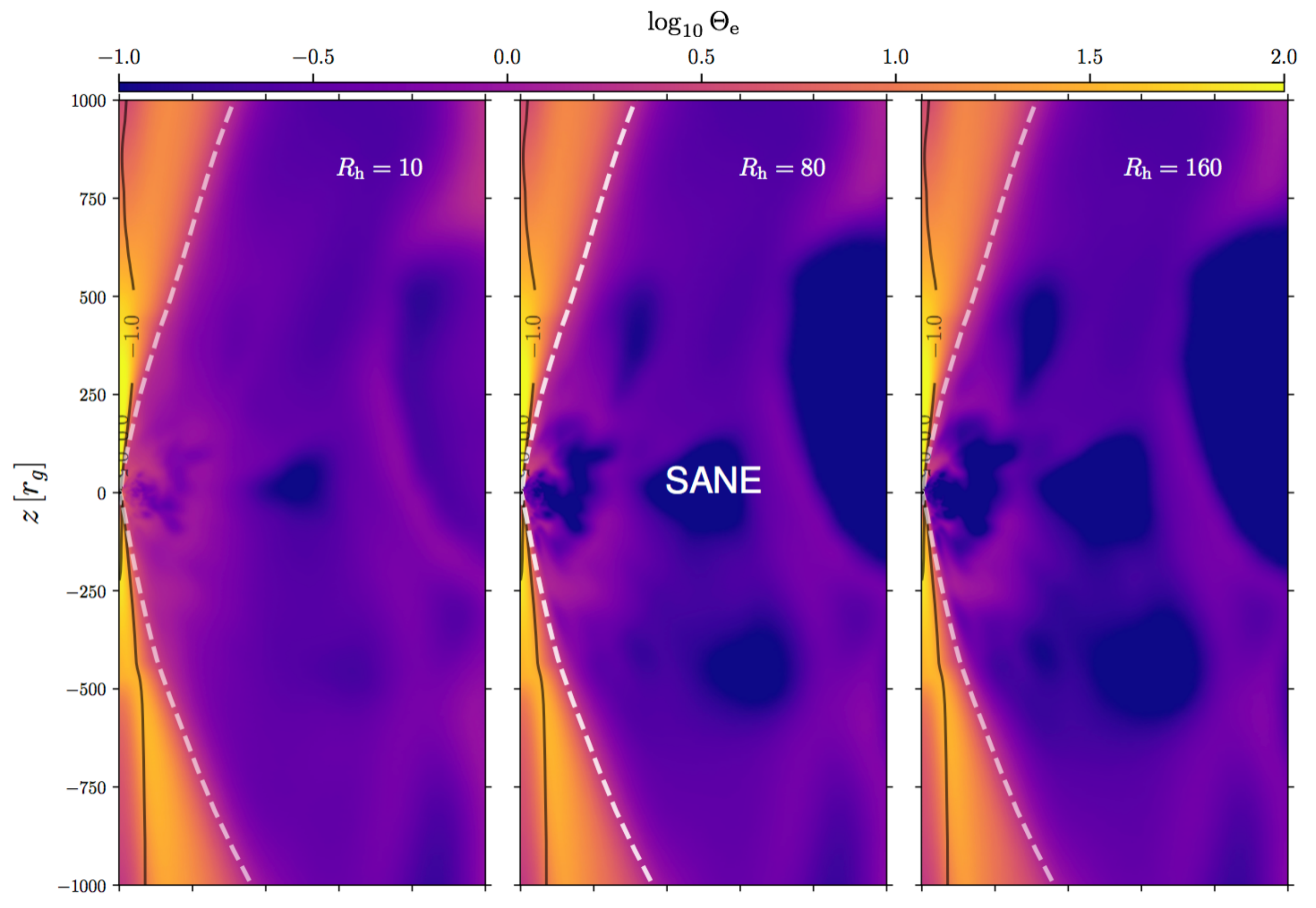}
    \vspace{-0.6cm}
     \includegraphics[width=0.65\textwidth]{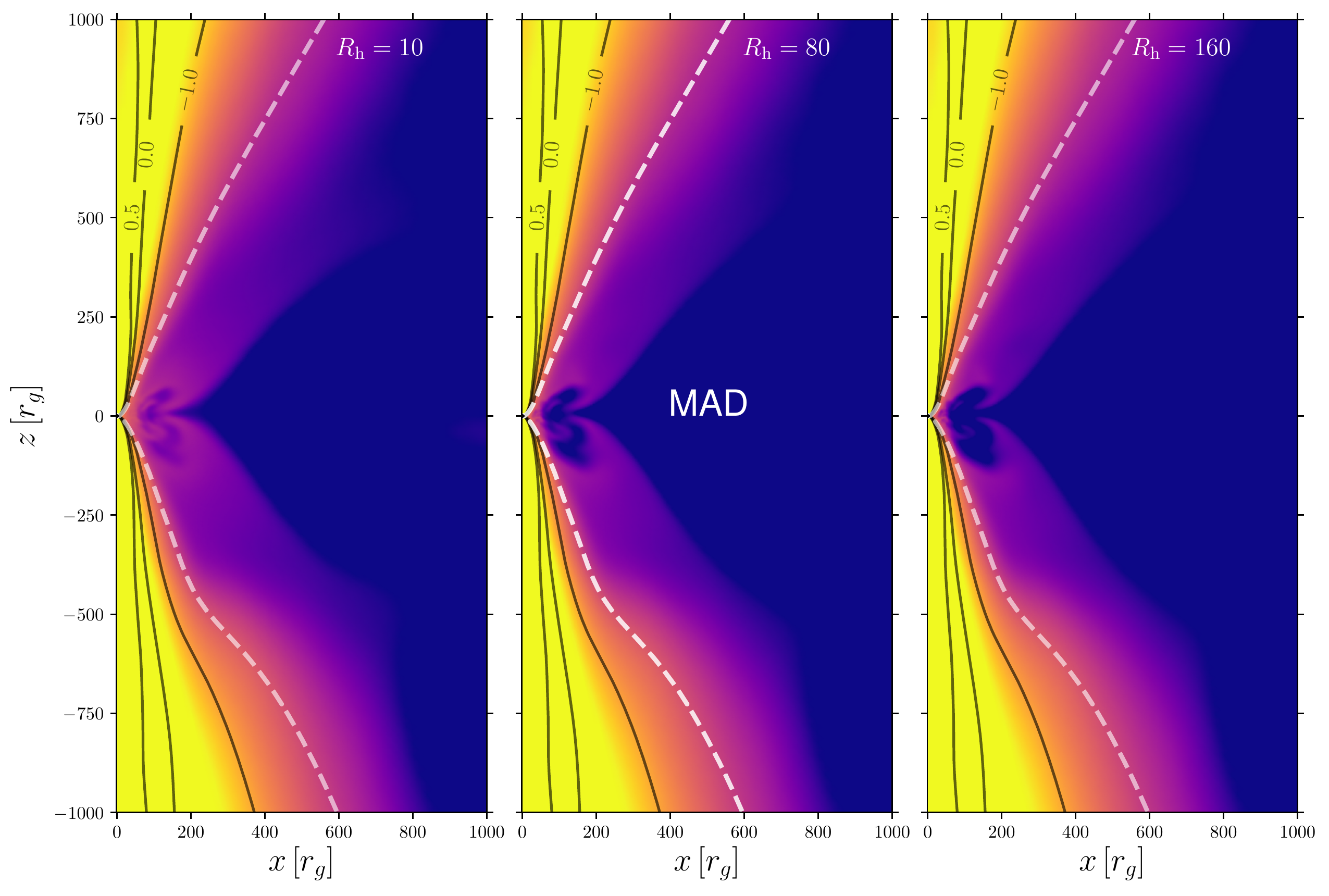}
     \vspace{0.5cm}
     \caption{Logarithm of the electron temperature, $\Theta_e$, for models \texttt{MT.S.5} (top) and \texttt{MT.M.5} (bottom) using different $R_{\rm high}$ values while keeping $R_{\rm low}=1$ fixed.}
        \label{fig:SANE_Te}
\end{figure*}

\begin{figure*}[h!]
\centering
\includegraphics[width=0.75\textwidth]{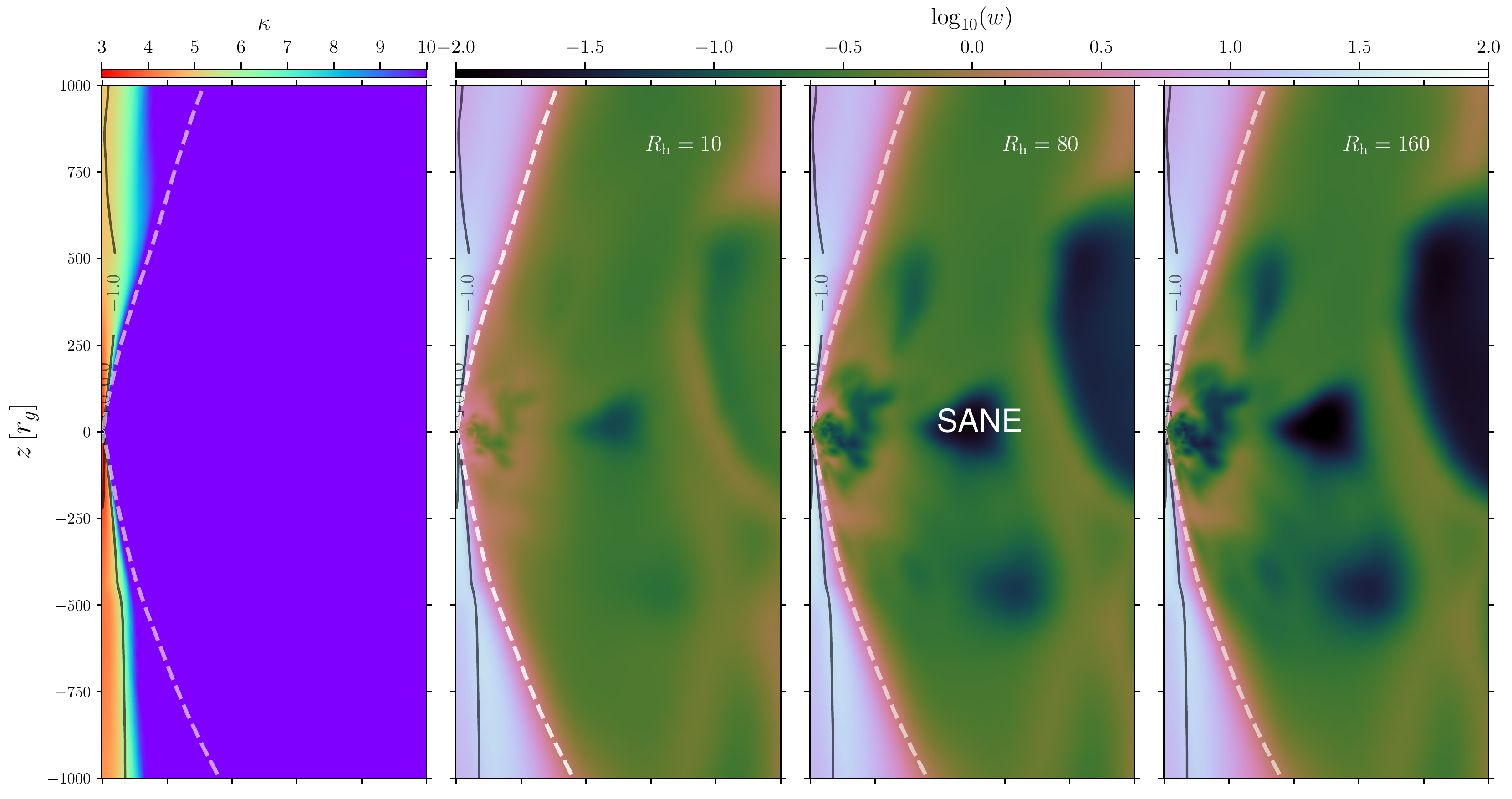}
\includegraphics[width=0.75\textwidth]{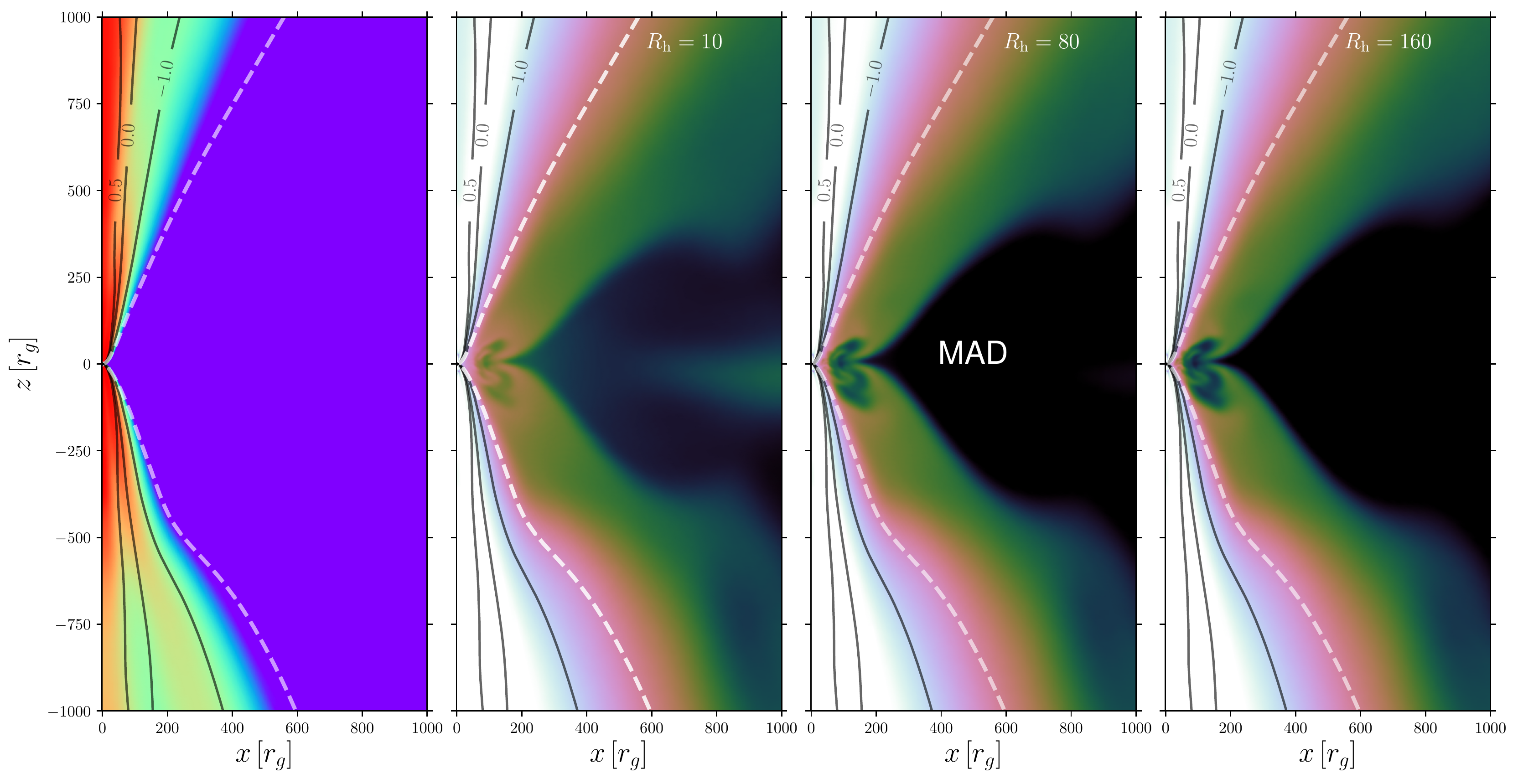}
\caption{Logarithm of the $\kappa$ value (first and fifth panel) and the width of the kappa eDF, $w$, for models \texttt{MT.S.5} (top) and \texttt{MT.M.5} (bottom) using different $R_{\rm high}$ 
values while keeping $R_{\rm low}=1$, $r_{\rm inj}=10\,M$, and $\varepsilon=0.5$ fixed.}
\label{fig:SANE_w}
\end{figure*} 

In order to interpret the results of the GRRT calculations it is important to understand effects of the adjustable parameters of the temperature description, e.g. $R_{\rm high}$, in combination with the plasma properties ($\sigma$, plasma-$\beta$) on the distribution of the defining parameters of the eDFs, i.e. the electron temperature, $\Theta_{\rm e}$, the width, $w$, as well as the $\kappa$ value. In Fig. 
\ref{fig:SANE_Te} we show the distribution of the electron temperature, $\Theta_{\rm e}$, for models \texttt{MT.S.5} (top) and \texttt{MT.M.5} (bottom) for three different values of $R_{\rm high}$ while keeping 
$R_{\rm low}=1$ fixed. The distribution of the electron temperature can be understood in the following way: According to Eq. \ref{eq:Te} the electron temperature depends on the code temperature 
$T_{\rm code}\propto p/\rho$ and the temperature ratio, $T_{\rm ratio}=1+R_{\rm high}\beta^2/(1+\beta^2)$ setting $R_{\rm low}=1$. The distribution of the code temperature and the 
plasma-$\beta$ 
are shown for models \texttt{MT.S.5} and \texttt{MT.M.5} in Fig. \ref{fig:Morphology}. From the distribution of the code temperature it is clear that the largest electron temperatures will be 
located in the jet region and that the electron temperature in the jet will be larger for the MAD model than for the SANE one (compare top and bottom panels in Fig. \ref{fig:SANE_Te}). The variation 
of the electron temperature with the $R_{\rm high}$ value can be explained by the distribution of the plasma-$\beta$ (see second and sixth panel in Fig. \ref{fig:Morphology}) and the behaviour of 
the electron-to-proton temperature ration plotted in Fig. \ref{fig:Te}. Independent of the accretion model, low plasma-$\beta$ values are found in the jet region and high values in the wind and disk 
regions. Thus, increasing the $R_{\rm high} $ value will decrease the electron-to-proton temperature ratio in disk and wind regions  while keeping the ratio in the jet region nearly unchanged (see, 
solid lines in Fig. \ref{fig:Te}). This behaviour can be best seen in the bottom row in Fig. \ref{fig:SANE_Te} by the decrease of the electron temperature in the disk region ($100 r_g\leq x\leq 300r_g$ 
and $-100r_g\leq z \leq 100r_g$). 

In Fig. \ref{fig:SANE_w} we analyse, similar to the electron temperature,  the variation and distribution of the $\kappa$ value and the width, $w$. The $\kappa$ value depends on the magnetisation, 
$\sigma$, and on plasma-$\beta$ (see Eq.\ref{eq:kappa}) and their distribution for models \texttt{MT.S.5} and \texttt{MT.M.5} are shown in Fig. \ref{fig:Morphology}. As mentioned earlier MAD 
models exhibit higher magnetisation in the jet region than their SANE counterparts. This leads in combination with Eq. \ref{eq:kappa} to smaller $\kappa$ values in the jet for the MAD models as 
compared to the SANE ones. Notice, that smaller $\kappa$ values correspond to flatter slopes in the high energy part of the kappa eDF and thus more particles with large electron Lorentz factors, 
$\gamma_e$ (see Fig. \ref{fig:EF}). The second important parameter for the kappa eDF is its width, $w$. The width is computed from the electron temperature, $\Theta_{\rm e}$, the $\kappa$ value together 
with some fraction $\varepsilon$ of the magnetic energy (see Eq. \ref{eq:w}). Since the fraction of the magnetic energy is typically $<1$ the distribution of $w$ follows mainly the distribution of the 
electron temperature (compare Fig. \ref{fig:SANE_Te} and Fig. \ref{fig:SANE_w}). It is important to mention again that we only apply the kappa eDF in regions where $3<\kappa\leq 8$ due to the 
limitation of the numerical approximations for the emission and absorption coefficients. Outside of this range we switch back to a Maxwell-J\"uttner distribution and the corresponding emission and 
absorption coefficients \citep{Pandya2016}. This implies that the kappa eDF is mainly used in the jet sheath region.

\section{Parameter space study}\label{sec:results}
Given our large set of GRMHD models i.e. accretion type and spin (see Table \ref{tab:TorusID}) together with the different choices of the electron micro-physics and hyper-parameters, e.g., 
$\sigma_{\rm 
cut}$, during the GRRT it is necessary to explore the parameter space in a robust and agnostic manner. However, given the computational costs we restrict ourselves to a small set of 
parameters which allow us to explore the hyper-surfaces of the parameter space in adequate and scientifically sufficient way while keeping the computational costs within a affordable range. In addition to the different spins and accretion models (MAD and SANE) we list in 
Table \ref{tab:paraspace} the free GRRT parameters of our study and their explored range. 

\begin{table}[h!]
\caption{GRRT Parameters and their values explored during the parameter space study}   
\centering
\begin{tabular}{cc|cc}
\hline
 parameter &  values & parameter & values\\
\hline
\hline
$R_{\rm low}$ & 0.1,1 & $R_{\rm high}$ & 10, 80, 160 \\
$\sigma_{\rm cut}$ & 0.1,0.5,1,3,5,10 & $\varepsilon$ & 0, 0.25, 0.5, 0.75, 1\\
eDF & thermal, kappa &$i$ & 160$^\circ$ \\
$r_{\rm inj}$ & 10 M & & \\
\hline
\end{tabular}
\label{tab:paraspace}
\end{table} 

During the GRRT we compute both the broad-band spectrum from $10^9$ Hz to $10^{16}$ Hz and ray-traced 86\,GHz images. This allows us to compare the spectral properties of our models with 
multi-frequency observations of M\,87 (see Table \ref{Tab:obsSED}) and at the same time contrast the 86 GHz jet structure with the one obtained via Global Millimetre VLBI (GMVA) observations of 
M\,87 \citep{Kim2018a}. As mentioned earlier the GRMHD simulations are scale-free and we apply a black hole mass of  $6.5\times10^9\,\mathrm{M}_{\odot}$ and a distance of $16.8\,{\rm Mpc}$ 
\citep{EHT_M87_PaperI} during the radiative transfer to adjust our simulations to M\,87. In addition we use a fixed inclination angle of $160^\circ$ and iterate for each model the mass accretion 
rate until we obtain an average flux density of 1.0\,Jy at 230\,GHz \citep{Doeleman2012,Akiyama2015} within a time window of 2000\,M. This time window is taken between  $13000\,M\leq t\leq 
15000\,M$ for the MAD models and between $8000\,M\leq t\leq 10000\,M$ for the SANE ones ensuring that the GRMHD simulations are well within the the quasi-stable state (see Fig. 
\ref{fig:rates}).

In the following we will investigate the influence of:
\begin{itemize}
    \item the electron temperature via $R_{\rm low,\,high}$  
    \item the electron distribution function, i.e. Maxwell-J\"uttner (thermal) and kappa distribution (non-thermal)   
    \item the jet spine via the magnetisation cut-off $\sigma_{\rm cut}$ 
    \item the fraction of the magnetic energy contributing to the total energy contained in the kappa distribution via $\varepsilon$
    \item the black hole spin $a_{\star}$
\end{itemize}

on the spectral and the structural properties of our GRRT calculations. For the comparison between the theoretical and observed structure of M\,87 at 86\,GHz we used the super-resolved stacked 
86GHz GMVA image presented in \citep{Kim2018a}. In order to mimic the GMVA observations we rotated the average GRRT image to a position angle\footnote{We define the angles relative to 
North with positive sign to the East, i.e. counterclockwise, to be consistent with the notation of the observations.} of 288$^\circ$\citep{Walker2018} and convolved the image with a beam of 0.123 mas $\times$\,0.051 mas at a position angle of 0$^\circ$. After the convolution we rotated the image by -18$^\circ$, i.e., the jet axis coincide with the x-axis,  and sliced the jet perpendicular to the x-axis. During the slicing we limit the dynamical range to $10^4$ and only considered flux profiles with peak flux five times above the noise level for the jet diameter analysis. We define as the jet 
diameter the location where the flux density profile is above 50$\%$ of peak flux. The obtained profiles for the jet diameter, $D_{\rm jet}$ and the apparent opening angle, $\phi_{\rm app}$, along  the jet axis are compared to the ones from the GMVA observations. %\left(D_{\rm jet,\,obs}\propto r^{0.47,0.58}\right).

\subsection{Influence of the Electron Temperature}\label{sec:Te}
In our first scan of the parameter space we analyse the influence of the electron temperature, $\Theta_{\rm e}$ on the spectrum and the 86\,GHz image structure. 

\begin{figure*}[h!]
\centering
\includegraphics[width=0.49\textwidth]{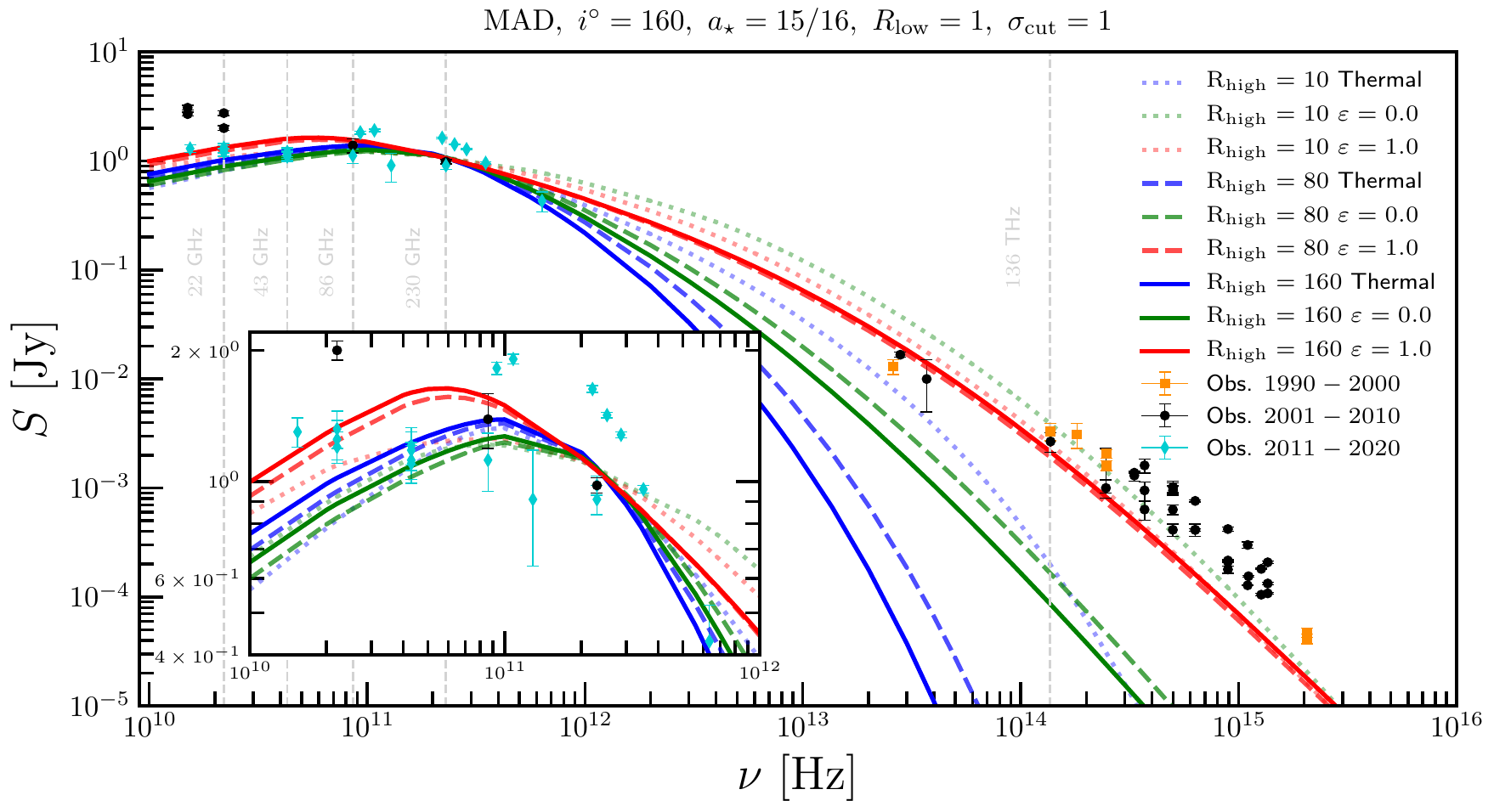}
\includegraphics[width=0.49\textwidth]{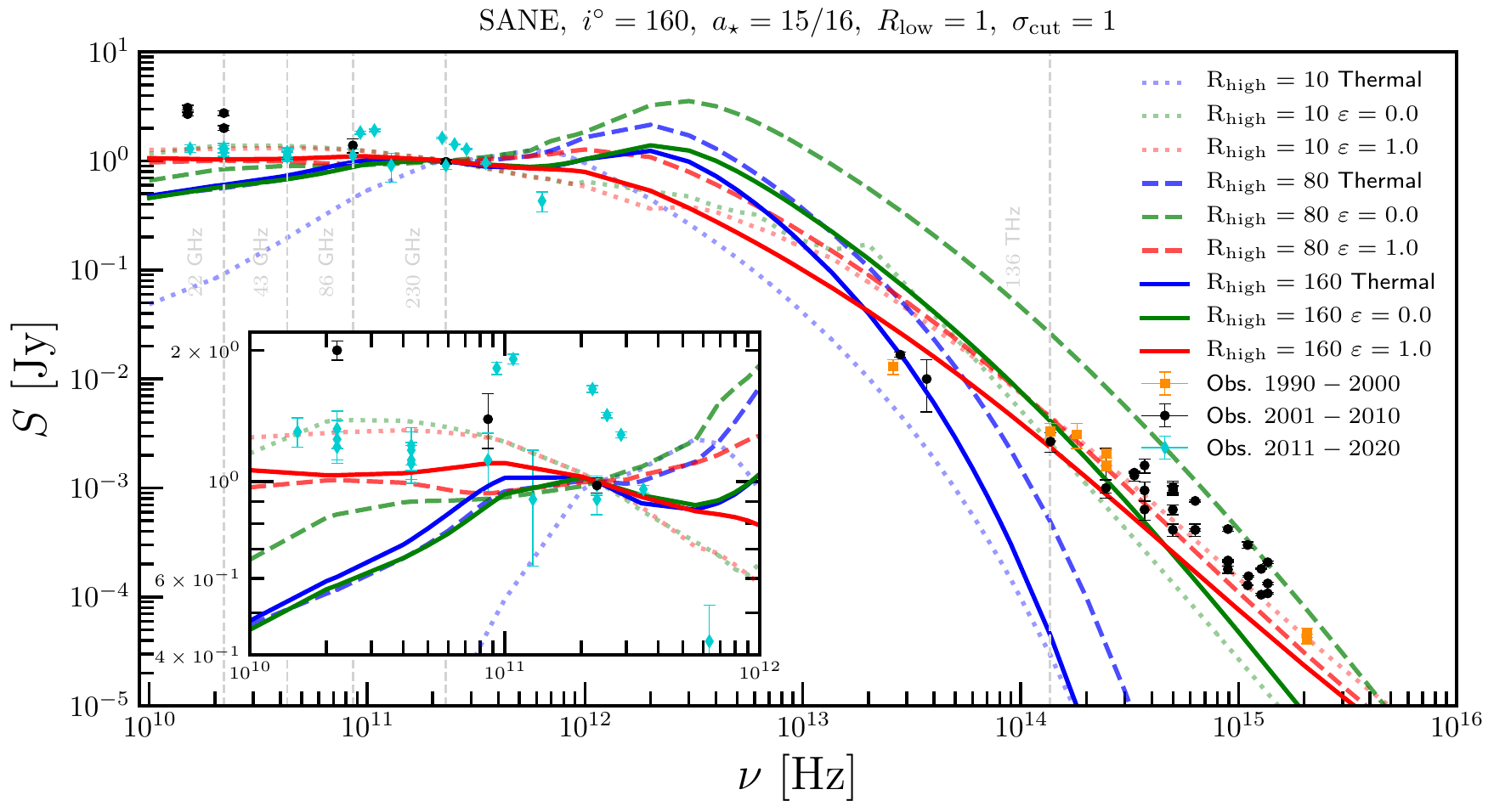}
\includegraphics[width=0.49\textwidth]{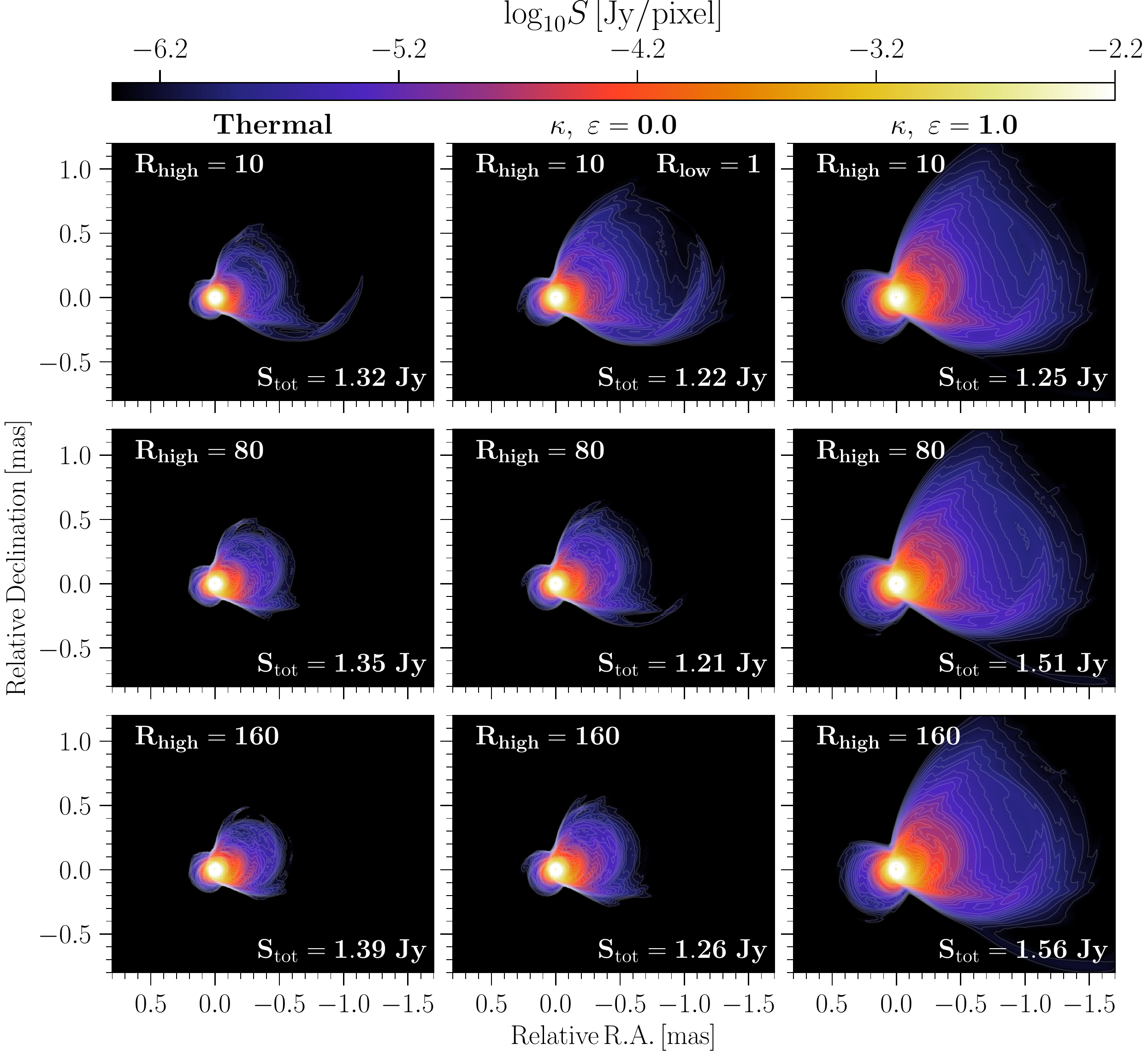}
\includegraphics[width=0.49\textwidth]{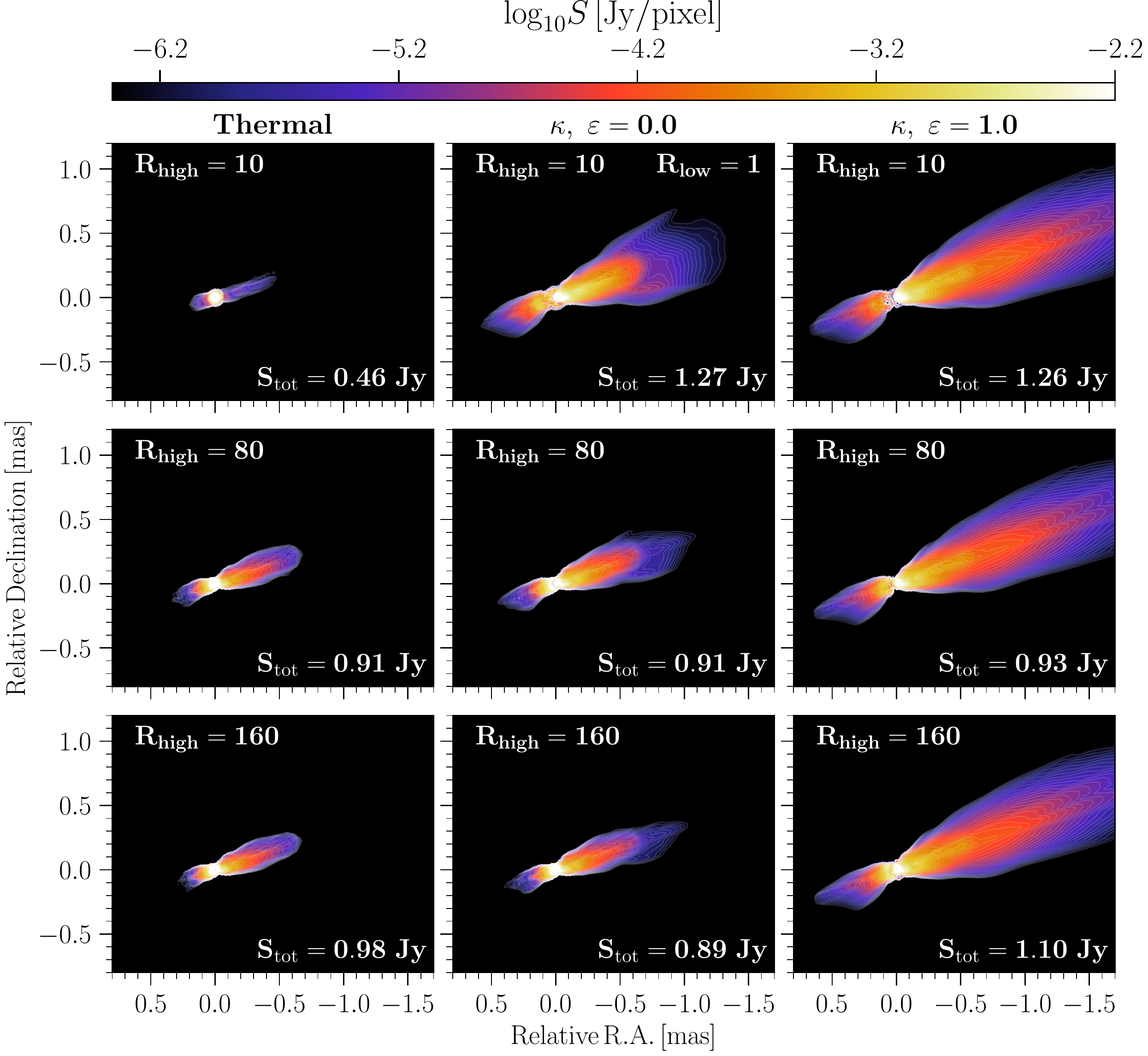}
\includegraphics[width=0.49\textwidth]{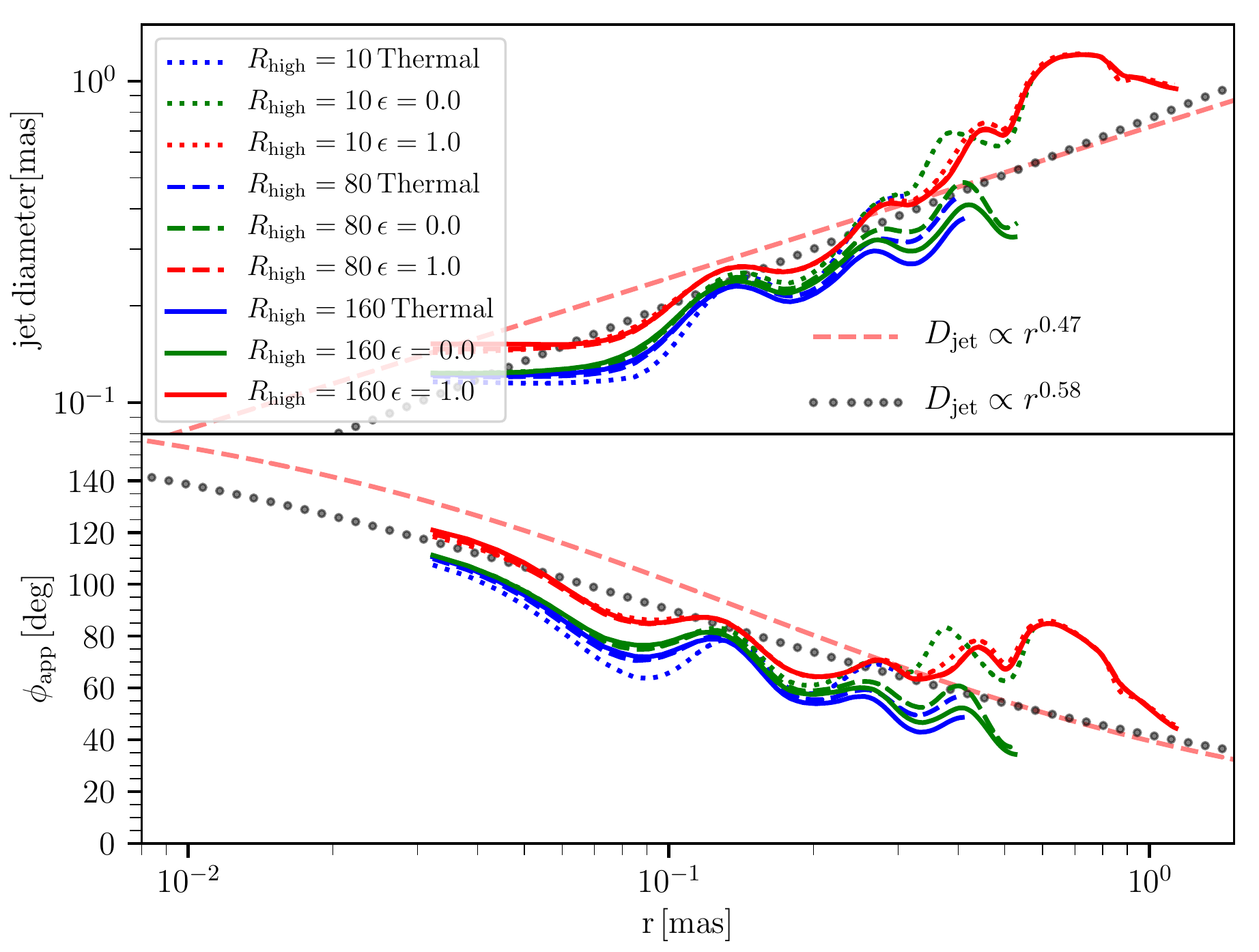}
\includegraphics[width=0.49\textwidth]{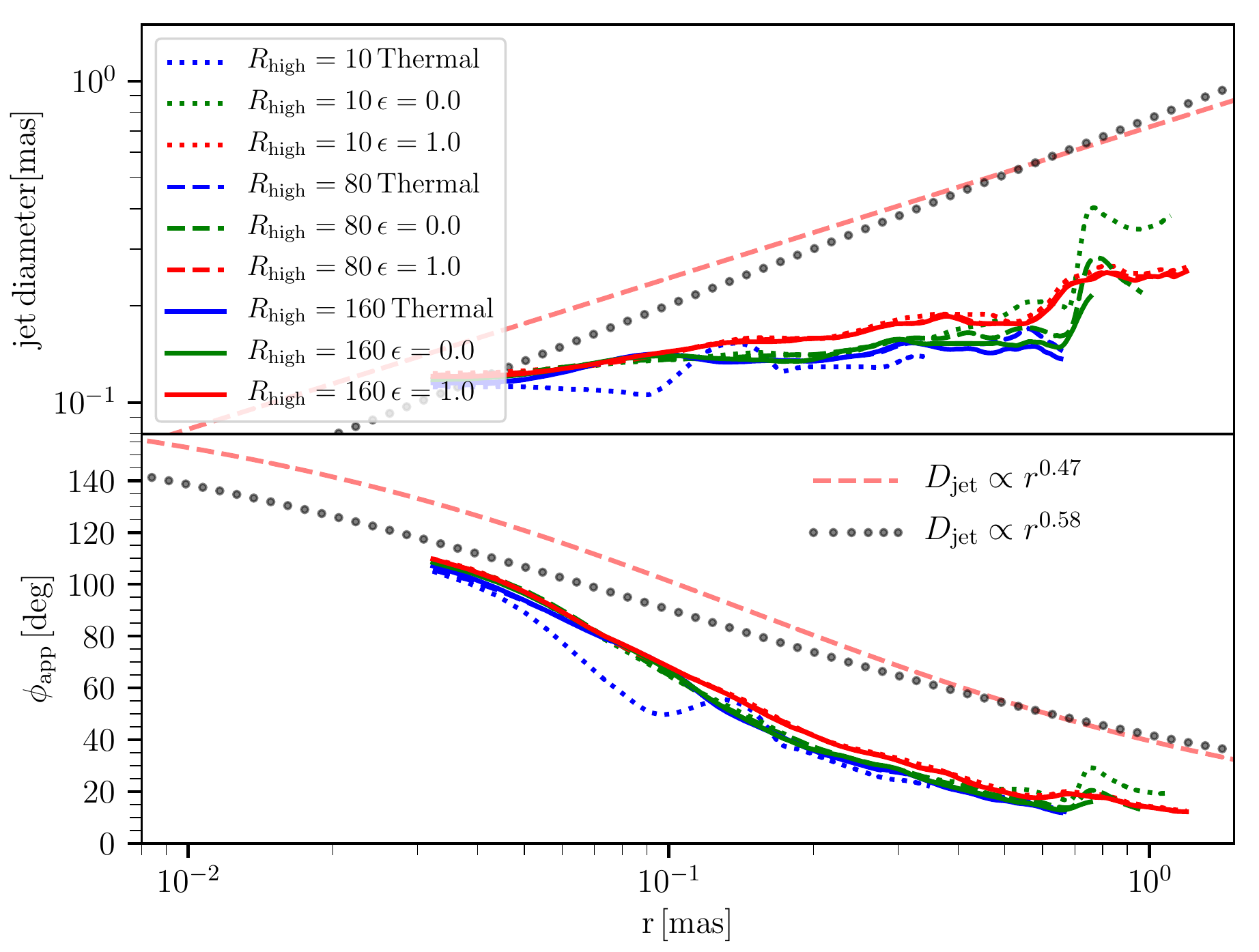}
\caption{The influence of the $R_{\rm high}$ parameter on the spectral and structural properties of our M\,87 simulations for a black hole with spin $a_{\star}=0.9375$. The left column corresponds to 
a MAD model and the right column to a SANE one. The top panels show the broad-band spectrum averaged for 2000\,M for $R_{\rm high}$=10 (dotted), $R_{\rm high}$=80 (dashed)  and $R_{\rm high}$=160 (solid) while keeping $R_{\rm low}=1$ 
fixed. The different eDFs are marked by different colors: thermal (blue), kappa with $\varepsilon=0$ (green) and kappa with $\varepsilon=1$ (red). The middle image shows the ray-traced 86\,GHz images for three different eDFs (columns) and 
different $R_{\rm high}$ (rows). The bottom panels show the jet diameter and the opening angles profiles using the same color and line styles as for the spectrum.} 
\label{fig:Rhsweep}
\end{figure*}

\subsubsection{Influence of the $R_{\rm high}$ parameter}
We first focus on the influence of the $R_{\rm high}$ parameter using values of 10, 80, and 160 while keeping $R_{\rm low}=1$ fixed. Notice, that the modelling of 230\,GHz EHT observations 
showed at $R_{\rm high}=1$ for both MAD and SANE models is not in agreement with observational constraints \citep[see Table 3 in ][]{EHT_M87_PaperV}. This statement was obtained for thermal models. However, within our GRRT setup we only apply the kappa-eDF in the jet region. Thus, the disk region includes a thermal eDF as in the case of \citet{EHT_M87_PaperV}. For this reason we excluded $R_{\rm 
high}=1$ in our parameter sweep. Furthermore, we use $\sigma_{\rm cut}=1$ and three different eDFs, namely thermal, kappa eDF with $\varepsilon=0$ and with $\varepsilon=1$. Together with 
the three $R_{\rm high}$ values and the two different accretion models (MAD and SANE) this leads to 18 models explored here and the results for a black hole with spin $a_{\star}=0.9375$ are presented 
in Fig. \ref{fig:Rhsweep}.

The influence of the $R_{\rm high}$ parameter on the broad band spectrum can be best seen for the thermal models (Maxwell-J\"uttner eDF) in the MAD case (top left panel, blue curves in Fig. 
\ref{fig:Rhsweep}). The high-frequency part of the spectrum ($\nu>10^{12}$ Hz) steepens significantly with increasing $R_{\rm high}$ while the low frequency part ($\nu\leq 10^{12}$ Hz) slightly 
increases in flux without a change in the spectral slope (see inset in top left panel). Given that the high energy emission is mainly emitted from the disk region predominated by high plasma-$\beta$ 
values, the electron temperature can be written as $\Theta_{\rm e}\propto 1/R_{\rm high}$\footnote{only listing the dependence on $R_{\rm high}$} (see Eq. \ref{eq:Te}). Thus increasing the $R_{\rm high}$ 
value lowers the electron temperature, $\Theta_{\rm e}$, in the disk (see Fig. \ref{fig:Te} and Fig. \ref{fig:SANE_Te}). Using the relation between the electron temperature and the $R_{\rm high}$ parameter, 
the thermal emissivity in the high frequency regime is given by
\begin{equation}
    j_{\nu,\rm th}\propto \exp\left(-R_{\rm high}^{2/3}\nu^{1/3}\right),
    \label{eq:jnuthapprox}
\end{equation} 
which finally explains both, the exponential decrease with frequency and the variation with $R_{\rm high}$ for the high frequency part of the spectrum.

In a similar manner we can explain the low frequency behaviour of the spectrum. Assuming that the low frequency emission is generated in the jet and disk wind regions which are characterised by 
low plasma-$\beta$ values the temperature ratio is not affected by the choice of $R_{\rm high}$ and thus the electron temperature is not altered (see Eq. \ref{eq:Te}, Fig. \ref{fig:Te} and Fig. 
\ref{fig:SANE_Te}). 

Next, we turn to the comparison between eDFs. Notice, that the kappa eDF is only applied within the jet and elsewhere we use a thermal eDF. Thus, we need to consider the emissivities and 
absorptivities from both eDFs during the radiative transfer depending on the regions the ray crosses.
Again, we focus first on the high energy part of the spectrum and use for the kappa emissivities the approximations of \citet{Pandya2016}. Thus we can write the kappa emissivity as: 
\begin{equation}
   j_{\nu,\,\kappa}\propto \nu^{-(\kappa-2)/2} (w\kappa)^{\kappa-2}
\end{equation}
Inserting the definition of the width of the kappa eDF (Eq. \ref{eq:w}) the kappa emissivity can be approximated by\footnote{ignoring the injection radius dependency for reasons of simplicity}:
\begin{equation}
    %j_{\nu,\,\kappa}\propto \nu^{-(\kappa-2)/2} \left[(\kappa -3)\Theta_{\rm e} + \varepsilon(\kappa -3) \sigma\right]^{\kappa-2}
    j_{\nu,\,\kappa}\propto \nu^{-(\kappa-2)/2} \left[\Theta_{\rm e} + \varepsilon \sigma\right]^{\kappa-2} 
\end{equation}
Together with the thermal emissivitiy the total emissivity has the simplified form:
\begin{equation}
    %j_{\nu,\rm tot}\propto\exp\left(-T_e^{-2/3}\nu^{1/3}\right)+\nu^{-(\kappa-2)/2} \left[(\kappa -3)T_{\rm e} + \varepsilon(\kappa -3) \sigma\right]^{\kappa-2}
    j_{\nu,\rm tot}\propto\exp\left(-T_e^{-2/3}\nu^{1/3}\right)+\nu^{-(\kappa-2)/2} \left[\Theta_{\rm e} + \varepsilon\sigma\right]^{\kappa-2}
    \label{eq:jnukapproxTe}
\end{equation}
If we consider that the high energy thermal emission originates mainly from the disk (high plasma-$\beta$ region) and that the jet sheath\footnote{using a $\sigma_{\rm cut}=1$ we exclude 
emission from jet spine and thus can neglect radiation from this region} has intermediate plasma-$\beta$ values the equation above can be further simplified and the dependence on the $R_{\rm 
high}$ can be included:
\begin{equation}
    %j_{\nu,\rm tot}\propto\exp\left(-R_{\rm high}^{2/3}\nu^{1/3}\right)+\nu^{-(\kappa-2)/2} \left[(\kappa -3)/R_{\rm high} + \varepsilon(\kappa -3) \sigma\right]^{\kappa-2}
    j_{\nu,\rm tot}\propto\exp\left(-R_{\rm high}^{2/3}\nu^{1/3}\right)+\nu^{-(\kappa-2)/2} \left[1/R_{\rm high} + \varepsilon\sigma\right]^{\kappa-2}
    \label{eq:jnukapprox}
\end{equation}
With this simplification of the emissivity at hand, we can understand the spectral behaviour in the high frequency regime for the thermal--non-thermal eDFs. First we concentrate on the 
$\varepsilon=0$ 
models (green curves in top panels of Fig. \ref{fig:Rhsweep}). Including the kappa eDF leads to a power-law tail in the high frequency emission as compared to exponential decay in 
the thermal model which is set by the second term in Eq. \ref{eq:jnukapprox}, i.e., the contribution of the non-thermal electrons with large $\gamma_e$ within the tail of the kappa eDF. For the 
$\varepsilon=0$ 
models the $R_{\rm high}$ dependence of the spectrum is still visible i.e., steepening of the spectrum with increasing $R_{\rm high}$. This effect is introduced by the $R_{\rm high}$ 
depending terms in Eq.~\ref{eq:jnukapprox}. For the $\varepsilon=1$ models there is no clear $R_{\rm high}$ dependence on the high frequency spectrum (see red curves in top left panel of Fig. 
\ref{fig:Rhsweep}. This can be explained by the second term in Eq.~\ref{eq:jnukapprox} which includes the contribution from the magnetic field, i.e. the non-thermal particles gain additional energy 
from the magnetic field. With $\varepsilon=1$ the second term in Eq.~\ref{eq:jnukapprox} is the dominating term in the emissivity and thus the dependence on $R_{\rm high}$ is no longer visible in 
the high frequency emission.

The relations and interpretation of the spectral behaviour above apply also to the SANE models. However, the trends are less clear as in the MAD case. The reason for this are connected to the less magnetised structure of the SANE simulations (see Fig. \ref{fig:Morphology}). In addition the required mass accretion rate during the flux normalisation process varies 
strongly for the SANE models across both, the different eDFs and $R_{\rm high}$ values. The turnover frequency, $\nu_{\rm t,th}$, for thermal emission can be approximated as $\nu_{\rm 
t,th}\propto 
T_e^2 B$  \citep[see, e.g.,][]{Zdziarski1998}. Inserting the scaling relations from code units to cgs units together with the $R_{\rm high}$ depending electron temperature leads to:
\begin{equation}
 \nu_{\rm t,th}\propto B_{\rm code}\sqrt{\dot{m}}/R_{\rm high}^2.
 \label{eq:turnoverfreq}
\end{equation} Thus, changes in the mass accretion rate and $R_{\rm high}$ value lead to shifts in the turnover position. The strong variations in the mass accretion rate are not found for the MAD 
models which explains the nearly identical turnover positions in the MAD cases. Notice, that the turnover frequency for the SANE models with $R_{\rm high}=10$ is located around $10^{11}$ Hz as 
compared to $\sim 10^{12}$ Hz for larger values of $R_{\rm high}$. Taking the dependence of the turnover frequency on $\dot{m}$ and $R_{\rm high}$ in account together with 
Eqs.\ref{eq:jnuthapprox} 
and \ref{eq:jnukapprox} the spectral behaviour seen in the SANE models can be explained in a similar manner as for the MAD ones (top right panel in Fig. \ref{fig:Rhsweep}).

So far our discussion was focused on the spectrum and in the following we will elaborate the influence of the $R_{\rm high}$ parameter and the eDF on the 86\,GHz image structure. In the middle panel of 
Fig. \ref{fig:Rhsweep} we present the 86\,GHz images from our models for M\,87 for different $R_{\rm high}$ values (columns) and different eDFs (rows). The overall structure of the 86\,GHz 
images can be described by a bright innermost region and a fainter jet. The innermost emission ($r<0.1\,{\rm mas}$ for the MAD models is mainly generated in the disk region and  is nearly independent of the 
choice of $R_{\rm high}$. In case of the SANE models the innermost emission for low $R_{\rm high}$ values is produced in the disk and for large $R_{\rm high}$ values the counter-jet contributes the majority of the emission  \citep[see also Fig. 11 in][]{EHT_M87_PaperV}. On the other hand, the extend of the jet depends on the eDF and the $R_{\rm high}$ parameter. From the spectrum we 
can see that all MAD models are optically thin at 86\,GHz (see inset in left top panel) and we can use the approximation of the thermal emissivity (Eq. \ref{eq:jnuthapprox}) to qualitatively 
understand the obtained emission structure. The jet emission could be understood as a blend of radiation generated in the jet sheath and disk wind. These two regions are characterised by 
plasma-$\beta$ values $\sim 0.1$ (see Fig. \ref{fig:Morphology}) and emission roughly follows Eq.\ref{eq:jnuthapprox}. Thus, the shortening of the jet and the steepening of the broad-band 
spectrum with increased $R_{\rm high}$ values can be explained.

Including non-thermal electrons via the kappa eDF in the jet sheath leads to wider and more extended jets (see first row in middle panel of Fig. \ref{fig:Rhsweep}). Similar to the change in the spectral 
behaviour between thermal and non-thermal eDF we can explain this behaviour by Eq. \ref{eq:jnukapprox}. The addition of non-thermal electrons with large electron Lorentz factors, $\gamma_e$, 
increases the emission in the jet sheath at large distances from the black hole. For the model with $\varepsilon=0$ increasing the $R_{\rm high}$ value leads to shorter and narrower jet. This is 
expected from Eq. \ref{eq:jnukapprox} and is in agreement with our explanation for the spectral properties for this model (green curves in the top left panel of Fig. \ref{fig:Rhsweep}). Adding the 
magnetic energy of plasma as additional energy source for the non-thermal particles further enhances the emission of the jet on large scales and increases the jet width. Analog to the discussion of 
the spectral behaviour for the $\varepsilon=1$ models, there is no significant dependence on the $R_{\rm high}$ parameter (see red curves in top left panel and right column in left middle panel in 
Fig. \ref{fig:Rhsweep}). The 86\,GHz image structure of the SANE models (right middle panel in Fig. \ref{fig:Rhsweep}) follow the same discussion as above for the MAD models. The only 
exception is the 86\, GHz thermal image for $R_{\rm high}=10$. In this case at source is optically thick at 86\,GHz as compared to thermal $R_{\rm high}>10$ (see dotted blue curve in the inset in top right panel of Fig. \ref{fig:Rhsweep}) and characterised by a low total flux density as compared to the other models.

In the bottom panels of Fig. \ref{fig:Rhsweep} we compare the jet diameter and opening angle, $\phi_{\rm app}$, from our models with the results from the GMVA observation of M\,87. The faint red and black dotted curves 
corresponds to jet diameter fits to the GMVA observations \citep{Kim2018a}. Independent of the electron temperature and eDF all MAD models exhibit opening angle profiles which are in 
agreement with the observations. Whereas the opening angle for the SANE models decreases faster than the observations with distance from the black hole. As mentioned above the inclusion of 
non-thermal particles slightly increases the jet opening angle and significantly enhances the emission on large scales. 

\begin{figure*}[h!]
\centering
\includegraphics[width=0.49\textwidth]{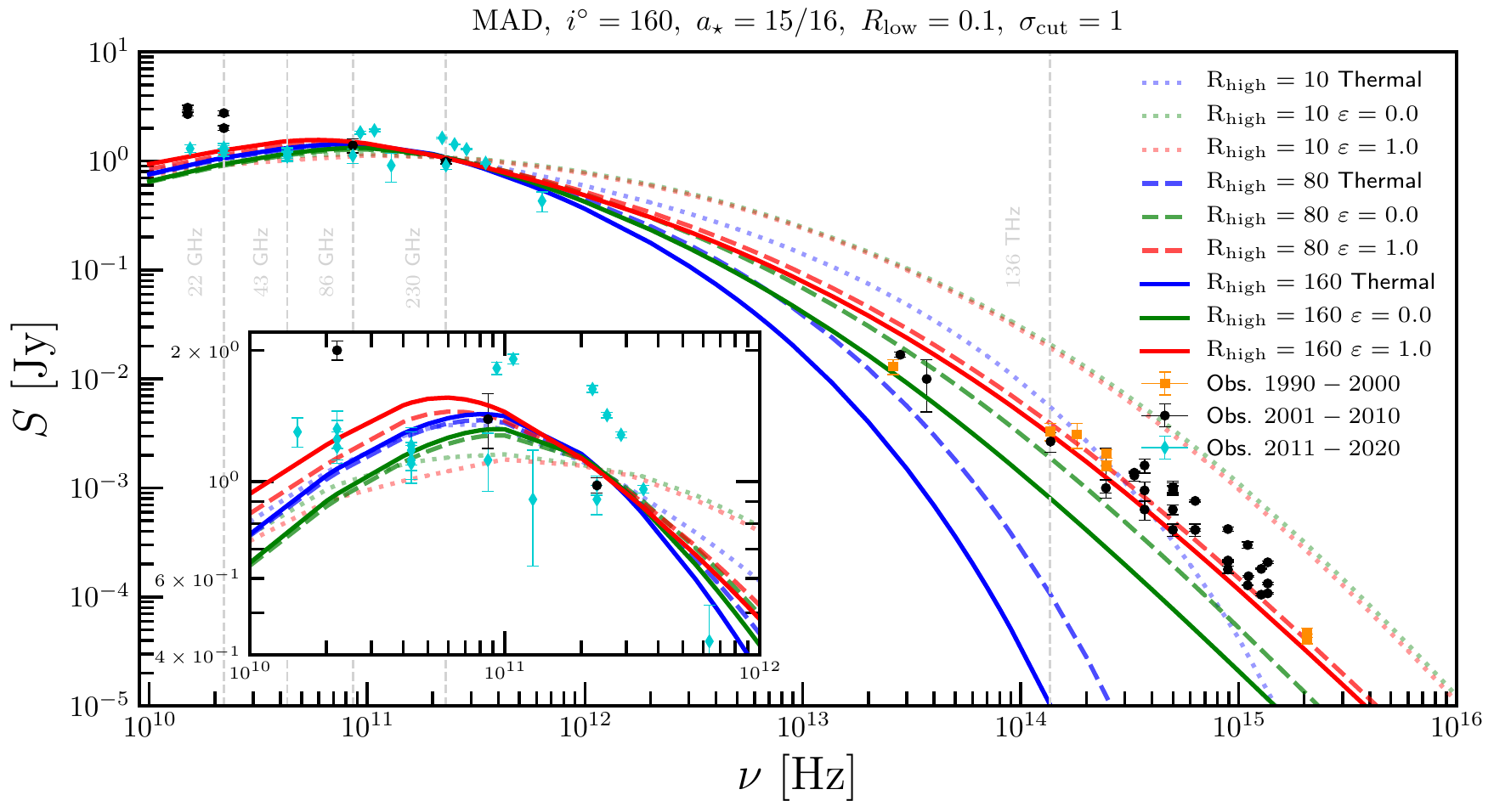}
\includegraphics[width=0.49\textwidth]{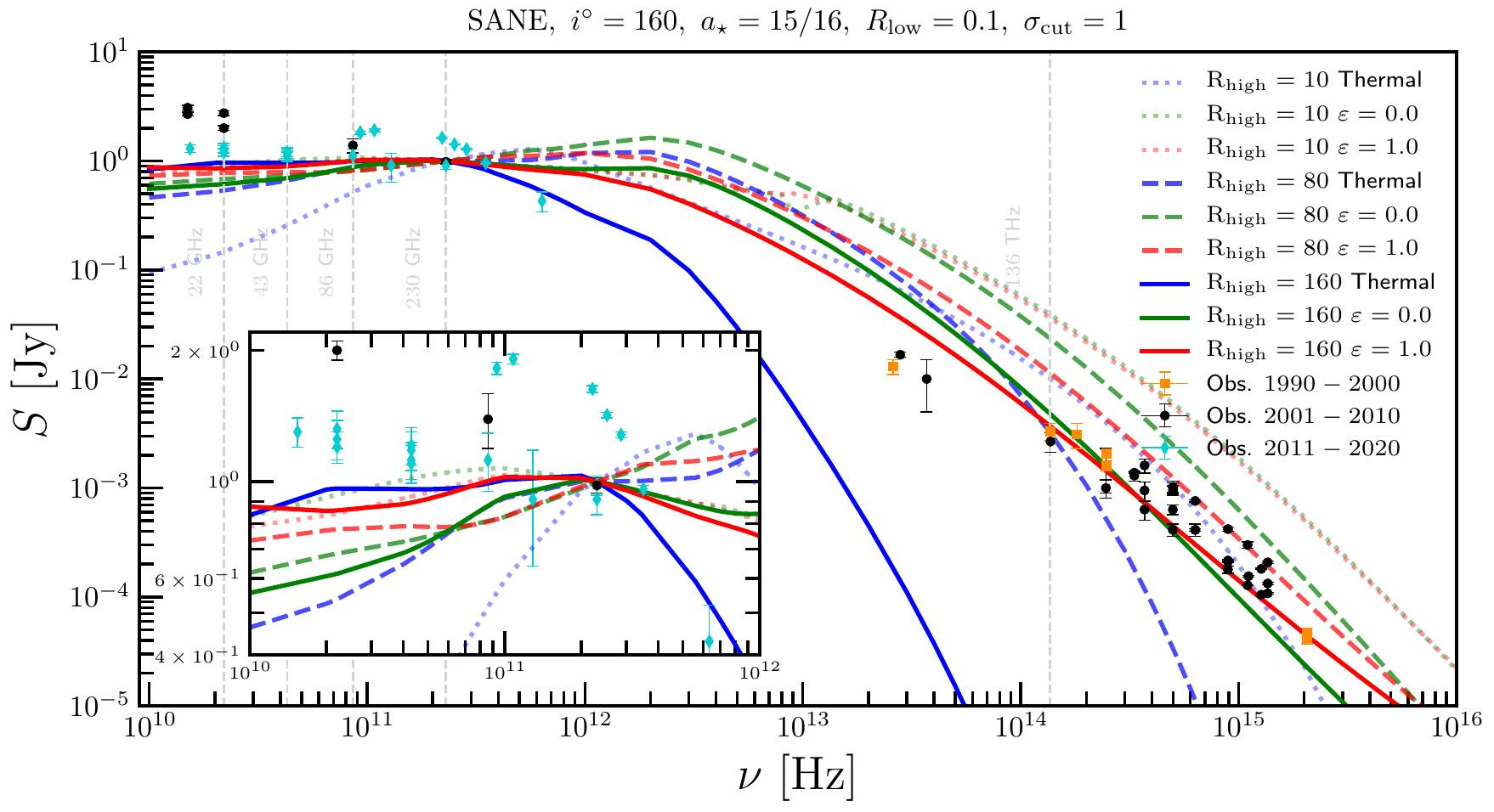}
\includegraphics[width=0.49\textwidth]{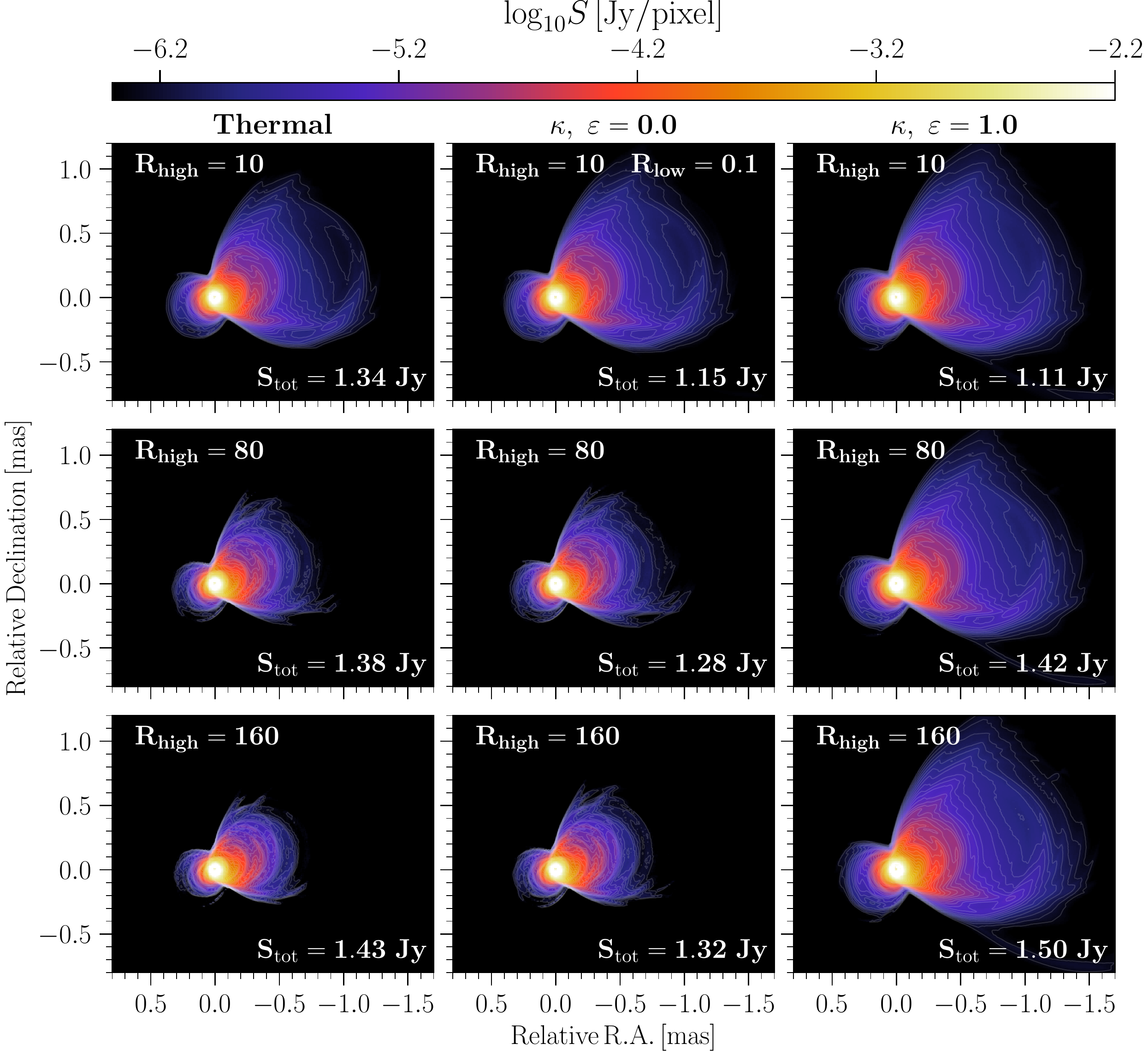}
\includegraphics[width=0.49\textwidth]{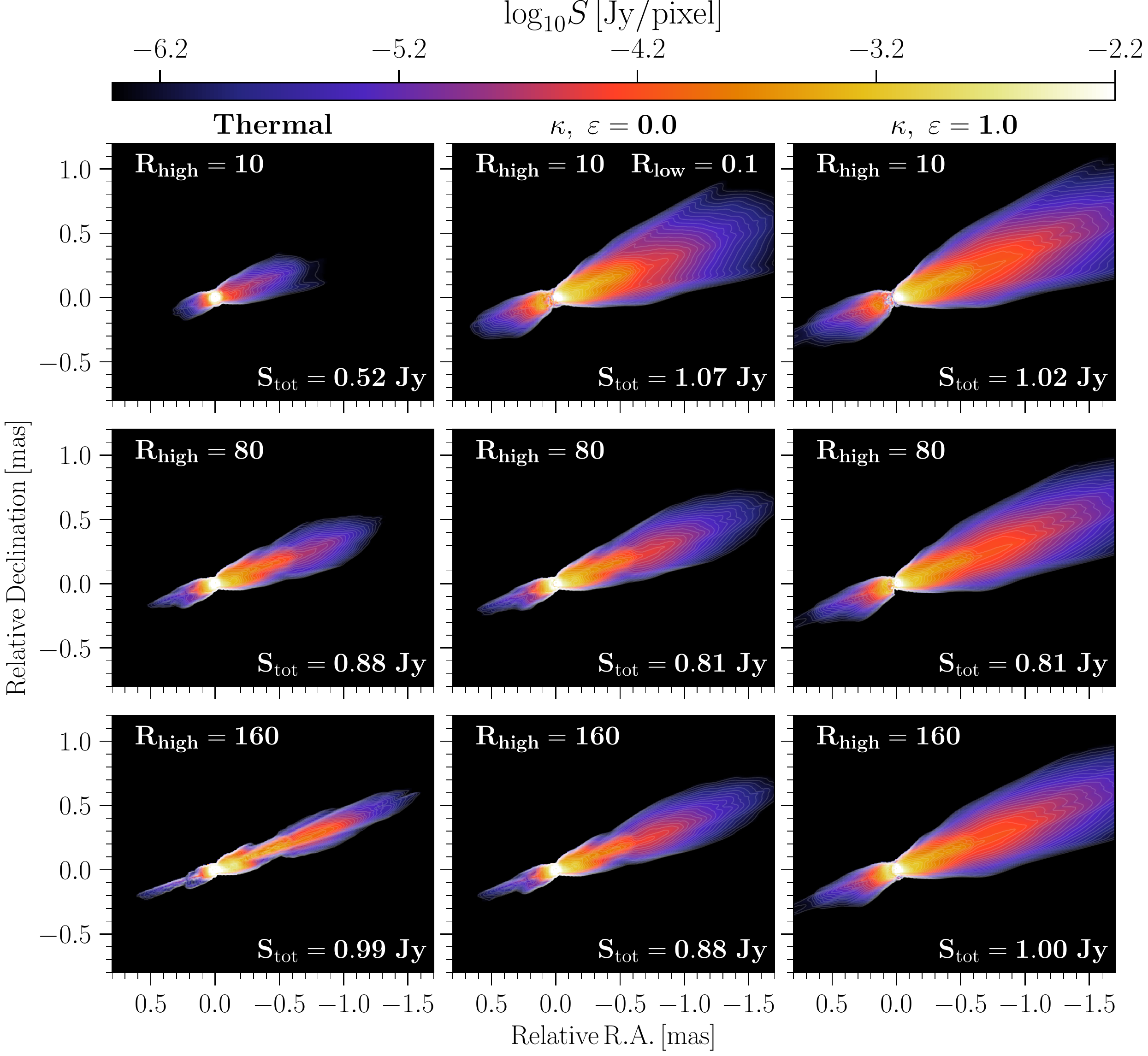}
\includegraphics[width=0.49\textwidth]{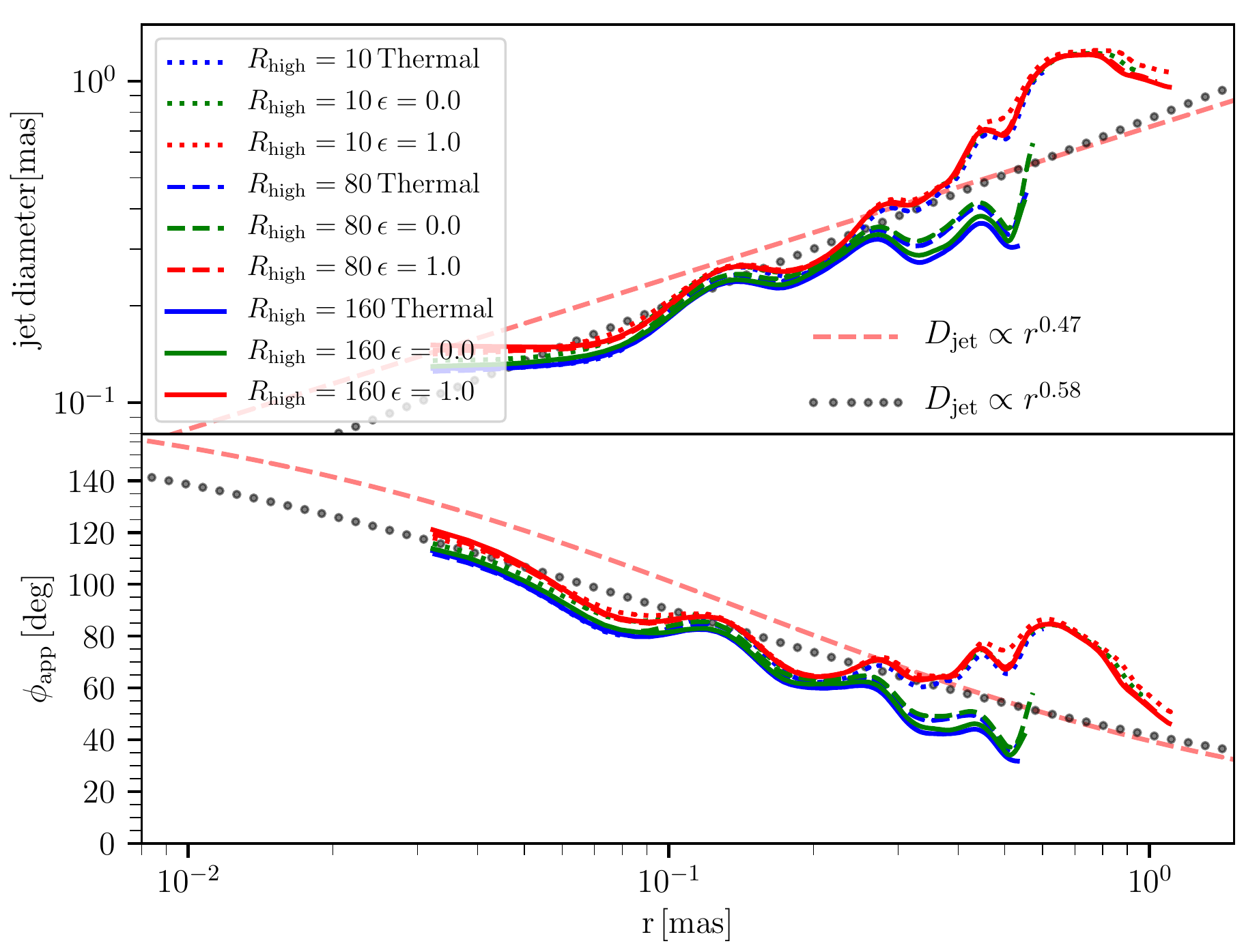}
\includegraphics[width=0.49\textwidth]{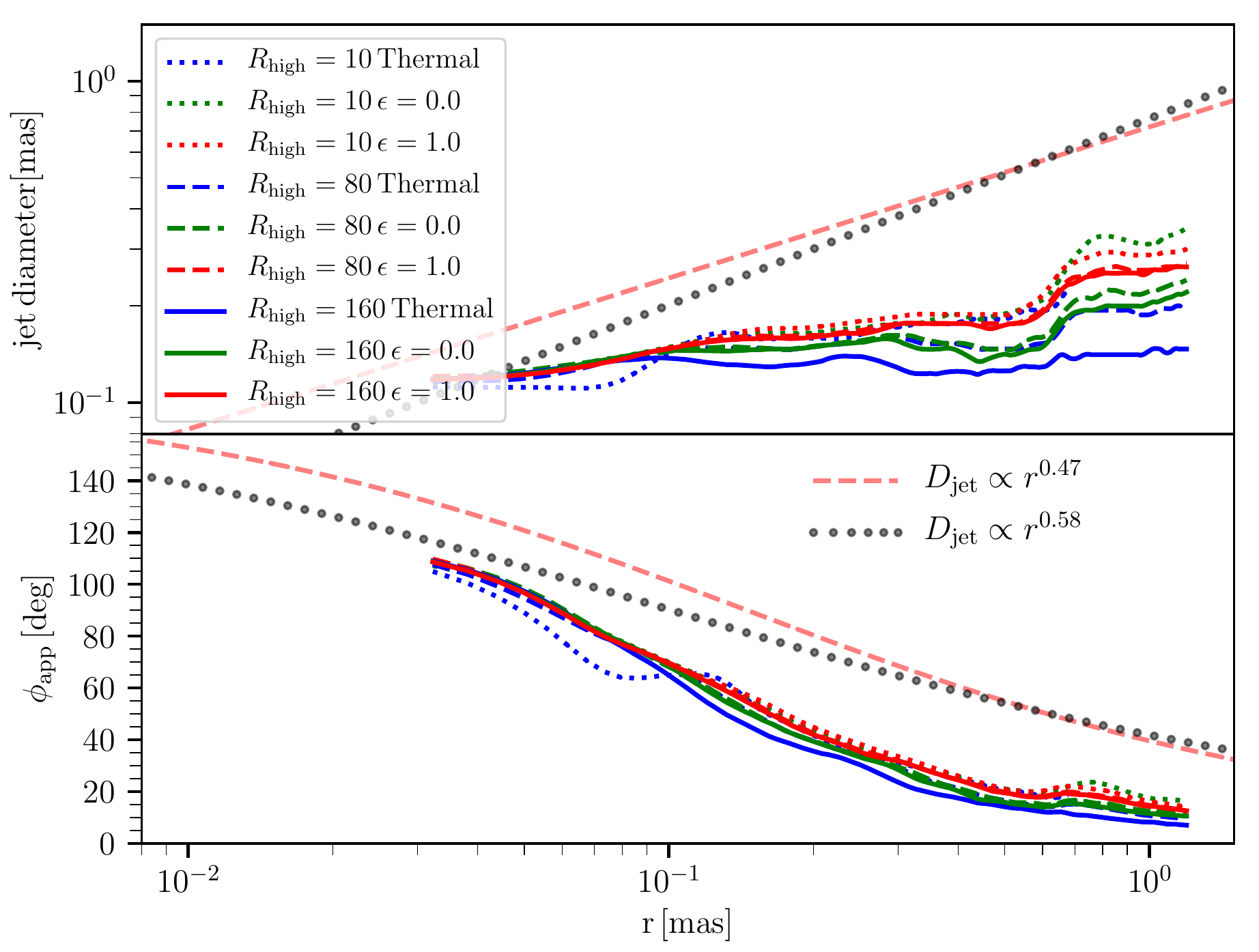}
\caption{The influence of the $R_{\rm low}$ parameter on the spectral and structural properties of our M\,87 simulations for a black hole with spin $a_{\star}=0.9375$. The left column corresponds to 
a MAD model and the right column to a SANE one. The top panels show the broad-band spectrum averaged for 2000\,M for $R_{\rm high}$=10 (dotted), $R_{\rm high}$=80 (dashed)  and $R_{\rm high}$=160 (solid) while keeping $R_{\rm low}=0.1$ 
fixed. The different eDFs are marked by different colors: thermal (blue), kappa with $\varepsilon=0$ (green) and kappa with $\varepsilon=1$ (red). The middle image shows the ray-traced 86\,GHz images for three different eDFs (columns) and 
different $R_{\rm high}$ (rows). The bottom panels show the jet diameter and the opening angles profiles using the same color and line styles as for the spectrum.}
\label{fig:Rlowsweep}
\end{figure*}
\subsubsection{Influence of the $R_{\rm low}$ parameter}

The second variable parameter of the electron temperature description (Eq.~\ref{eq:Te}) is the  $R_{\rm low}$ value. In the previous analysis we used $R_{\rm low}=1$. In this section we 
investigate the changes in the spectrum and the 86\,GHz image structure if $R_{\rm low}=0.1$ is used and compare the results to the one from the $R_{\rm low}=1$ models. The resulting 
broad-band 
spectrum and 86\,GHz images are presented in Fig. \ref{fig:Rlowsweep}. The comparison between the $R_{\rm low}=1$ (Fig. \ref{fig:Rhsweep}) and the $R_{\rm low}=0.1$ models show that 
the latter ones are brighter at high frequencies $\nu>10^{14}$\,Hz and show more extended jets for the thermal and $\varepsilon=0$ models. In order to understand this behaviour we need to 
analyse the dependence of the electron temperature with $R_{\rm low}$ (see Eq. \ref{eq:Te}). Reducing the $R_{\rm low}$ value increases the electron temperature in regions where 
plasma-$\beta<1$ 
while keeping the electron temperature unchanged in high plasma-$\beta$ regions (see Fig. \ref{fig:Te}). This implies that electrons in the jet sheath are heated up as compared to 
$R_{\rm low}=1$ models. The heating of the jet sheath is also clearly visible in the 86\,GHz images. The jets are more extended and wider than for the models with $R_{\rm low}=1$ (compare the 
bottom panels in Fig.~\ref{fig:Rhsweep} and Fig.~\ref{fig:Rlowsweep}). The steepening of the spectrum at high frequencies ($\nu>10^{14}$\,Hz) with increased $R_{\rm high}$ can be explained by the 
inverse proportionality of the electron temperature and the $R_{\rm high}$ values ($T_{\rm e}\propto 1/R_{\rm high}$) in the disk region ( high plasma-$\beta$ region). However, from the middle 
panel in Fig.~\ref{fig:Rlowsweep} we also see that the jets are shortened with $R_{\rm high}$. In case of $R_{\rm low} <1$ we can use Eq.~\ref{eq:Te} and ask for which combination of $R_{\rm 
high}$ and plasma-$\beta$ the electrons in the jet will be cooled, i.e., $T_{\rm ratio}>1$:
\begin{equation}
    \beta>\sqrt{\frac{1-R_{\rm low}}{R_{\rm high}-1}}
\end{equation}
Using our values of $R_{\rm high}=$10, 80, and 160 this leads to plasma-$\beta$ values of 0.3, 0.1 and 0.07. Given that the typical plasma-$\beta$ in the jet sheath $\sim 0.1$ we can finally 
understand the shortening of the jet in the 86\,GHz images with increased $R_{\rm high}$: In order to cool the electrons in the jet sheath if $R_{\rm low}=0.1$ and $R_{\rm high}=10$ typical 
plasma-$\beta$ values around 0.3 are required. Since the typical values are $\sim 0.1$ the sheath electrons will be heated instead of being cooled and the jet will be more extended. If we increase 
the $R_{\rm high}$ values the required plasma-$\beta$ for cooling the sheath electrons drops below the typical plasma-$\beta$ found in this region and the electrons will be cooled. As a result the 
jet will be shortened with increased $R_{\rm high}$. This effect also contributes to the steepening of the broad-band spectrum. In the SANE case only the thermal model for $R_{\rm high}=10$ 
which is optically thick at 86\,GHz does not follow the above explanation (see dotted blue curves in the inset of the top right panel in Fig. \ref{fig:Rlowsweep}). Similar as in the previous analysis, the 
spectrum and image structure of the $\varepsilon=1$ models are independent on the choice of the $R_{\rm high}$ value.  
\begin{figure*}[h!]
\centering
 \includegraphics[width=0.49\textwidth]{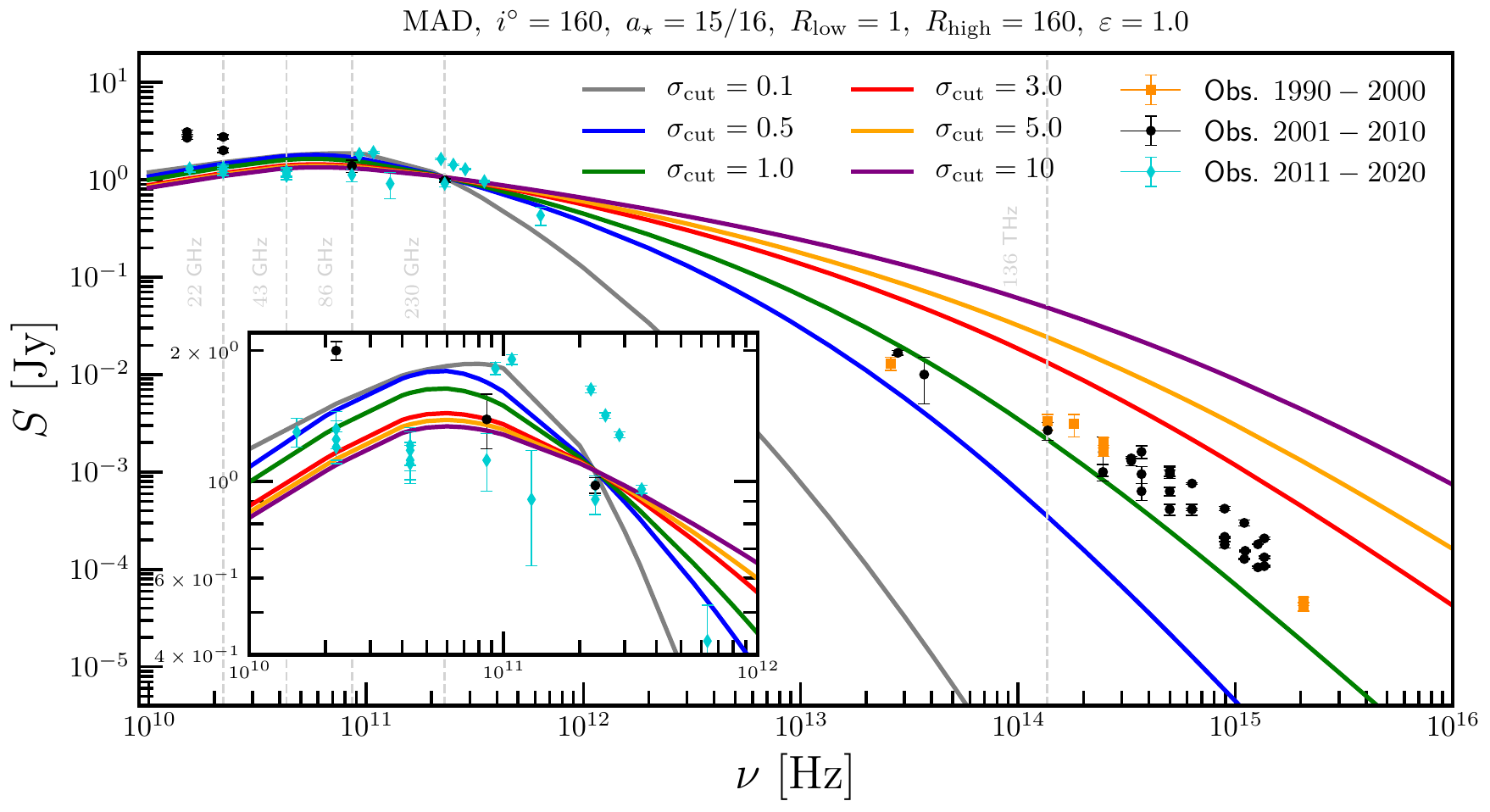}
 \includegraphics[width=0.49\textwidth]{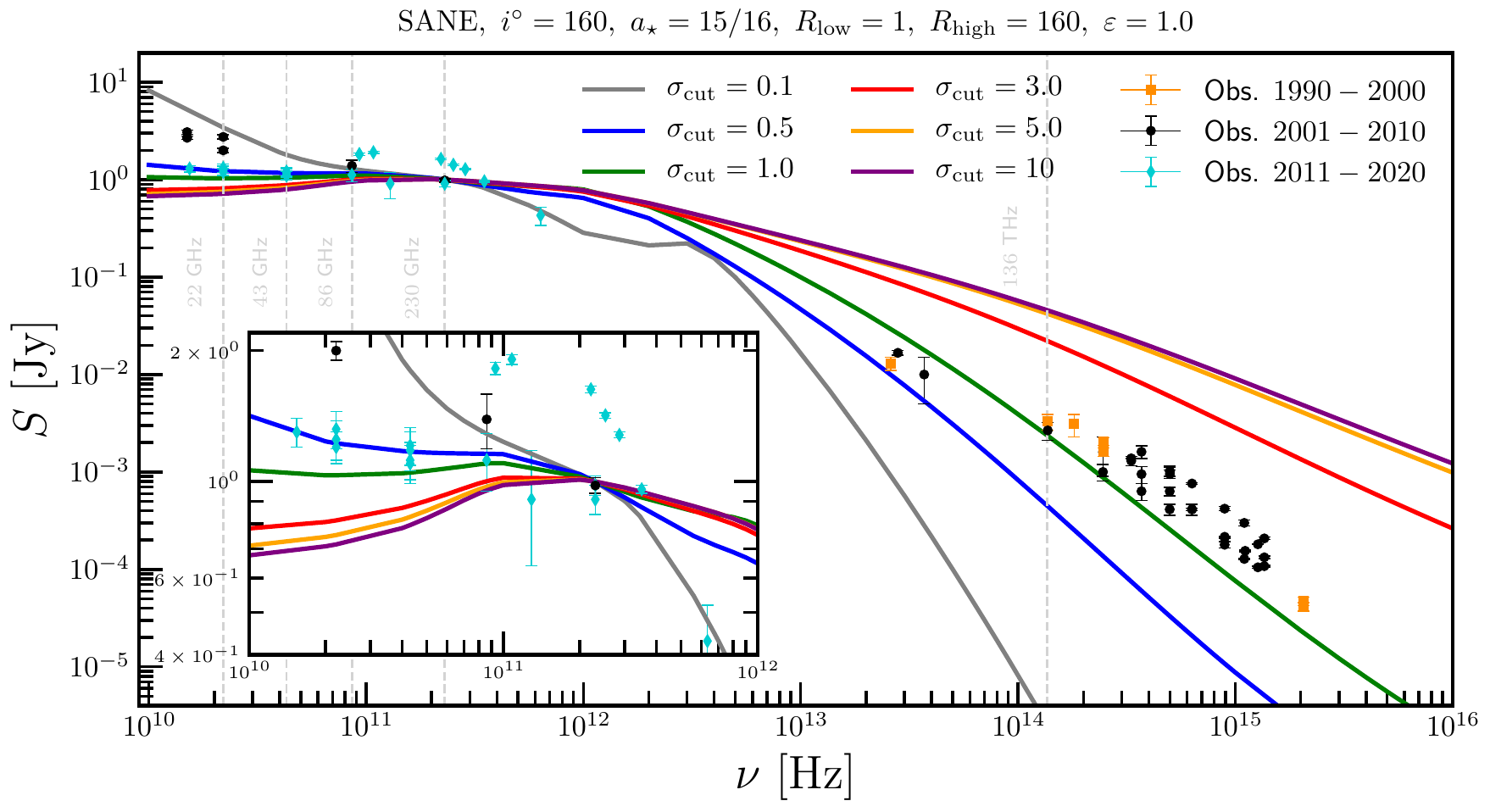}
 \includegraphics[width=0.49\textwidth]{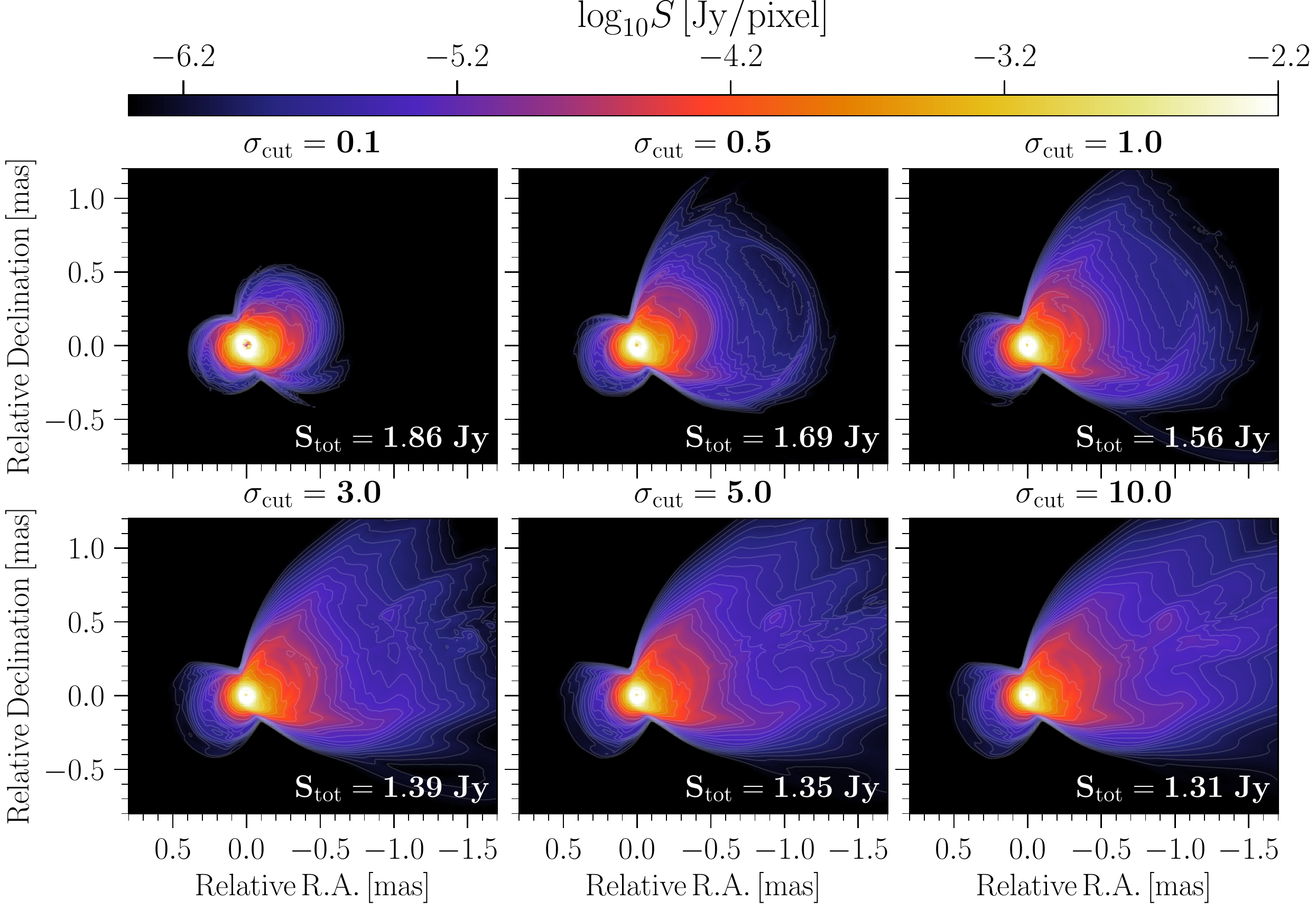}
 \includegraphics[width=0.49\textwidth]{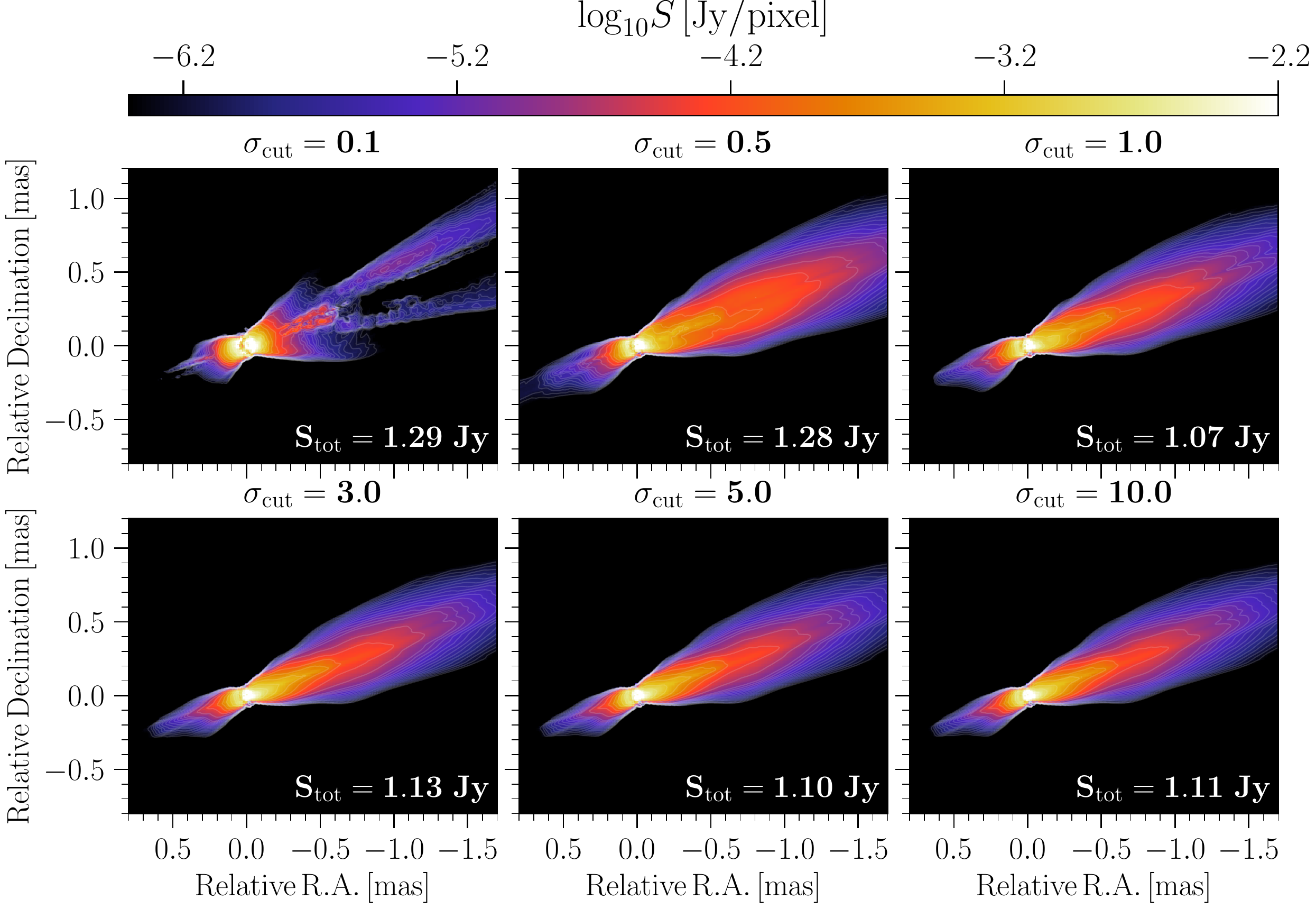}
 \includegraphics[width=0.49\textwidth]{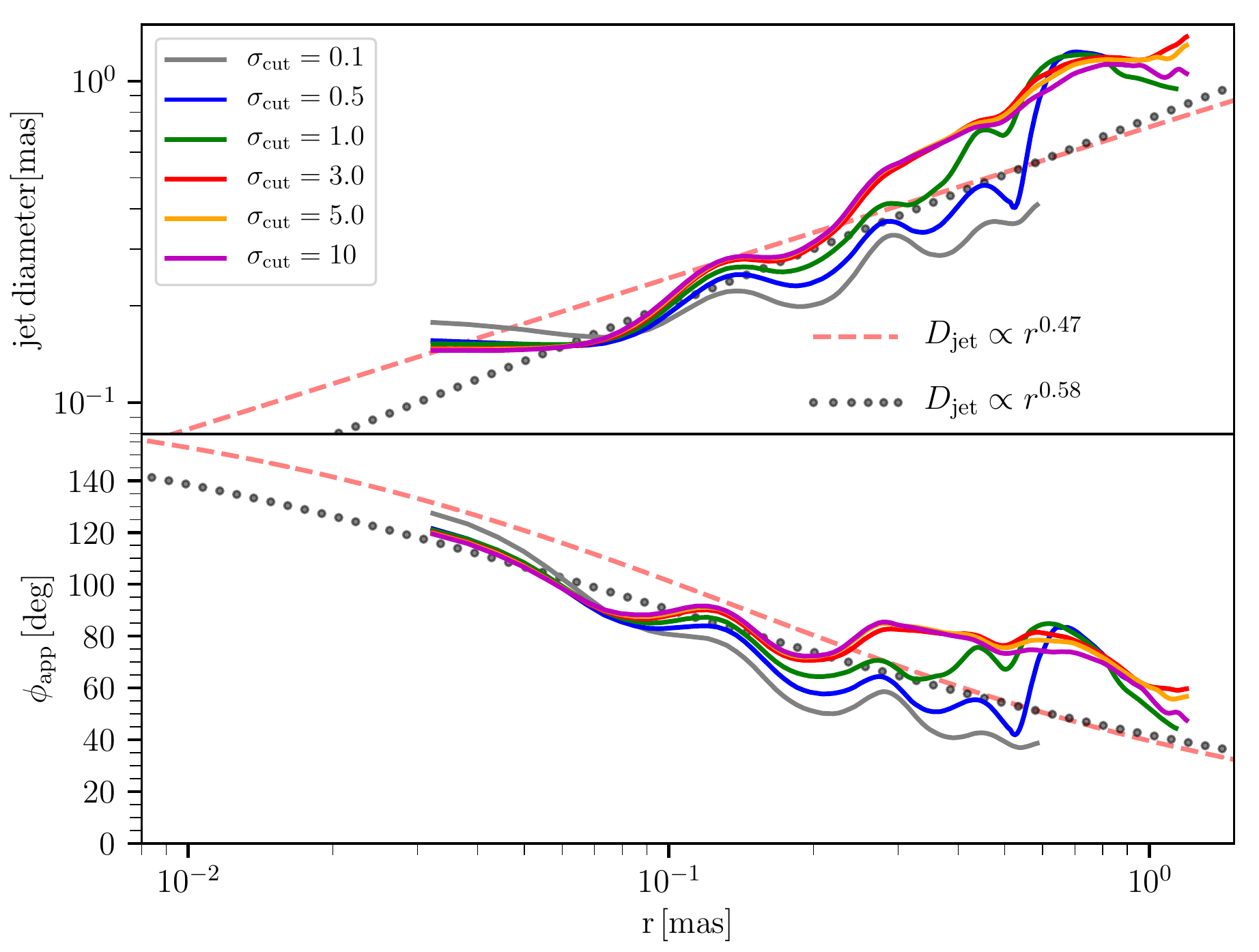}
\includegraphics[width=0.49\textwidth]{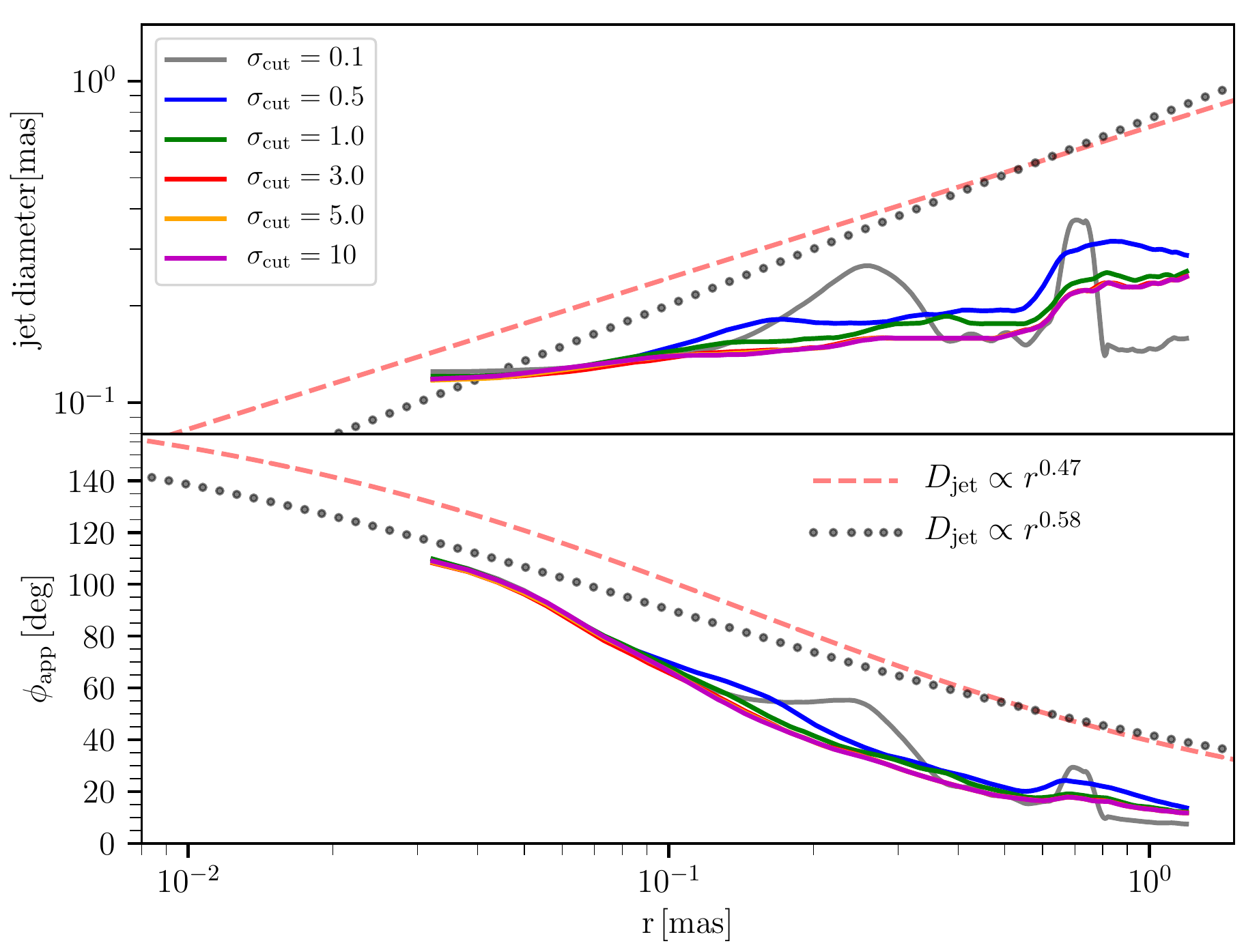}
\caption{The influence of the $\sigma_{\rm cut}$ parameter on the spectral and structural properties of our M\,87 simulations for a black hole with spin $a_{\star}=0.9375$. The left column 
corresponds to a MAD model and the right column to a SANE one. The top panels show the broad-band spectrum averaged for 2000\,M for six different values of $\sigma_{\rm cut}$ (color-coded) 
while keeping $R_{\rm low}=1$, $R_{\rm high}=160$ and $\varepsilon=1.0$ fixed. The middle image shows the ray-traced 86\,GHz images for different $\sigma_{\rm cut}$ values. The bottom 
panels show the jet diameter and the opening angles profiles using the same color-coding as for the spectrum.}
\label{fig:jetspine}
\end{figure*}

\begin{figure*}[h!]
\centering
\includegraphics[width=0.49\textwidth]{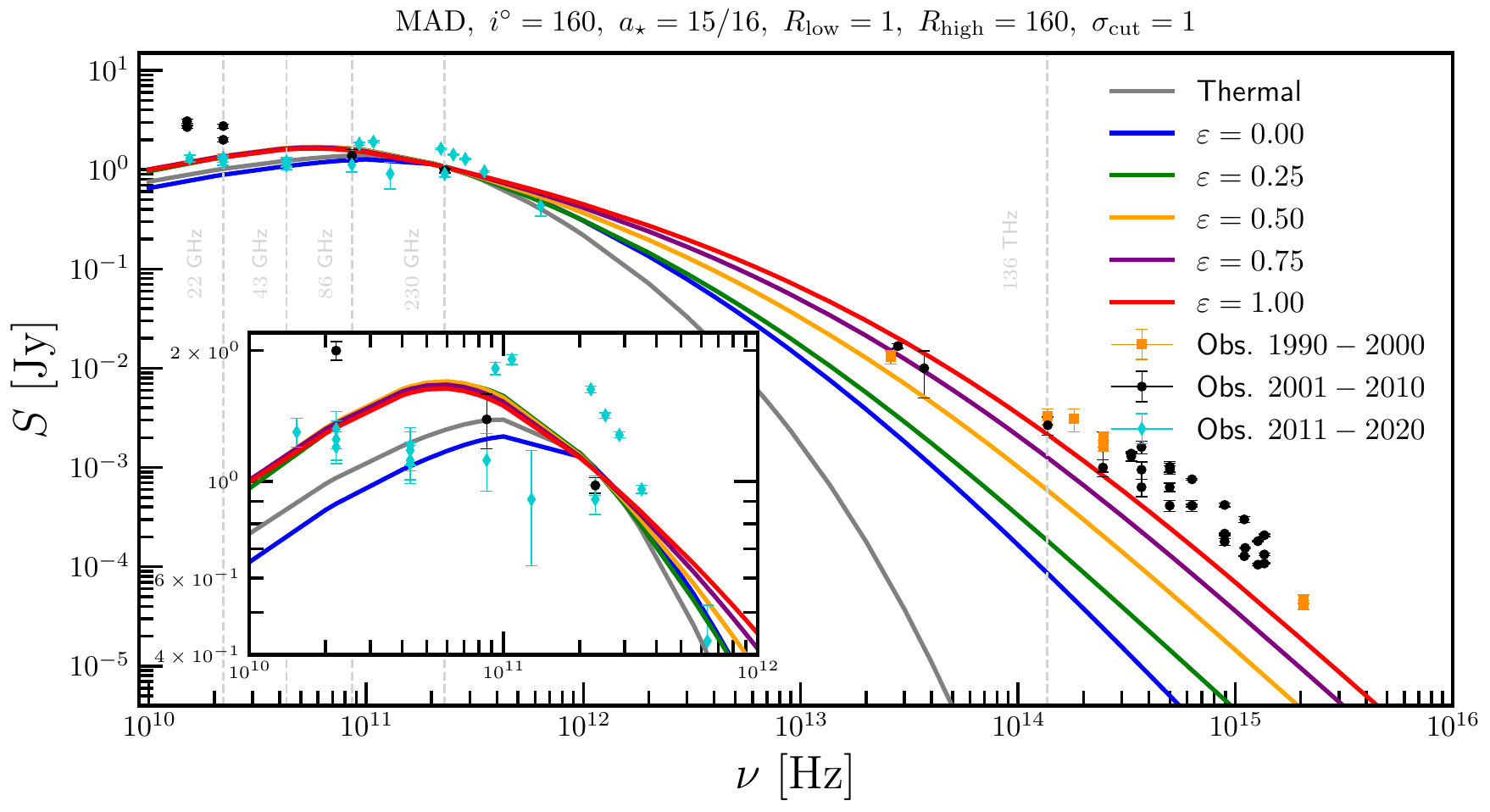}
\includegraphics[width=0.49\textwidth]{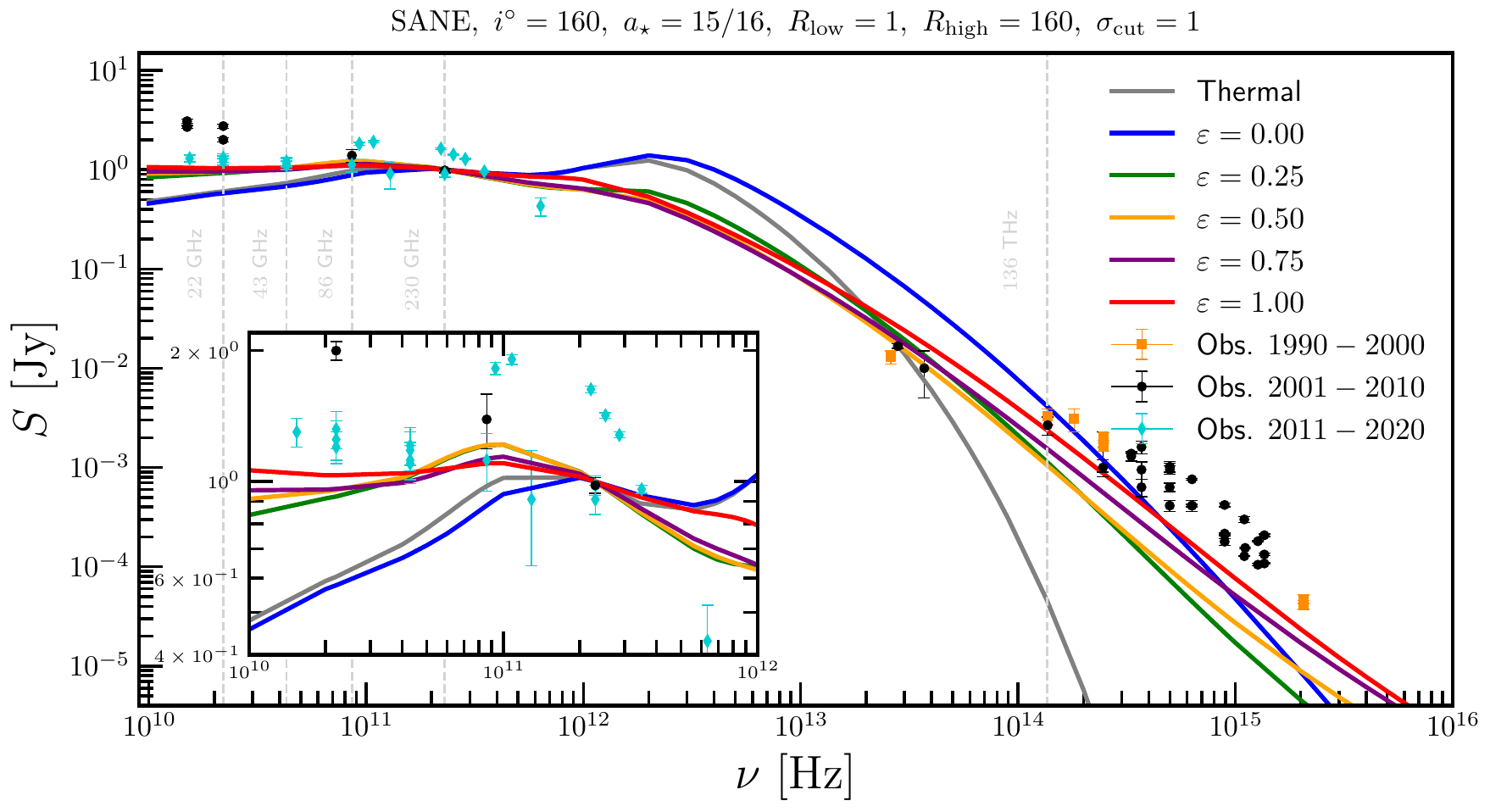}
\includegraphics[width=0.49\textwidth]{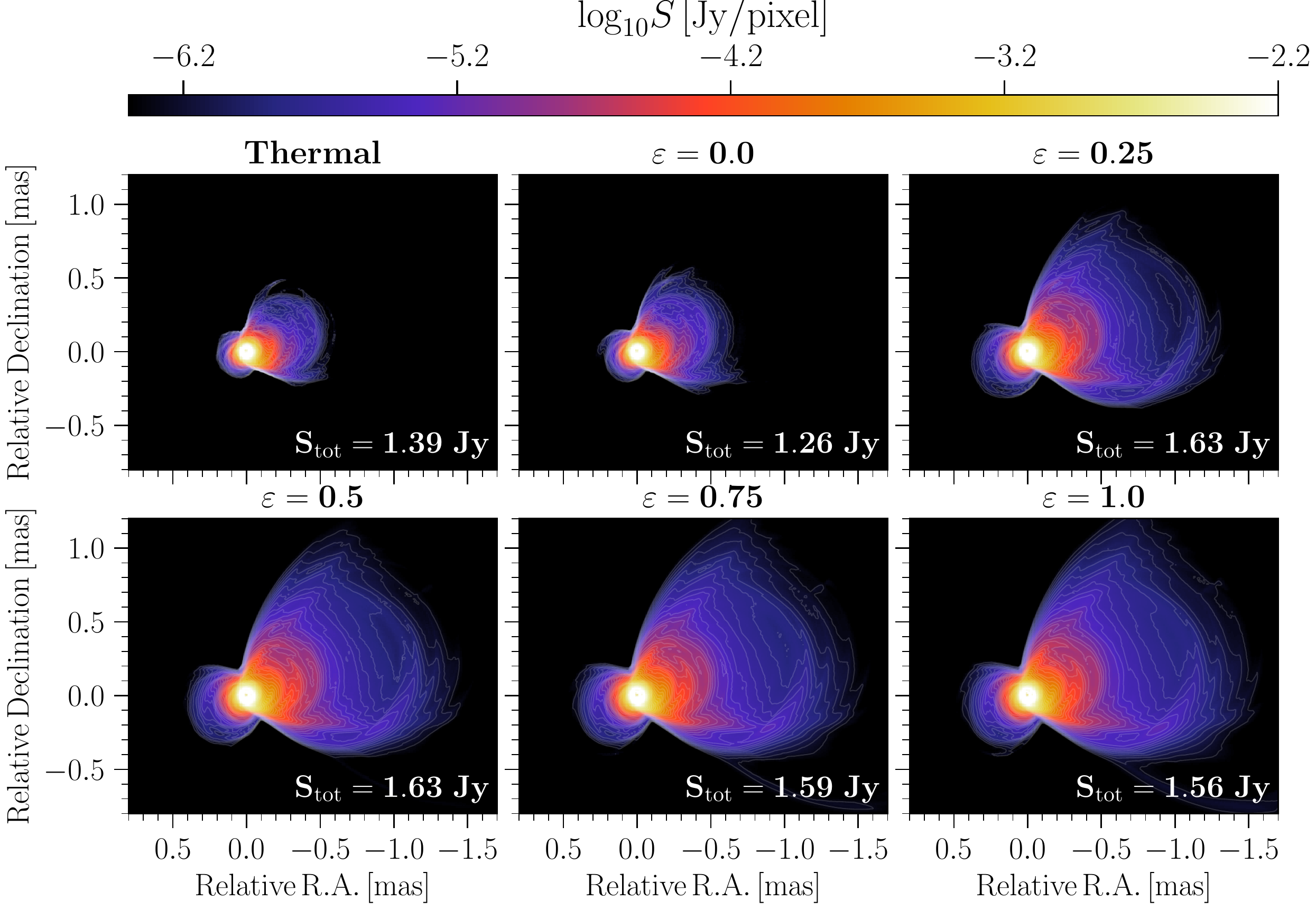}
\includegraphics[width=0.49\textwidth]{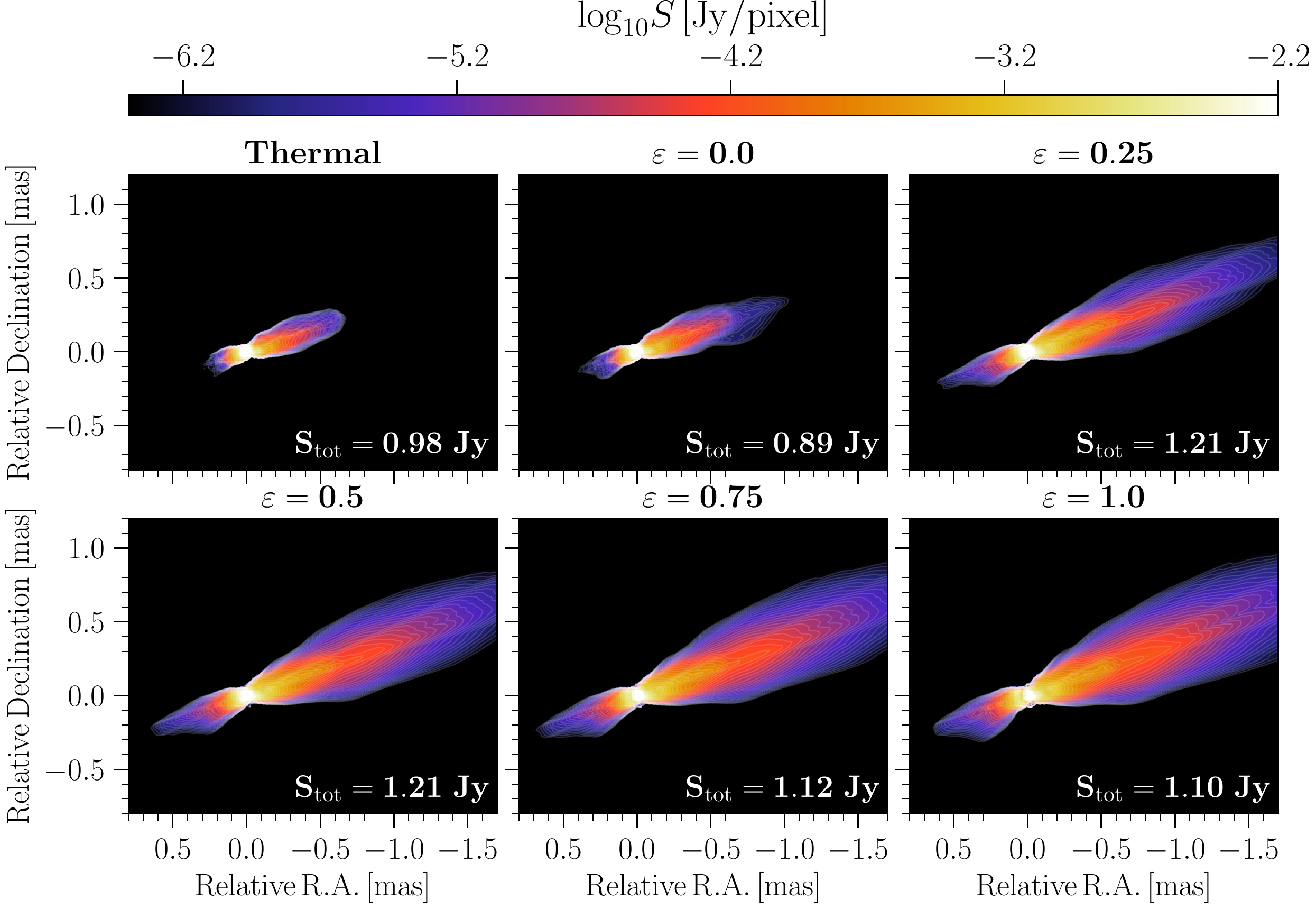}
\includegraphics[width=0.49\textwidth]{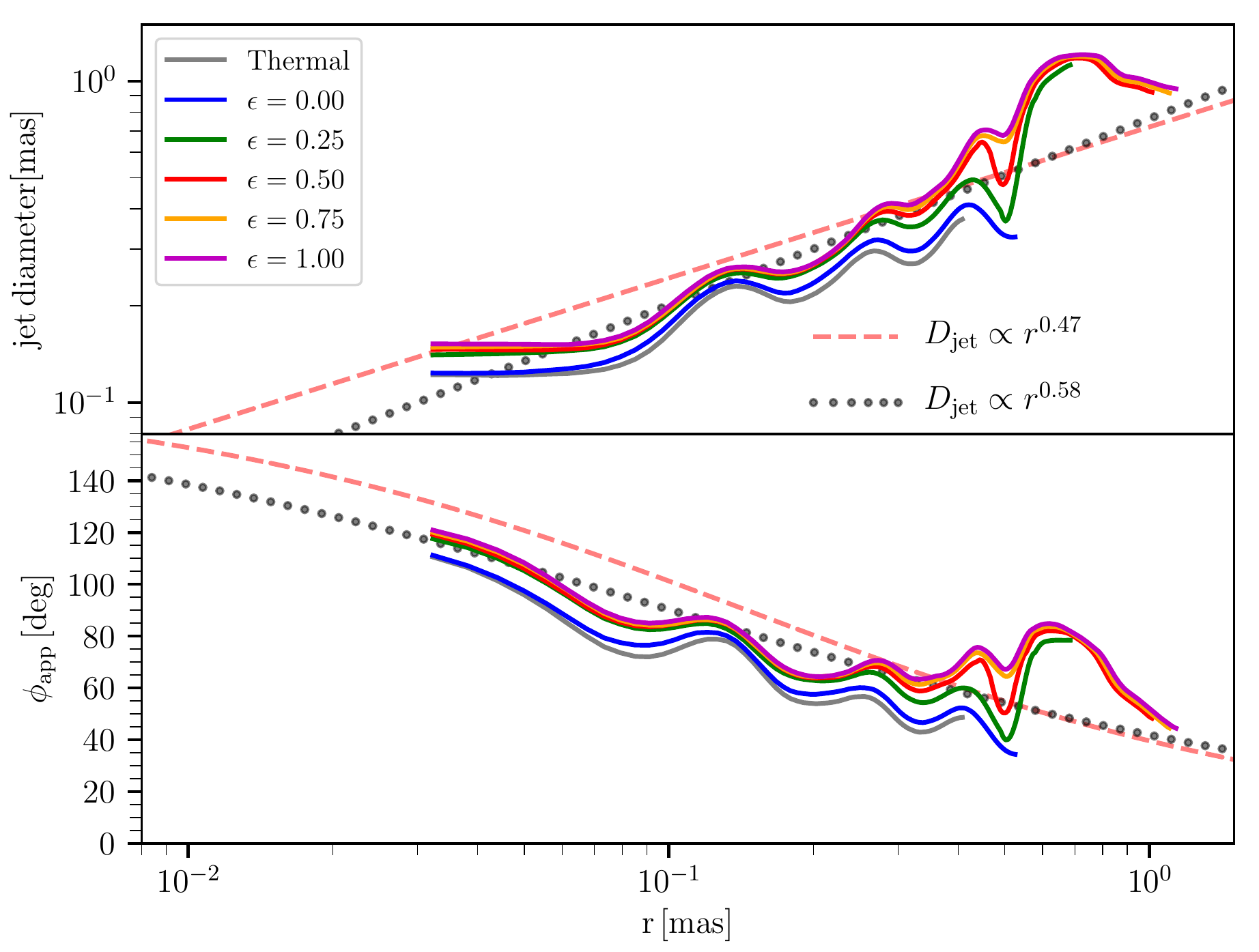}
\includegraphics[width=0.49\textwidth]{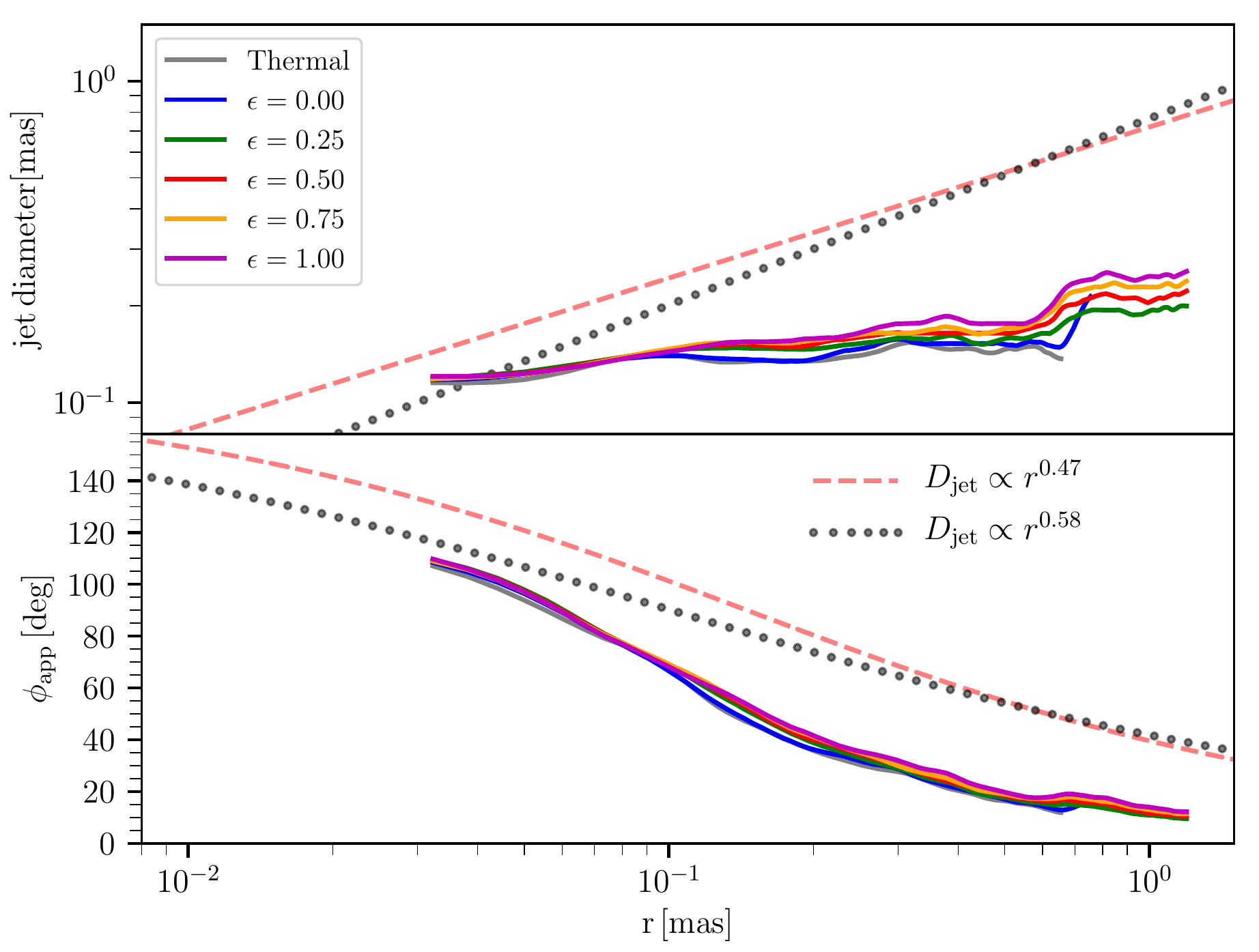}

\caption{The influence of the $\varepsilon$ parameter (fraction of magnetic energy used for the acceleration of non-thermal particles)  on the spectral and structural properties of our M\,87 
simulations for a black hole with spin $a_{\star}=0.9375$. The left column corresponds to a MAD model and the right column to a SANE one. The top panels show the broad-band spectrum averaged 
for 2000\,M for thermal and five different values of $\varepsilon$ (color-coded) while keeping $R_{\rm low}=1$, $R_{\rm high}=160$ and $\sigma_{\rm cut}=1.0$ fixed. The middle image shows the 
ray-traced 86\,GHz images for different $\varepsilon$ values. The bottom panels show the jet diameter and the opening angles profiles using the same color-coding as for the spectrum.} %\aco{CF 
%please use $\varepsilon$ instead of $\epsilon$, and frameon=False to have consistent notation and style.}}
\label{fig:epsilon}
\end{figure*}
\subsection{Influence of the jet spine}\label{sec:JetSpine}

In our next parameter sweep we explore the influence of the position of the jet spine on the broad-band spectrum and the 86\,GHz image structure. Since the jet spine is not well defined in the 
literature, commonly the magnetization of the plasma above a certain value, $\sigma_{\rm cut}$, is used to define the boundary between the jet sheath and jet spine. Here we use the parameters 
from the first analysis $R_{\rm low}=1.0$, $R_{\rm high}=160$ and we apply a non-thermal eDF with $\varepsilon=1$ in the jet sheath while varying the jet spine location via $\sigma_{\rm cut}=0.1,\ 
0.5,\ 1.0,\ 5.0$, and 10.

Figure \ref{fig:jetspine} shows the broad-band spectrum (top) the 86\,GHz jet structure (middle) and the jet width and opening angle profiles (bottom) for five different values of the $\sigma_{\rm 
cut}$ 
for a MAD model (left) and a SANE model with black hole spin $a_{\star}=0.9375$.
In the broad-band spectrum the increase in the $\sigma_{\rm cut}$ value leads to higher flux density and flatter spectra in the high frequency regime ($\nu> 10^{12}$\,Hz). This behaviour could be 
explained in the following way: Increasing the $\sigma_{\rm cut}$ will include more magnetised plasma from the jet spine in the jet sheath (see contour lines in first and fifth panel 
Fig.~\ref{fig:Morphology}). At the same time these regions have lower plasma-$\beta$ which will lead, according to Eq. \ref{eq:Te}, to higher electron temperatures in the jet sheath (see second and 
sixth panel in Fig.~\ref{fig:Morphology}). In addition to the increase of the electron temperature in the jet sheath with increased $\sigma_{\rm cut}$, the power-law slope of the non-thermal eDF will 
be flattened (see Eq. \ref{eq:kappa} and Fig.~\ref{fig:SANE_w}). Combined these two effects, namely the increased electron temperatrue and the decrease in the $\kappa$ value will lead to a 
larger width of the kappa-eDF (see Eq. \ref{eq:w}), i.e. more energetic non-thermal particles. Since changes in the $\sigma_{\rm cut}$ value will not alter the temperature of the disk region the 
changes in the emission are entirely due to the jet sheath emission (second term in Eq. \ref{eq:jnukapproxTe}). From this equation the increase of the flux density and the flattening of the spectrum 
with increased $\sigma_{\rm cut}$ can be explained. For the SANE model there is nearly no difference between the spectrum for $\sigma_{\rm cut}=5$ and $\sigma_{\rm cut}=10$ which is an 
indication that the largest magnetisation in this model is around $\sigma\sim 3$.

\begin{figure*}[h!]
\centering
 \includegraphics[width=0.48\textwidth]{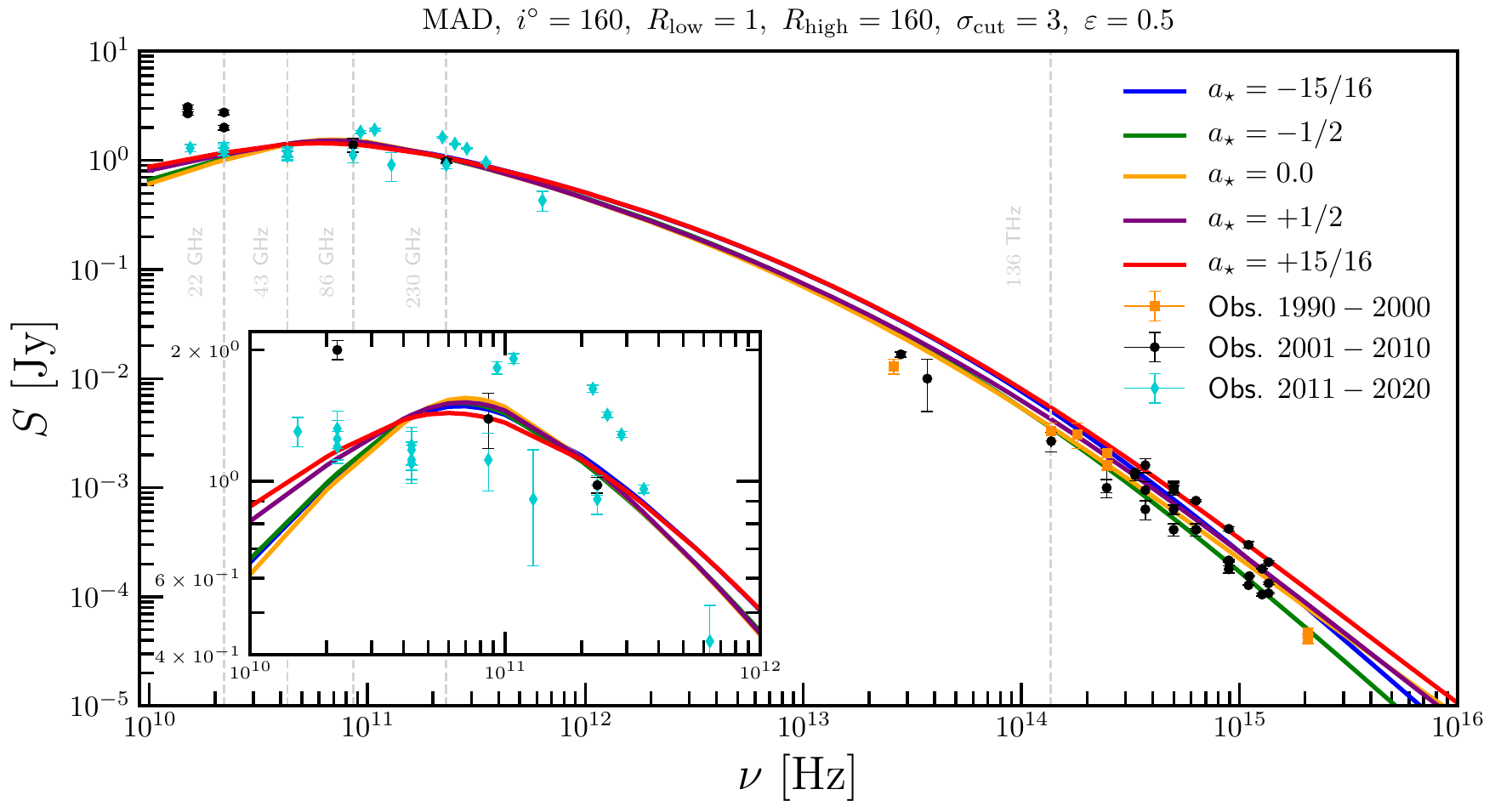}
  \includegraphics[width=0.48\textwidth]{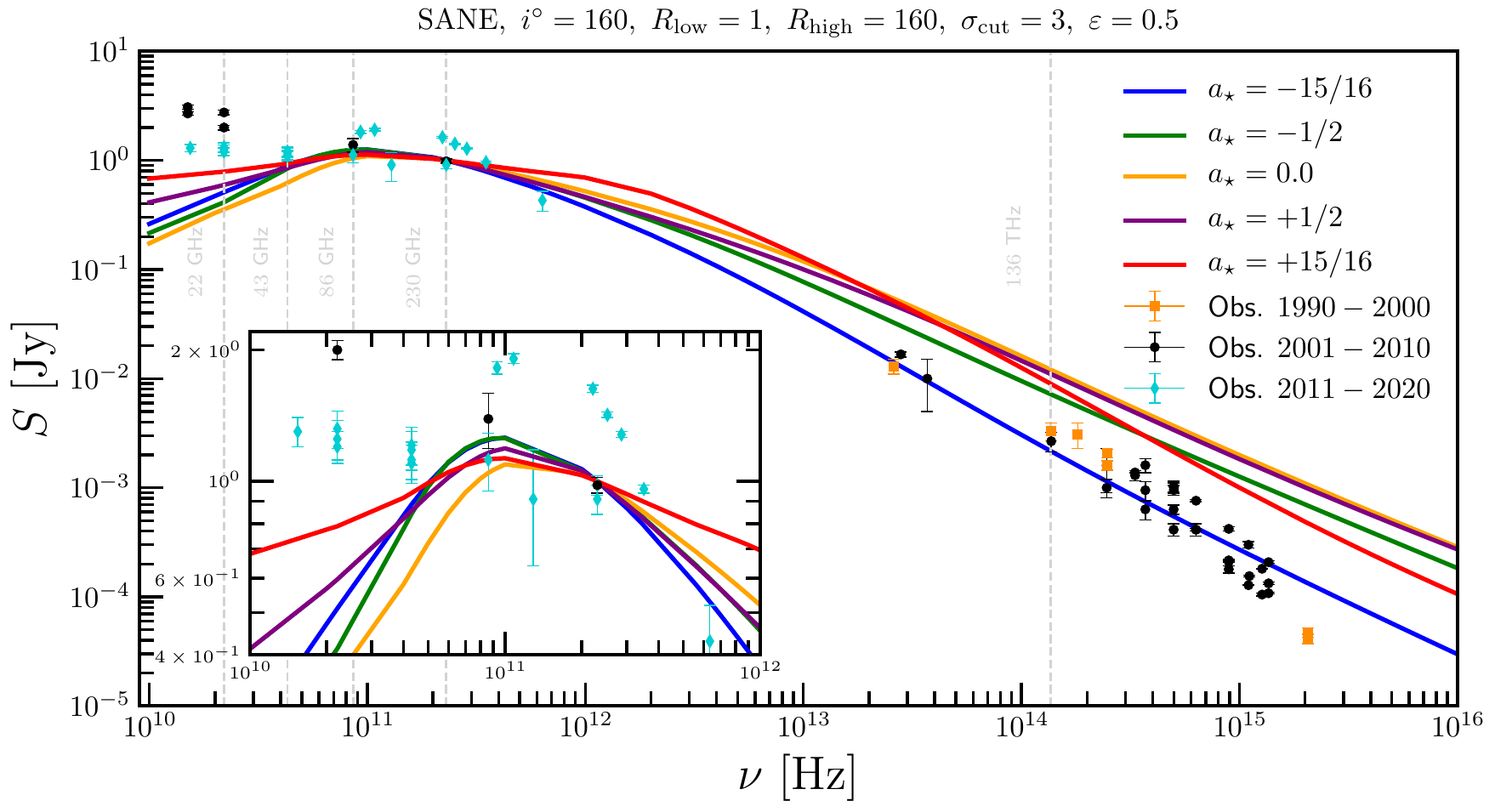}
 \includegraphics[width=0.95\textwidth]{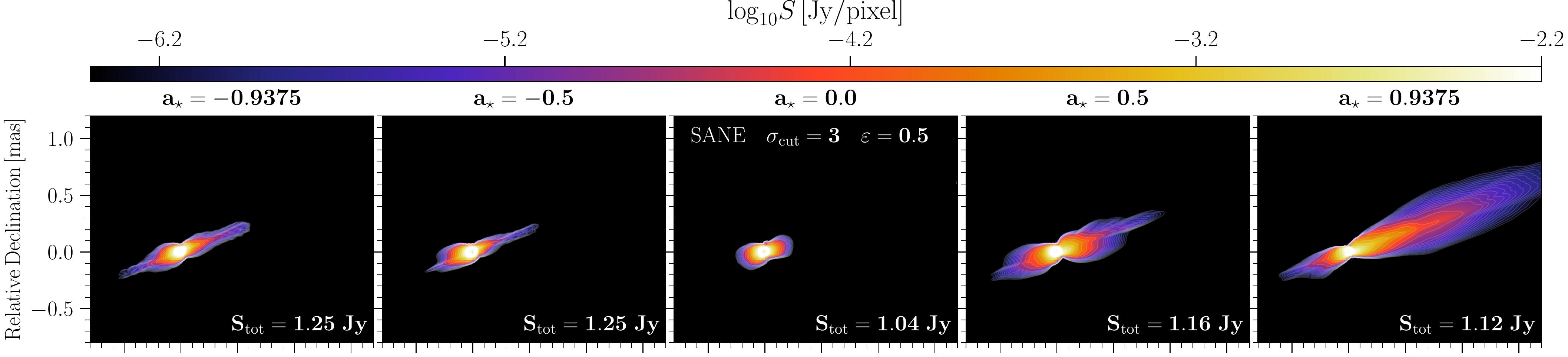} 
 \includegraphics[width=0.95\textwidth]{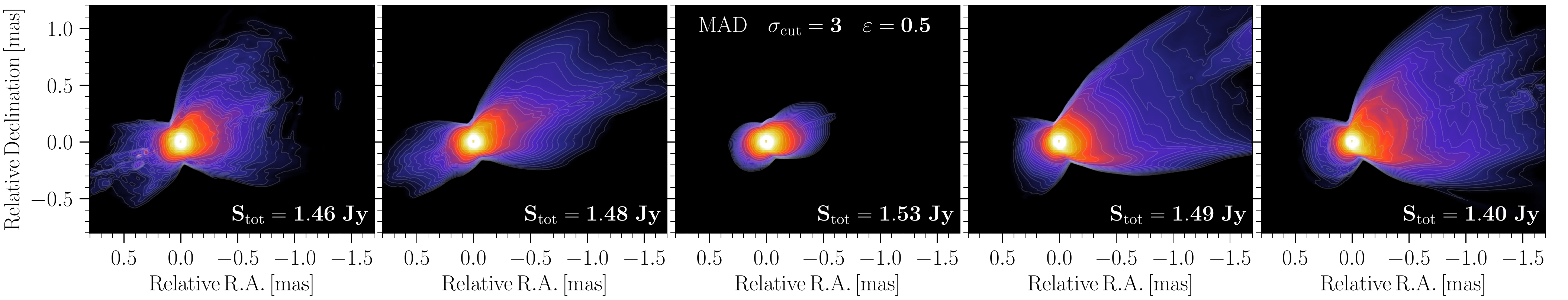}
 \includegraphics[width=0.49\textwidth]{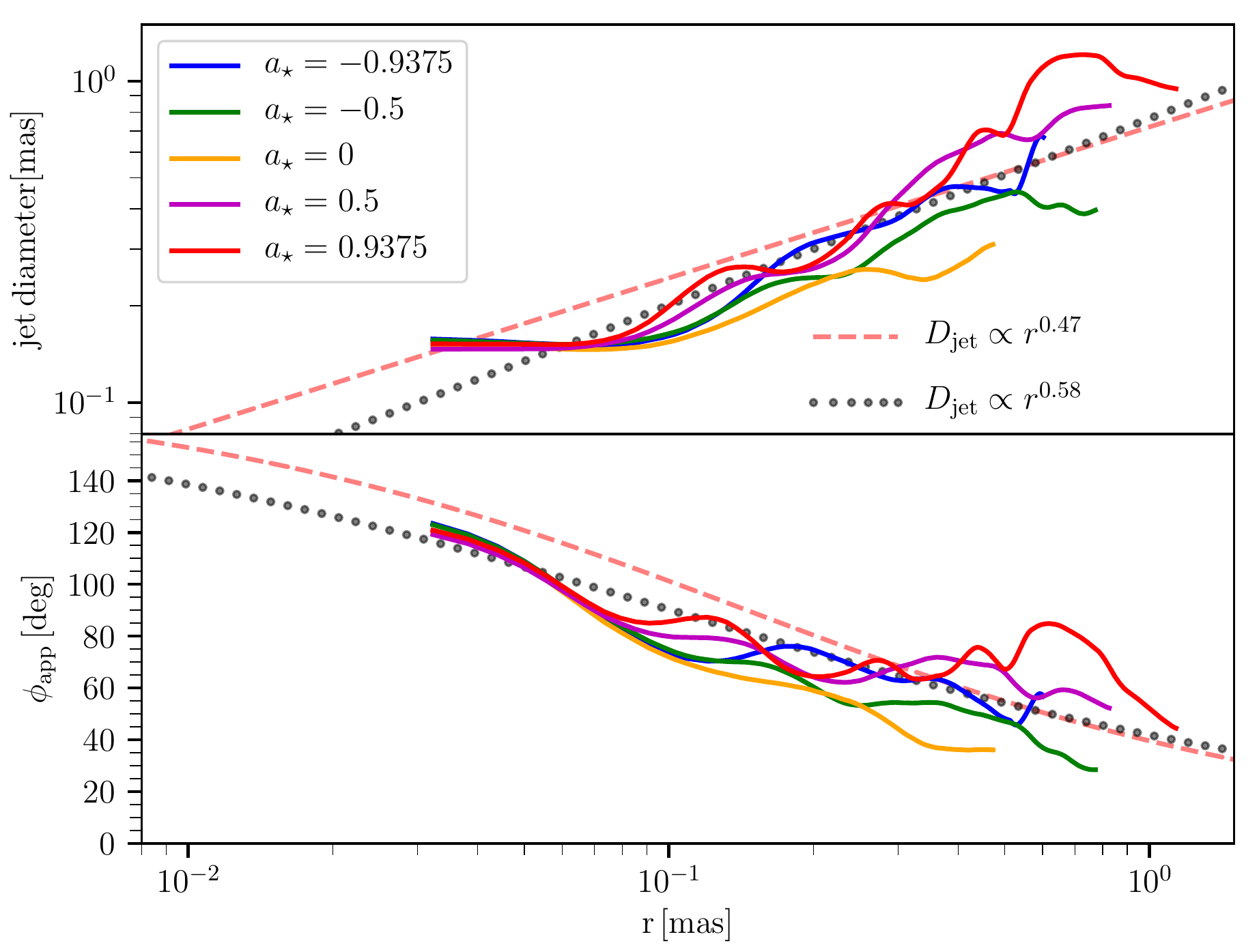}
\includegraphics[width=0.49\textwidth]{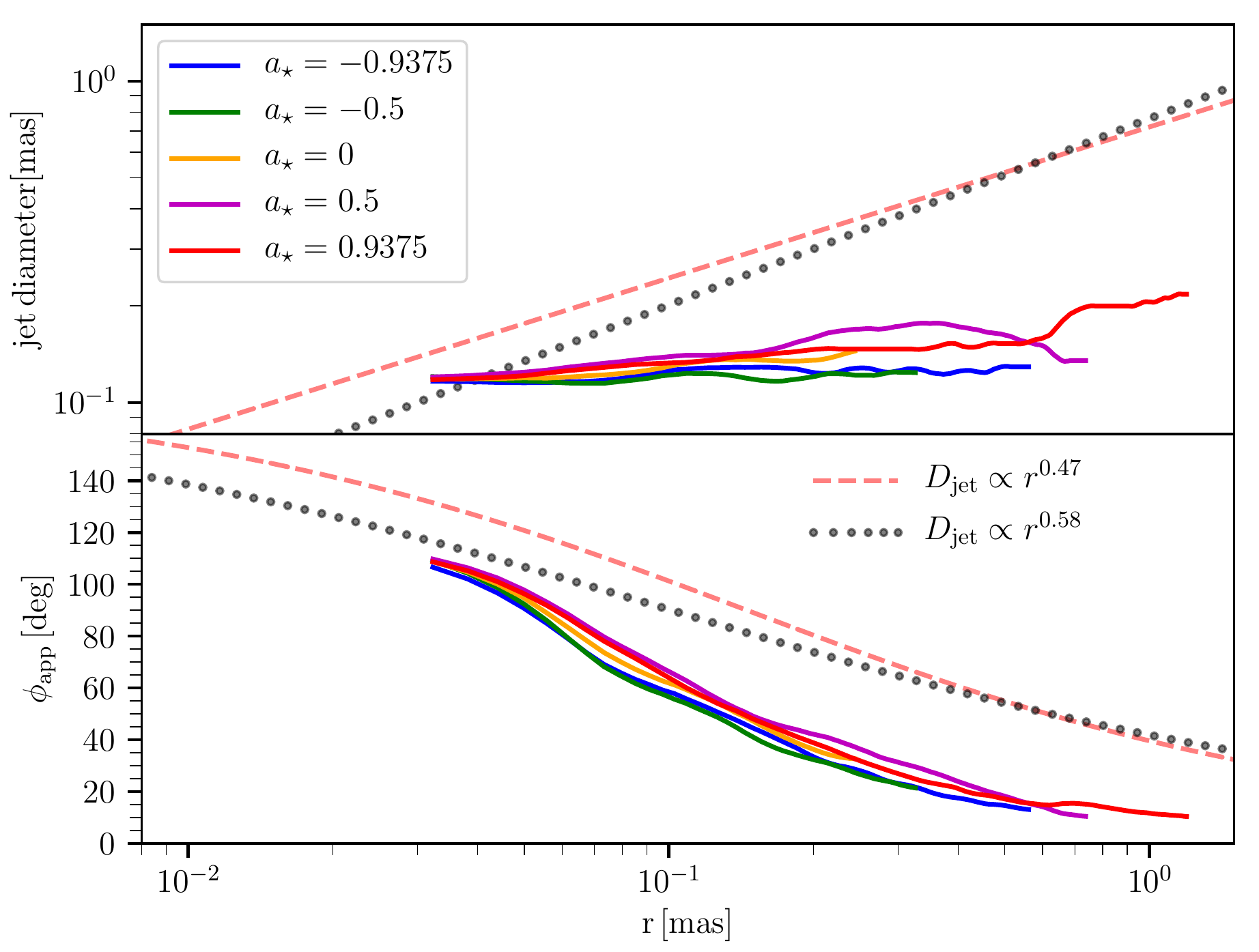}
\caption{The influence of the black hole spin, $a_{\star}$, on the spectral and structural properties of our M\,87 simulations. The left column corresponds to a MAD model and the right column to 
a SANE one. The top panels show the broad-band spectrum averaged for 2000\,M for thermal and five different spin values $a_{\rm star}=-0.9375,\ -0.5,\ 0,\ 0.5$, and $0.9375$ (color-coded) while 
keeping $R_{\rm low}=1$, $R_{\rm high}=160$ $\sigma_{\rm cut}=3.0$ and $\varepsilon=0.5$ fixed. The middle image shows the ray-traced 86\,GHz images for different spin values. The bottom 
panels show the jet diameter and the opening angles profiles using the same color-coding as for the spectrum.}
\label{fig:spin}
\end{figure*}

With increasing $\sigma_{\rm cut}$ value the 86\,GHz jet becomes brighter at larger distances from the black hole. For the MAD models the jet width increases (see bottom left panels) and the 
emission signature of the jet spine becomes visible as a central ridge for $\sigma_{\rm cut}\geq 3$. In contrast to the MAD models, the SANE ones show the largest jet width for $\sigma_{\rm 
cut}=0.5$ 
and larger values of $\sigma_{\rm cut}$ do not affect the jet width and opening angle (see bottom right panels). As mentioned above, with increased $\sigma_{\rm cut}$ value more 
magnetised and hotter plasma is included in the jet sheath and at the same time the power-law slope of the non-thermal eDF flattens. Thus, more radiating electrons with large $\gamma_e$ are 
included in the jet sheath and the jet becomes brighter on large distances. Similar as in the broad-band spectrum, the independence of the jet width and opening angle for $\sigma_{\rm cut}\geq 3$ 
for the SANE models, is a clear signal that we include the entire jet (jet spine and jet sheath) in the GRRT. 

\subsection{Influence of the magnetic energy}\label{sec:Benergy}
The last free parameter we consider in our radiative transfer calculation is the fraction of magnetic energy, $\varepsilon$, which is added to the non-thermal electrons (see Eq. \ref{eq:w}). So far we 
considered two extreme values, namely $\varepsilon=0$ and $\varepsilon=1.0$ and we include here three additional values for $\varepsilon=0.25,\,0.5$, and 0.75 and for comparison reasons a 
pure thermal model. During this sweep in the parameter space we use $R_{\rm low}=1.0$, $R_{\rm high}=160$ and $\sigma_{\rm cut}=1$ and the results are presented in Fig.~\ref{fig:epsilon}. As 
can be seen in the broad-band spectrum (top panels) increasing the $\varepsilon$ value increases the flux density at high frequencies ($\nu>10^{12}$\, Hz) and flattens the spectrum. By increasing 
the $\varepsilon$ value more energy from the magnetic field is included in the non-thermal particles (see Eq. \ref{eq:w}) and the non-thermal term in the combined emission coefficient becomes 
larger and dominates for large values of $\varepsilon$ (see Eq. \ref{eq:jnukapproxTe}). As a result the flux density at high frequencies ($\nu>10^{12}$\,Hz) increases and the exponential decay (due 
to the thermal disk emission) is transitioning into a power-law. This behaviour is independent of the accretion model. Notice that due to the different turnover positions in the SANE models this trend 
is not as clear as in the MAD models. The shift in the turnover position is as mentioned before connected to the variation in mass accretion rate during the flux normalisation (see Eq. 
\ref{eq:turnoverfreq}).

Increasing the fraction of magnetic energy in the kappa eDF leads to more extended jets (see middle panel in Fig. \ref{fig:epsilon}). For $\varepsilon\geq 0.5$ no further changes in the jet width and 
opening angle are obtained (see bottom panels in Fig.~\ref{fig:epsilon}). Increasing $\varepsilon$ will increase the number of electrons with large $\gamma_{\rm e}$, i.e., more high energy 
electrons in the tail of the kappa distribution (see Eq. \ref{eq:w}). Since we apply the kappa eDF in the jet sheath region this region will become brighter. For $\varepsilon\sim 0.5$ the entire jet 
sheath region includes electrons with large $\gamma_{\rm e}$. The geometry of the jet sheath is defined by the spin of the black hole and the accretion model (see Fig.~\ref{fig:spinsigma}). Thus, 
increasing the $\varepsilon$ value above 0.5 has no effect on the jet structure and will only increase the emission (best seen at high frequencies in the broad-band spectrum). 

\subsection{Influence of the black hole spin, $a_\star$}\label{sec:spin}
The black hole spin parameter $a_{\star}$ is a fundamental parameter of our GRMHD simulations since it modifies the event horizon size, the gravitational potential, and characteristic radii such as 
the last marginally stable keplerian orbit and innermost stable circular orbit \cite{Font02b,Daigne04, Rezzolla_book:2013}. In Table \ref{tab:TorusID} we summarise the black hole spin parameters 
and the corresponding initial data for the magnetized torus. Figure \ref{fig:rates} shows the evolution of the accretion rates and the magnetic flux. The differences between MAD and SANE models 
are clearly visible but also between co-rotating and counter-rotating black holes within the same accretion model. This behaviour is best seen in the magnetic flux (third and sixth panel in 
Fig.~\ref{fig:rates}) and in the average values presented in Table \ref{tab:rates}).

As mentioned in the last paragraph, the spin of the black hole has major impact on the geometry of the jet. Therefore, we analyse in this section the influence of the black hole spin on the broad-band spectrum and the 86\,GHz image structure. Figure \ref{fig:spin} shows the average spectra and averaged 86\,GHz images using $R_{\rm low}=1$, $R_{\rm high}=160$, and $\sigma_{\rm cut}=3.0$ together with a kappa electron distribution with  $\varepsilon=0.5$ for five different spin values $a_{\star}=-0.9375,\ -0.5,\ 0.0,\ 0.5$, and $0.9375$. The used values for $\sigma_{\rm cut}$ and $\varepsilon$ are motivated from the previous slices through the parameter space and lead to an improved fit to the broad-band spectrum (see top panels in Fig. \ref{fig:spin}). Interestingly, changes in the black hole spin for this set of best-fit parameter introduce only a minor scatter in the broad-band spectrum especially in the MAD case. The low frequency part of the spectrum ($\nu<10^{11}$\,Hz)  is flatter for the MAD models than for their SANE counterparts (see also total flux at 86\,GHz given in the second panel of Fig.~\ref{fig:spin}). This can be explained the lack of a highly magnetised jet in the SANE models (see Fig.~\ref{fig:spinsigma}) which leads to larger $\kappa$ values (see Eq.~\ref{eq:kappa} and Fig.~\ref{fig:spinkappa}). Therefore, less energetic particles are in the tail of the eDF leading to smaller emissivities (see Eq.~\ref{eq:jnukapprox}) resulting in shorter jets (see second panel in Fig.~\ref{fig:spin}). 

The morphology of the 86\,GHz images presented in the middle panel of Fig.~\ref{fig:spin} shows the already mentioned differences between SANE (top) and MAD models (bottom): MAD models produce wide and highly magnetised jets which are lit up by the inclusion of non-thermal particles via the kappa-eDF. Independent of the accretion model Schwarzschild black holes ($a_{\star}=0$) do not launch a well established jet and the bulk of the emission is produced in the disk with some minor contribution from the disk wind. In the MAD case the differences between the jet width and the jet length for the different spins can be related to the variability of the accretion flow and the accreted magnetic flux. Following \citet{Tchekhovskoy2011,Tchekhovskoy2012}, the jet efficiency, $\eta$, 
for MAD models is proportional to $\phi_{\rm BH}^2$. Together with the average values for $\phi_{\rm BH}$ given in Table \ref{tab:rates} the differences in the jet morphology between pro- and retrograde models can be explained.

\begin{figure*}[h!]
\centering
\includegraphics[width=0.95\textwidth]{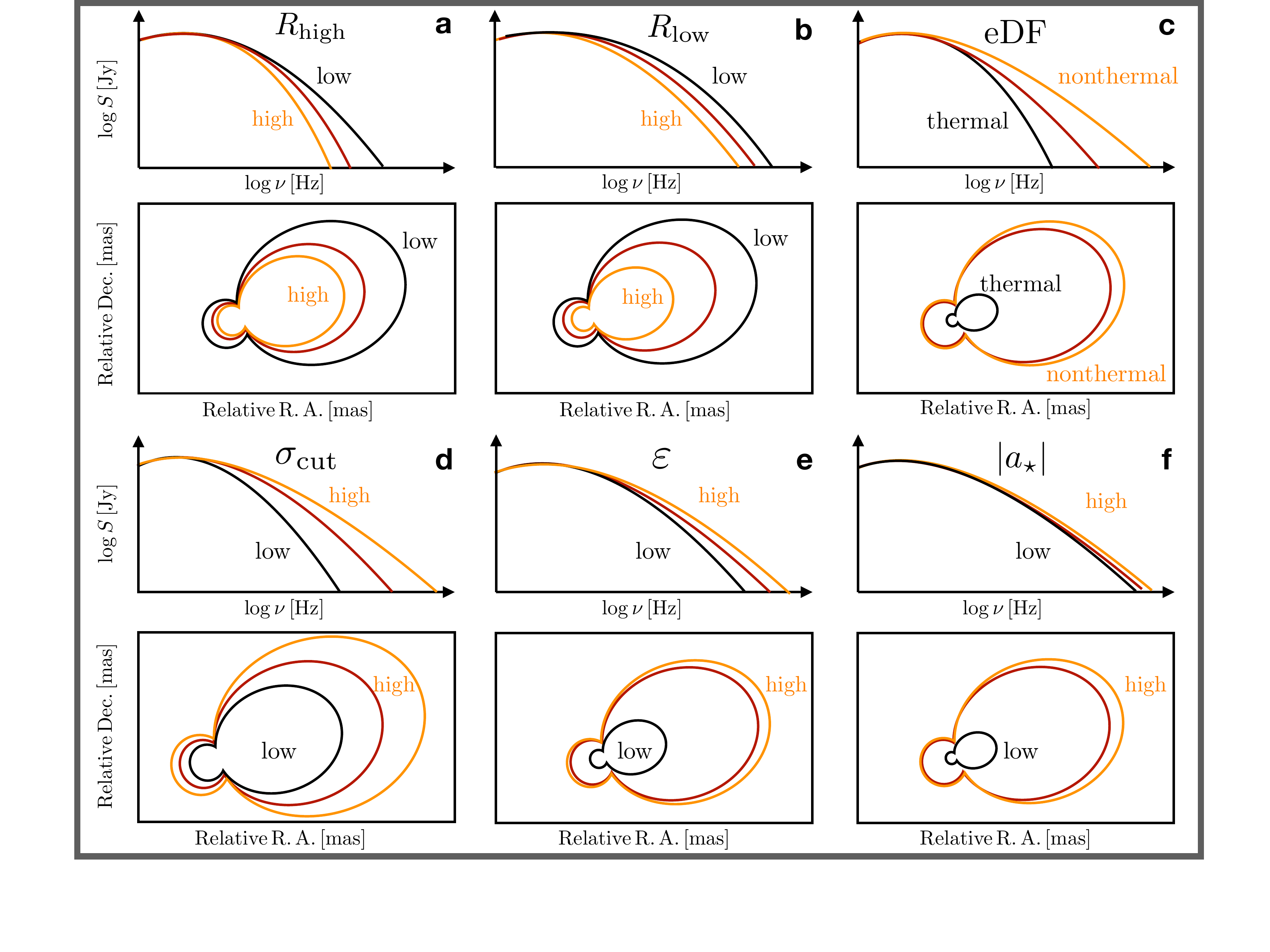}
 \caption{Schematic variation of the broad-band spectrum and image structure as a function various model parameters.}
\label{fig:pspacesummary}
\end{figure*}

\section{Summary and discussion}\label{sec:discussion}
M87 is the perfect laboratory to challenge our current understanding of jet launching and plasma physics. Therefore, we investigated which models are able to reproduce simultaneously the observed broad-band spectrum and image structure of M\,87. The workflow for this analysis can be divided into three main branches: 
\begin{itemize}
    \item 3D GRMHD simulations
    \item multi-frequency GRRT calculations
    \item comparison with observations
\end{itemize} 
Since our models are based on initial assumptions and require a set of adjustable parameters, we performed a detailed parameter space survey to better understand their impact on the spectrum 
and image structure.

\subsection{GRMHD simulations}
Given the a-priori unknown spin and accretion model for M\,87, our 3D GRMHD simulations cover five different spin values including pro- and retrograde black holes and two accretion 
models. This set of numerical simulations allows us to cover a fraction of the possible configuration of disk-jet system in M\,87. The simulations are performed in modified Kerr-Schild coordinates together with 
three AMR levels which capture the accretion process close the black hole with high resolution and ensures to resolve the fastest growing MRI mode throughout the course of the simulation 
\citep[see][] {Porth2019}. All of our GRMHD simulations are evolved until a quasi steady-state in the mass accretion is achieved. In the case of the MAD models the all models, expect for the fastest 
spinning retrograte black hole, are above the MAD parameter of $\Psi>10$. In agreement with the literature our MAD models show wide and highly magnetised jets as compared to the SANE 
models \citep[see, e.g.,][]{Tchekhovskoy2011,Chael2019}. These high-resolution 3D GRMHD simulation provide accurate plasma dynamics for our radiative transfer calculations and our 
investigation of the physical conditions in M\,87.

\subsection{GRRT calculations}
We performed a detailed parameter survey to understand the influence of the different parameters including the eDF on the broad-band spectrum and the 86\,GHz image structure. To ensure that 
the compact flux of our models at 230\,GHz are in agreement with the EHT observations we iterated the mass accretion rate to obtain an average flux of 1\,Jy at 230\,GHz within an time window of 
2000\,M. Besides the black hole spin, $a_\star$ and accretion model (MAD or SANE) there are five additional parameters namely, $R_{\rm high}$, $R_{\rm low}$, the eDF (thermal or nonthermal), 
$\sigma_{\rm cut}$ and $\varepsilon$. During the parameter survey we allow one parameter to vary while keeping the other parameters fixed. In Fig.~\ref{fig:pspacesummary} we present a 
qualitative summary of the variation of the spectrum and image structure for the different parameters under investigation. In each of the panels we show for the indicated parameter the evolution of 
the spectrum (top) and the image structure, i.e., outermost contour line (bottom). The black, red, and orange lines indicate a low, medium and high value for the parameter under investigation and 
its impact on the spectrum and on the 86\,GHz image structure. The separation between the contour lines indicate the sensitivity of the jet structure on changes in a certain parameter. For example, 
switching from a thermal to a non-thermal eDF significantly alters the jet structure (see panel c) while changes in the $R_{\rm high}$ value for a given eDF modifies the jet structure in a less strong 
way. 

The the distribution of the plasma-$\beta$ in the GRMHD simulations together with the R-$\beta$ description for the temperature ratio (Eq.~\ref{eq:Te}) leads during the GRRT to lower NIR fluxes for increasing $R_{\rm high}$ (see panel a and Fig~\ref{fig:pspacesummary}) while the NIR fluxes increase if $R_{\rm low}$ is decreased.
In addition, our parameter space survey shows, independent of the spin and accretion model, that the eDF has the largest impact on broad band spectrum (best see in the slope at high frequencies 
($\nu>10^{12}$\,Hz) in panel c in Fig.~\ref{fig:pspacesummary} and Fig.~\ref{fig:Rhsweep}) in agreement with the findings of \citet{Davelaar2019} for a high spinning SANE model. The second 
important outcome of our parameter survey is the influence of the $\sigma_{\rm cut}$ on the spectrum. For both accretion models, excluding most of the magnetised jet region via $\sigma_{\rm 
cut}<1$ 
during the GRRT highly underproduces the NIR fluxes while $\sigma_{\rm cut}>3$ overproduces the NIR flux and leads to too flat spectral slopes (see panel d in Fig.~\ref{fig:pspacesummary} 
and Fig.~\ref{fig:jetspine}) . A similar behaviour was found by \citet{Chael2019} for a MAD models with $a_{\star}=0.9375$ including electron heating.
It is interesting to mention, that one can find a combination of emission parameters which lead to very similar broad-band spectra for different spins (see panel f in Fig.~\ref{fig:pspacesummary} and 
Fig.~\ref{fig:spin}). Thus, we conclude that fitting the broad-band spectrum alone will not provide constraints on the black hole spin.

\begin{figure*}[h!]
\centering
\includegraphics[width=0.95\textwidth]{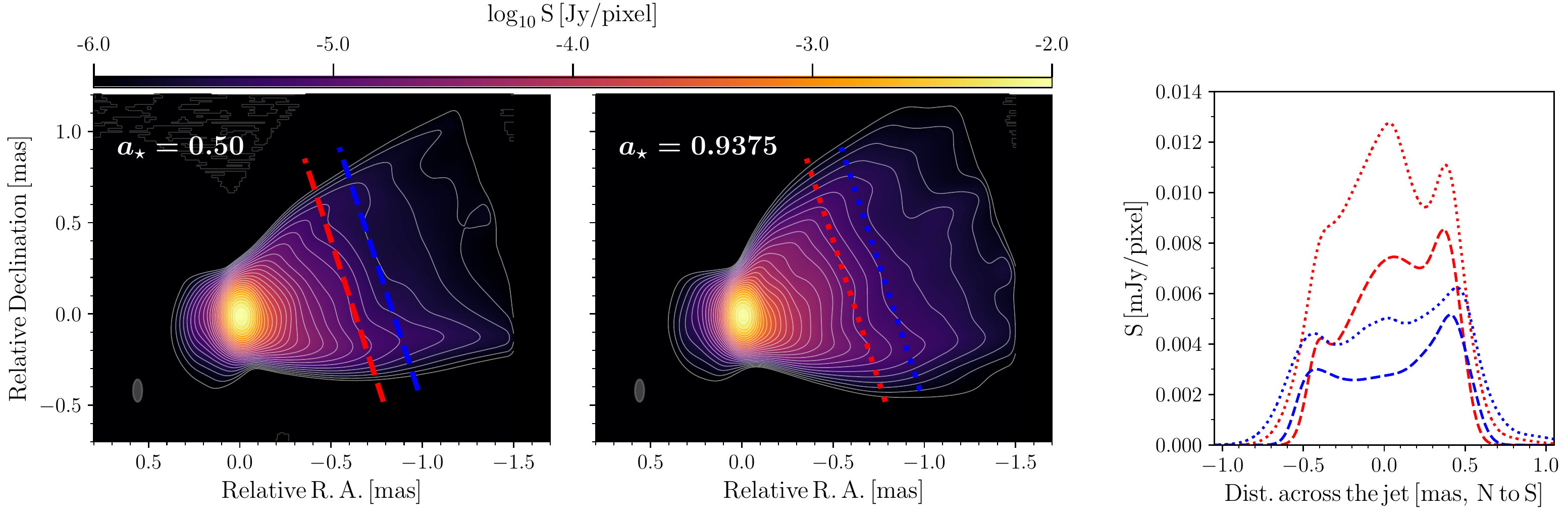}
 \caption{Best jet models (left panel for MAD $a_{\star}=0.50$ and middle panel for MAD $a_{\star}=0.9375$) convolved with a beam of $0.123\,\rm{mas}\times 0.051\,{\rm mas}$ (plotted as grey 
ellipse at the bottom left of the panels) in order to mimic the averaged and super-resolved 86\,GHz GMVA image of \citet{Kim2018a}. The right panel presents slices through the jet a $r\sim 
0.6\,\rm{mas}$ 
(red curves) and at $r\sim0.8\,\rm{mas}$.}
\label{fig:bestsliced}
\end{figure*}

\begin{table*}
\caption{Averaged quantities for the best fits MAD models. The columns show the model name, black hole spin, $a_\star$, mass accretion rate $\dot{M}$, magnetic field  and bulk Lorentz factor, 
where the last two quantities are split into the spine, sheath and disc/wind region.}
\begin{adjustbox}{max width=1.2\textwidth}
\centering
\begin{tabular}{lcccccccc}
\hline
\\[\dimexpr-\normalbaselineskip+2pt]
 Model   &  $a_\star$               &$\rm  \langle \dot{M}\rangle[M_{\odot}\,\mathrm{yr}^{-1}]$& $\rm  \langle B \rangle_{spine}[Gauss]$& $\rm  \langle B \rangle _{sheath}[Gauss]$& $\rm  \langle B 
\rangle _{disc}[Gauss]$ & $ \langle \Gamma \rangle_{\rm spine}$ & $ \langle \Gamma \rangle_{\rm sheath}$ &$ \langle \Gamma \rangle_{\rm disc+wind}$\\ [2pt]
\hline
                   & &  max\,/\,min                                  &  max\,/\,min                                       &  max\,/\,min &  max\,/\,min                                  &  max\,/\,min                                       &  max\,/\,min    &max\,/\,min    \\
\hline
\hline
\\[\dimexpr-\normalbaselineskip+2pt]
$\texttt{MT.M.1}$& $-15/16$& $3.99\times 10^{-4}$ & $1.5\times 10^{2}/2.9\times 10^{1}$ & $1.2\times 10^{2}/2.8\times 10^{1}$ & $2.9\times 10^{1}/2.8\times 10^{1}$ & $9.9/1.0$ & $9.4/1.0$ & 
$1.1/1.0$ \\ [2pt]
$\texttt{MT.M.2}$& $-1/2$       & $3.42\times 10^{-4}$ & $1.1\times 10^{2}/2.3\times 10^{1}$ & $8.4\times 10^{1}/2.6\times 10^{1}$ & $2.7\times 10^{1}/2.3\times 10^{1}$ & $4.3/1.0$ & $4.2/1.0$ & 
$1.2/1.0$ \\[2pt]
$\texttt{MT.M.3}$& $0$   & $2.50\times 10^{-4}$ & $9.3\times 10^{1}/6.9\times 10^{0}$ & $3.2\times 10^{1}/2.2\times 10^{1}$ & $2.3\times 10^{1}/6.9\times 10^{0}$ & $4.2/1.0$ & $1.6/1.0$ & 
$1.8/1.0$ \\[2pt]
$\texttt{MT.M.4}$& $+1/2$    & $2.32\times 10^{-4}$ & $9.0\times 10^{1}/1.2\times 10^{1}$ & $9.4\times 10^{1}/2.2\times 10^{1}$ & $2.2\times 10^{1}/1.2\times 10^{1}$ & $5.9/1.0$ & $5.3/1.0$ & 
$2.0/1.0$ \\[2pt]
$\texttt{MT.M.5}$&$+15/16$& $1.06\times 10^{-4}$ & $6.9\times 10^{1}/1.4\times 10^{1}$ & $7.7\times 10^{1}/1.5\times 10^{1}$ & $1.5\times 10^{1}/1.4\times 10^{1}$ & $9.3/1.0$ & $9.4/1.0$ & 
$1.3/1.0$ \\[2pt]
%\hline
%\hline
%\\[\dimexpr-\normalbaselineskip+2pt]
%$\texttt{MT.S.1}$& $-15/16$& $3.37 \times 10^{-2}$& $3.1\times 10^{2}/2.3\times 10^{2}$ & $3.2\times 10^{2}/2.4\times 10^{2}$ & $2.7\times 10^{2}/2.3\times 10^{2}$ & $1.6/1.0$ & $2.6/1.0$ & 
%$2.6/1.0$ \\[2pt]
%$\texttt{MT.S.2}$& $-1/2$    & $3.28\times 10^{-2}$ & $4.6\times 10^{2}/2.4\times 10^{2}$ & $3.0\times 10^{2}/2.5\times 10^{2}$ & $2.6\times 10^{2}/2.4\times 10^{2}$ & $2.0/1.0$ & $2.8/1.0$ & 
%$2.8/1.0$ \\[2pt]
%$\texttt{MT.S.3}$& $0$& $1.99\times 10^{-2}$ & $5.2\times 10^{2}/1.3\times 10^{2}$ & $2.1\times 10^{2}/2.0\times 10^{2}$ & $2.0\times 10^{2}/1.3\times 10^{2}$ & $3.6/1.0$ & $3.0/1.0$ & 
%$3.0/1.0$ \\[2pt]
%$\texttt{MT.S.4}$& $+1/2$   & $1.40\times 10^{-2}$ & $4.5\times 10^{2}/1.0\times 10^{2}$ & $2.1\times 10^{2}/1.5\times 10^{2}$ & $1.7\times 10^{2}/1.0\times 10^{2}$ & $3.9/1.0$ & $3.3/1.0$ & 
%$3.3/1.0$ \\[2pt]
%$\texttt{MT.S.5}$&$+15/16$&$9.48\times 10^{-3}$ & $2.3\times 10^{2}/7.6\times 10^{1}$ & $2.2\times 10^{2}/7.0\times 10^{1}$ & $1.4\times 10^{2}/7.0\times 10^{1}$ & $3.3/1.0$ & $4.4/1.0$ & 
%$4.4/1.0$ \\[2pt]
\hline
\hline
\end{tabular}
\label{tab:averagebestfit}
\end{adjustbox}
\end{table*} 

The above mentioned variations in the broad-band spectrum are mainly independent of the black hole spin and accretion model. However, the structure of the jet strongly depends on these two 
parameters. As mentioned earlier, the black hole spin and the accretion model are initial conditions of the GRMHD simulations and determine the plasma dynamics including the properties of the 
jets, e.g., width and its magnetisation. MAD models produce wide and highly magnetised jets as compared to their SANE counterparts. Thus, in the numerical simulations used in this work, the jet diameter and jet opening angle of MADs will always be 
larger than for SANE models irrespective of the adjustable parameters during the GRRT (see bottom panels in Figs. \ref{fig:Rhsweep} - \ref{fig:spin}). Similar, fast rotating black holes generate more powerful jets and 
more extended jets than slow rotating black holes (see Fig. \ref{fig:spin}).

\subsection{Comparison with observations}
After understanding the impact of the initial conditions of the GRMHD and the free parameters during the GRRT on the broad-band spectrum and the image structure we can finally compare our 
models to the observations of M\,87. The broad-band spectrum for M\,87 is remarkably well fitted for the MAD models using non-thermal particles via the kappa eDF with $\varepsilon=0.5$, 
$R_{\rm low}=1$, $R_{\rm high}=160$ and $\sigma_{\rm cut}=3$ (see top left panel in Fig.~\ref{fig:spin}). However, as mentioned above fitting the SED alone does not provide constraints on the 
black hole spin. Therefore, we extracted the jet collimation profiles from our 86\,GHz images and compared them to ones obtained from 86\,GHz GMVA observations \citep{Kim2018a}. This analysis 
shows that the high-spin prograde MAD models produce collimation profiles which are in agreement with the currently highest resolution VLBI images\footnote{the 2017 EHT observations do not 
detect a clear jet signature due the limited SNR} of the jet in M\,87 (see bottom panels in Fig.~\ref{fig:spin}).  In Figure \ref{fig:bestsliced} we present our best fit images convolved with a 0.123\,mas 
$\times$ 0.051\,mas beam and two slices through the jet at $r\sim 0.6$\,mas and at $r\sim0.8$\,mas which could be compared to Fig.~4 in \citet{Kim2018a}. Both models show a large jet opening 
angle and a significant signature of a counter jet. Furthermore, the jet slices show edge-brightening together with a third flux density maxima enclosed by the outer ridges. This is best seen  for the MAD $a_\star=0.9375$ model at $r=0.8$\,mas (dotted blue line in the right panel of Fig.~\ref{fig:bestsliced}). 

In addition to the convolved images presented in Fig. \ref{fig:bestsliced} we list in Table~\ref{tab:averagebestfit} the average magnetic field and bulk Lorentz factors for the best fit models. Our favoured accretion model (MAD) together with the estimates for the magnetic field and mass accretion rate agree with the results obtained from the modelling of the horizon scale polarisation structure of M\,87 ($1-30$\,G and $\dot{m}=3-20\times10^{-4}M_{\astrosun}/\rm yr$) \citep{EHT_M87pol}. It is important to mention, that these two set of values are obtained from independent  studies using different methods and underlying observations, confirming the physical conditions in the jet launching region of M\,87 presented here. Furthermore, our values for the bulk Lorentz factors, $\Gamma$, are roughly in agreement with the values derived from 43\,GHz VLBA and 86\,GHz GMVA observations \citep{Mertens2016,Walker2018,Kim2018a}. Notice the difference in the 
nomenclature between our work and the observations: Since we exclude the jet spine from our GRRT calculations our definition of the jet sheath is similar to definition of the jet spine in the observations. With this in mind, the observations suggest $\Gamma>6$ in the jet spine (our jet sheath) depending on the viewing angle and $\Gamma\leq 2$ in the jet wind region 
\citep{Mertens2016,Walker2018,Kim2018a} similar to the listed values in Table~\ref{tab:averagebestfit}.

\subsection{Limitations of the models}
In this work we use ideal GRMHD which does not include resistivity nor heating or cooling of the plasma. Including resistivity could lead production of hotter plasmoids in the disk and jet sheath 
region which could lead to more pronounced edge-brightening effects \citep{Ripperda2020} and stronger disk winds \citep{Vourellis2019}. By using electron heating models such as turbulent and 
reconnection heating the electron temperature can be directly obtained from the GRMHD simulations and the post-processing step via Eq. \ref{eq:Te} is not required \citep[see, e.g.,][]{Chael2019}. 
\newline Despite these limitations we do not expect large impacts on the results presented here: \citet{Nathanail2020}, \citet{Dihingia2021} and \citet{Chashkina2021} showed that the formation of plasmoids using ideal 
GRMHD is similar to the ones from resistive GRMHD. In addition, the recent work by \citet{Mizuno2021} revealed that the $R-\beta$ model for the electron temperature is well matched to heating 
models for $R_{\rm low}=1$ and $R_{\rm high}\leq 160$. However, future studies should include both resistivity and electron heating which will provide deeper insights into the jet launching 
process and while reducing the number of free parameters in the modelling.

\section{Conclusion and outlook}\label{sec:conc}
In this paper we presented long-term and high-resolution 3D GRMHD simulations of "standard and normal evolution" disks and "magnetically arrested disks" accretion around spinning black hole to 
investigate the launching of the M\,87 jet. From the 3D GRMHD simulations we computed the broad-band spectrum ($10^9\leq \nu\leq 10^{16}$\,Hz) and the jet collimation profiles. During the 
GRRT we investigated the influence of the electron temperature via the $R-\beta$ model, the electron distribution function (thermal and non-thermal particles) and the jet sheath on the radiative signatures. From our detailed parameter survey we found that the best fit to the broad-band spectrum (see Table \ref{Tab:obsSED} for the used observational data) is obtained if non-thermal  particles energetically supported by the magnetic field ($\varepsilon=0.5$) are included in a magnetised jet sheath ($\sigma_{\rm cut}=3$). Interestingly, the broad-band spectrum could be equally 
well fitted independent of the spin of the black hole and the accretion model. To break this degeneracy in the models we included the jet collimation profile in our analysis and compared it to the observed one extracted from 86\,GHz GMVA observations \citep{Kim2018a}.  The combined constrain of broad-band spectrum and collimation profiles favoured the MAD accretion model together with fast spinning prograde black holes ($a_\star\geq0.5$). We show that our results are in agreement with the values obtained from polarisation modelling and from estimates extracted from VLBI observations of M\,87. In addition to the provided best fit models one of the major results of this work is that the combined modelling of broad-band spectrum and additional constraints such as the image structure can help to further constrain possible models for M\,87. 
\newline Future work on modelling the jet launching and propagation M\,87 should include more advanced, non-standard GRMHD simulations including resistivity and electron heating models. In 
order to further investigate the physical conditions in M\,87 additional observational constraints such as the polarisation structure and high-energy radiation (x-rays) should be computed during the 
radiative transfer. Finally, to reduce the influence of hyper-parameter included in the image reconstructions and to account for realistic observational conditions ( e.g., sparse sampling of the u-v plane and antenna calibration uncertainties) a direct comparison between synthetic visibilities and measured ones should be performed. However, such a detailed direct fitting of the GRMHD simulations to the observations is computationally challenging and requires specialised numerical methods and codes currently under development \citep[see, e.g.,][]{Fromm2019}.

\section{Acknowledgments}
This research is supported by the ERC synergy
  grant ``BlackHoleCam: Imaging the Event Horizon of Black Holes" (Grant
  No. 610058). CMF is supported by the Black Hole Initiative at Harvard
  University, which is supported by a grant from the John Templeton
  Foundation. AN was supported by the Hellenic Foundation for 
  Research and Innovation (H.F.R.I.) under the “2nd Call for H.F.R.I. Research 
  Projects to support Post-Doctoral Researchers” (Project Number: 00634).  
  ZY is supported by a UKRI Stephen Hawking Fellowship and 
  acknowledges support from a Leverhulme Trust Early Career Fellowship.
  JD is supported by NASA grant NNX17AL82 and a Joint
  Columbia/Flatiron Postdoctoral Fellowship. Research at the Flatiron
  Institute is supported by the Simons Foundation. The simulations were
  performed on GOETHE-HLR LOEWE at the CSC-Frankfurt, Iboga at ITP Frankfurt
  and Pi2.0 at Shanghai Jiao Tong University.
 
{\it Software:} {\tt BHAC}\footnote{\href{https://bhac.science/}{https://bhac.science/}} \citep{Porth2017}, {\tt BHOSS} \citep{Younsi2020}, {\tt 
ehtim}\footnote{\href{https://achael.github.io/eht-imaging/}{https://achael.github.io/eht-imaging/}} 
\citep{Chael2018}

\bibliographystyle{aa}
\bibliography{biblio}

\begin{thebibliography}{63}
\expandafter\ifx\csname natexlab\endcsname\relax\def\natexlab#1{#1}\fi

\bibitem[{{Akiyama} {et~al.}(2015){Akiyama}, {Lu}, {Fish}, {Doeleman},
  {Broderick}, {Dexter}, {Hada}, {Kino}, {Honma}, {Johnson}, {Algaba}, {Asada},
  {Brinkerink}, {Blundell}, \& et~al.}]{Akiyama2015}
{Akiyama}, K., {Lu}, R.-S., {Fish}, V.~L., {et~al.} 2015, Astrophys. J., 807,
  150

\bibitem[{{An} {et~al.}(2018){An}, {Sohn}, \& {Imai}}]{An2018}
{An}, T., {Sohn}, B.~W., \& {Imai}, H. 2018, Nature Astronomy, 2, 118

\bibitem[{{Ant{\'o}n} {et~al.}(2006){Ant{\'o}n}, {Zanotti}, {Miralles},
  {Mart{\'{\i}}}, {Ib{\'a}{\~n}ez}, {Font}, \& {Pons}}]{Anton06}
{Ant{\'o}n}, L., {Zanotti}, O., {Miralles}, J.~A., {et~al.} 2006, Astrophys.
  J., 637, 296

\bibitem[{{Asada} \& {Nakamura}(2012)}]{Asada2012}
{Asada}, K. \& {Nakamura}, M. 2012, Astrophys. J. l, 745, L28

\bibitem[{{Ball} {et~al.}(2018){Ball}, {Sironi}, \& {{\"O}zel}}]{Ball2018a}
{Ball}, D., {Sironi}, L., \& {{\"O}zel}, F. 2018, Astrophys. J., 862, 80

\bibitem[{{Blandford} {et~al.}(2019){Blandford}, {Meier}, \&
  {Readhead}}]{Blandford2019}
{Blandford}, R., {Meier}, D., \& {Readhead}, A. 2019, \araa, 57, 467

\bibitem[{Blandford \& Payne(1982)}]{Blandford:1982di}
Blandford, R.~D. \& Payne, D.~G. 1982, Mon. Not. R. Astron. Soc., 199, 883

\bibitem[{{Blandford} \& {Znajek}(1977)}]{Blandford1977}
{Blandford}, R.~D. \& {Znajek}, R.~L. 1977, Mon. Not. R. Astron. Soc., 179, 433

\bibitem[{{Broderick} \& {Loeb}(2009)}]{Broderick2009}
{Broderick}, A.~E. \& {Loeb}, A. 2009, Astrophys. J., 697, 1164

\bibitem[{{Chael} {et~al.}(2019){Chael}, {Narayan}, \& {Johnson}}]{Chael2019}
{Chael}, A., {Narayan}, R., \& {Johnson}, M.~D. 2019, \mnras, 486, 2873

\bibitem[{{Chael} {et~al.}(2018){Chael}, {Johnson}, {Bouman}, {Blackburn},
  {Akiyama}, \& {Narayan}}]{Chael2018}
{Chael}, A.~A., {Johnson}, M.~D., {Bouman}, K.~L., {et~al.} 2018, \apj, 857, 23

\bibitem[{{Chashkina} {et~al.}(2021){Chashkina}, {Bromberg}, \&
  {Levinson}}]{Chashkina2021}
{Chashkina}, A., {Bromberg}, O., \& {Levinson}, A. 2021, arXiv e-prints,
  arXiv:2106.15738

\bibitem[{{Cruz-Osorio} {et~al.}(2021){Cruz-Osorio}, {Fromm}, {Mizuno},
  {Nathanail}, {Younsi}, {Porth}, {Davelaar}, {Falke}, {Kramer}, , \&
  {Rezzolla}}]{Cruz2021b}
{Cruz-Osorio}, A., {Fromm}, C.~M., {Mizuno}, Y., {et~al.} 2021, Nature,
  accepted

\bibitem[{{Cruz-Osorio} {et~al.}(2020){Cruz-Osorio}, {Gimeno-Soler}, \&
  {Font}}]{Cruz2020}
{Cruz-Osorio}, A., {Gimeno-Soler}, S., \& {Font}, J.~A. 2020, Mon. Not. R.
  Astron. Soc., 492, 5730

\bibitem[{{Daigne} \& {Font}(2004)}]{Daigne04}
{Daigne}, F. \& {Font}, J.~A. 2004, Mon. Not. R. Astron. Soc., 349, 841

\bibitem[{{Davelaar} {et~al.}(2019){Davelaar}, {Olivares}, {Porth},
  {Bronzwaer}, {Janssen}, {Roelofs}, {Mizuno}, {Fromm}, {Falcke}, \&
  {Rezzolla}}]{Davelaar2019}
{Davelaar}, J., {Olivares}, H., {Porth}, O., {et~al.} 2019, Astron. Astrophys.,
  632, A2

\bibitem[{{Dexter} {et~al.}(2012){Dexter}, {McKinney}, \& {Agol}}]{Dexter2012}
{Dexter}, J., {McKinney}, J.~C., \& {Agol}, E. 2012, Mon. Not. R. Astron. Soc.,
  421, 1517

\bibitem[{{Di Matteo} {et~al.}(2003){Di Matteo}, {Allen}, {Fabian}, {Wilson},
  \& {Young}}]{DiMatteo2003}
{Di Matteo}, T., {Allen}, S.~W., {Fabian}, A.~C., {Wilson}, A.~S., \& {Young},
  A.~J. 2003, Astrophys. J., 582, 133

\bibitem[{{Dihingia} {et~al.}(2021){Dihingia}, {Vaidya}, \&
  {Fendt}}]{Dihingia2021}
{Dihingia}, I.~K., {Vaidya}, B., \& {Fendt}, C. 2021, \mnras, 505, 3596

\bibitem[{{Doeleman} {et~al.}(2012){Doeleman}, {Fish}, {Schenck}, {Beaudoin},
  {Blundell}, {Bower}, {Broderick}, {Chamberlin}, {Freund}, {Friberg},
  {Gurwell}, {Ho}, {Honma}, {Inoue}, {Krichbaum}, {Lamb}, {Loeb}, {Lonsdale},
  {Marrone}, {Moran}, {Oyama}, {Plambeck}, {Primiani}, {Rogers}, {Smythe},
  {SooHoo}, {Strittmatter}, {Tilanus}, {Titus}, {Weintroub}, {Wright}, {Young},
  \& {Ziurys}}]{Doeleman2012}
{Doeleman}, S.~S., {Fish}, V.~L., {Schenck}, D.~E., {et~al.} 2012, Science,
  338, 355

\bibitem[{{EHT MWL Science Working Group} {et~al.}(2021){EHT MWL Science
  Working Group}, {Algaba}, {Anczarski}, {Asada}, {Balokovi{\'c}}, {Chandra},
  {Cui}, {Falcone}, {Giroletti}, {Goddi}, {Hada}, {Haggard}, {Jorstad}, {Kaur},
  \& {Kawashima}}]{Algaba2021}
{EHT MWL Science Working Group}, {Algaba}, J.~C., {Anczarski}, J., {et~al.}
  2021, Astrophys. J. Let., 911, L11

\bibitem[{{Event Horizon Telescope Collaboration}
  {et~al.}(2019{\natexlab{a}}){Event Horizon Telescope Collaboration},
  {Akiyama}, {Alberdi}, {Alef}, {Asada}, {Azulay}, {Baczko}, {Ball},
  {Balokovi{\'c}}, {Barrett}, {et~al.}}]{EHT_M87_PaperI}
{Event Horizon Telescope Collaboration}, {Akiyama}, K., {Alberdi}, A., {et~al.}
  2019{\natexlab{a}}, Astrophys. J. Lett., 875, L1

\bibitem[{{Event Horizon Telescope Collaboration}
  {et~al.}(2019{\natexlab{b}}){Event Horizon Telescope Collaboration},
  {Akiyama}, {Alberdi}, {Alef}, {Asada}, {Azulay}, {Baczko}, {Ball},
  {Balokovi{\'c}}, {Barrett}, {et~al.}}]{EHT_M87_PaperV}
{Event Horizon Telescope Collaboration}, {Akiyama}, K., {Alberdi}, A., {et~al.}
  2019{\natexlab{b}}, Astrophys. J. Lett., 875, L5

\bibitem[{{Event Horizon Telescope Collaboration} {et~al.}(2021){Event Horizon
  Telescope Collaboration}, {Akiyama}, {Algaba}, {Alberdi}, {Alef}, {Anantua},
  {Asada}, {et~al.}}]{EHT_M87pol}
{Event Horizon Telescope Collaboration}, {Akiyama}, K., {Algaba}, J.~C.,
  {et~al.} 2021, \apjl, 910, L13

\bibitem[{{Fishbone} \& {Moncrief}(1976)}]{Fishbone76}
{Fishbone}, L.~G. \& {Moncrief}, V. 1976, Astrophys. J., 207, 962

\bibitem[{Font \& Daigne(2002)}]{Font02b}
Font, J.~A. \& Daigne, F. 2002, Astrophys. J, 581, L23

\bibitem[{{Fromm} {et~al.}(2019){Fromm}, {Younsi}, {Baczko}, {Mizuno}, {Porth},
  {Perucho}, {Olivares}, {Nathanail}, {Angelakis}, {Ros}, {Zensus}, \&
  {Rezzolla}}]{Fromm2019}
{Fromm}, C.~M., {Younsi}, Z., {Baczko}, A., {et~al.} 2019, \aap, 629, A4

\bibitem[{{Hada} {et~al.}(2013){Hada}, {Kino}, {Doi}, {Nagai}, {Honma},
  {Hagiwara}, {Giroletti}, {Giovannini}, \& {Kawaguchi}}]{Hada2013}
{Hada}, K., {Kino}, M., {Doi}, A., {et~al.} 2013, Astrophys. J., 775, 70

\bibitem[{{Hada} {et~al.}(2017){Hada}, {Park}, {Kino}, {Niinuma}, {Sohn}, {Ro},
  {Jung}, {Algaba}, {Zhao}, {Lee}, {Akiyama}, {Trippe}, {Wajima},
  {Sawada-Satoh}, {Tazaki}, {Cho}, {Hodgson}, {Lee}, {Hagiwara}, {Honma},
  {Koyama}, {Oh}, {Lee}, {Yoo}, {Kawaguchi}, {Roh}, {Oh}, {Yeom}, {Jung}, {Oh},
  {Kim}, {Hwang}, {Byun}, {Cho}, {Kim}, {Kobayashi}, \& {Shibata}}]{Hada2017}
{Hada}, K., {Park}, J.~H., {Kino}, M., {et~al.} 2017, Publications of the ASJ,
  69, 71

\bibitem[{{Kim} {et~al.}(2018{\natexlab{a}}){Kim}, {Krichbaum}, {Lu}, {Ros},
  {Bach}, {Bremer}, {de Vicente}, {Lindqvist}, \& {Zensus}}]{Kim2018a}
{Kim}, J.~Y., {Krichbaum}, T.~P., {Lu}, R.~S., {et~al.} 2018{\natexlab{a}},
  Astron. Astrophys., 616, A188

\bibitem[{{Kim} {et~al.}(2018{\natexlab{b}}){Kim}, {Lee}, {Hodgson}, {Algaba},
  {Zhao}, {Kino}, {Byun}, \& {Kang}}]{Kim2018b}
{Kim}, J.-Y., {Lee}, S.-S., {Hodgson}, J.~A., {et~al.} 2018{\natexlab{b}},
  Astron. Astrophys., 610, L5

\bibitem[{{Lister} {et~al.}(2018){Lister}, {Aller}, {Aller}, {Hodge}, {Homan},
  {Kovalev}, {Pushkarev}, \& {Savolainen}}]{Lister2018}
{Lister}, M.~L., {Aller}, M.~F., {Aller}, H.~D., {et~al.} 2018, Astrophys. J.,
  Supp., 234, 12

\bibitem[{{Lora-Clavijo} {et~al.}(2015){Lora-Clavijo}, {Cruz-Osorio}, \&
  {Guzm{\'a}n}}]{Lora2015}
{Lora-Clavijo}, F.~D., {Cruz-Osorio}, A., \& {Guzm{\'a}n}, F.~S. 2015,
  Astrophys. J., Supp., 218, 24

\bibitem[{{Lynden-Bell}(2006)}]{LyndenBell2006}
{Lynden-Bell}, D. 2006, \mnras, 369, 1167

\bibitem[{{MAGIC Collaboration} {et~al.}(2020){MAGIC Collaboration}, {Acciari},
  {Ansoldi}, {Antonelli}, {Arbet Engels}, {Arcaro}, {Baack}, {Babi{\'c}},
  {Banerjee}, {Bangale}, {Barres de Almeida}, {Barrio}, {Becerra Gonz{\'a}lez},
  \& {Bednarek}}]{MAGIC2020}
{MAGIC Collaboration}, {Acciari}, V.~A., {Ansoldi}, S., {et~al.} 2020, Mon.
  Not. R. Astron. Soc., 492, 5354

\bibitem[{{Mertens} {et~al.}(2016){Mertens}, {Lobanov}, {Walker}, \&
  {Hardee}}]{Mertens2016}
{Mertens}, F., {Lobanov}, A.~P., {Walker}, R.~C., \& {Hardee}, P.~E. 2016,
  Astron. Astrophys., 595, A54

\bibitem[{{Mizuno} {et~al.}(2021){Mizuno}, {Fromm}, {Younsi}, {Porth},
  {Olivares}, \& {Rezzolla}}]{Mizuno2021}
{Mizuno}, Y., {Fromm}, C.~M., {Younsi}, Z., {et~al.} 2021, \mnras, 506, 741

\bibitem[{{Mo{\'s}cibrodzka} {et~al.}(2016){Mo{\'s}cibrodzka}, {Falcke}, \&
  {Shiokawa}}]{Moscibrodzka2016}
{Mo{\'s}cibrodzka}, M., {Falcke}, H., \& {Shiokawa}, H. 2016, Astron.
  Astrophys., 586, A38

\bibitem[{{Nakamura} {et~al.}(2018){Nakamura}, {Asada}, {Hada}, {Pu}, {Noble},
  {Tseng}, {Toma}, {Kino}, {Nagai}, {Takahashi}, {Algaba}, {Orienti},
  {Akiyama}, {Doi}, {Giovannini}, {Giroletti}, {Honma}, {Koyama}, {Lico},
  {Niinuma}, \& {Tazaki}}]{Nakamura2018}
{Nakamura}, M., {Asada}, K., {Hada}, K., {et~al.} 2018, Astrophys. J., 868, 146

\bibitem[{{Nathanail} {et~al.}(2020){Nathanail}, {Fromm}, {Porth}, {Olivares},
  {Younsi}, {Mizuno}, \& {Rezzolla}}]{Nathanail2020}
{Nathanail}, A., {Fromm}, C.~M., {Porth}, O., {et~al.} 2020, \mnras, 495, 1549

\bibitem[{{Olivares} {et~al.}(2020){Olivares}, {Younsi}, {Fromm}, {De
  Laurentis}, {Porth}, {Mizuno}, {Falcke}, {Kramer}, \&
  {Rezzolla}}]{Olivares2020}
{Olivares}, H., {Younsi}, Z., {Fromm}, C.~M., {et~al.} 2020, Mon. Not. R.
  Astron. Soc., 497, 521

\bibitem[{{Olivares S{\'{a}}nchez} {et~al.}(2018){Olivares S{\'{a}}nchez},
  {Porth}, \& {Mizuno}}]{Olivares2018a}
{Olivares S{\'{a}}nchez}, H., {Porth}, O., \& {Mizuno}, Y. 2018, J. Phys. Conf.
  Ser., 1031, 012008

\bibitem[{{Pandya} {et~al.}(2016){Pandya}, {Zhang}, {Chandra}, \&
  {Gammie}}]{Pandya2016}
{Pandya}, A., {Zhang}, Z., {Chandra}, M., \& {Gammie}, C.~F. 2016, Astrophys.
  J., 822, 34

\bibitem[{{Perlman} {et~al.}(2001){Perlman}, {Sparks}, {Radomski}, {Packham},
  {Fisher}, {Pi{\~n}a}, \& {Biretta}}]{Perlman2001}
{Perlman}, E.~S., {Sparks}, W.~B., {Radomski}, J., {et~al.} 2001, Astrophys. J.
  Lett., 561, L51

\bibitem[{{Pierrard} \& {Lazar}(2010)}]{Pierrard2010}
{Pierrard}, V. \& {Lazar}, M. 2010, \solphys, 267, 153

\bibitem[{{Porth} {et~al.}(2019){Porth}, {Chatterjee}, {Narayan}, {Gammie},
  {Mizuno}, {Anninos}, {Baker}, {Bugli}, {Chan}, {Davelaar}, {Del Zanna},
  {Etienne}, {Fragile}, {Kelly}, {Liska}, {Markoff}, {McKinney}, {Mishra},
  {Noble}, {Olivares}, {Prather}, {Rezzolla}, {Ryan}, {Stone}, {Tomei},
  {White}, {Younsi}, \& {The Event Horizon Telescope
  Collaboration}}]{Porth2019}
{Porth}, O., {Chatterjee}, K., {Narayan}, R., {et~al.} 2019, arXiv e-prints,
  arXiv:1904.04923

\bibitem[{{Porth} {et~al.}(2017){Porth}, {Olivares}, {Mizuno}, {Younsi},
  {Rezzolla}, {Moscibrodzka}, {Falcke}, \& {Kramer}}]{Porth2017}
{Porth}, O., {Olivares}, H., {Mizuno}, Y., {et~al.} 2017, Computational
  Astrophysics and Cosmology, 4, 1

\bibitem[{{Prieto} {et~al.}(2016){Prieto}, {Fern{\'a}ndez-Ontiveros},
  {Markoff}, {Espada}, \& {Gonz{\'a}lez-Mart{\'\i}n}}]{Prieto2016}
{Prieto}, M.~A., {Fern{\'a}ndez-Ontiveros}, J.~A., {Markoff}, S., {Espada}, D.,
  \& {Gonz{\'a}lez-Mart{\'\i}n}, O. 2016, Mon. Not. R. Astron. Soc., 457, 3801

\bibitem[{{Reid} {et~al.}(1982){Reid}, {Schmitt}, {Owen}, {Booth}, {Wilkinson},
  {Shaffer}, {Johnston}, \& {Hardee}}]{Reid1982}
{Reid}, M.~J., {Schmitt}, J.~H.~M.~M., {Owen}, F.~N., {et~al.} 1982, \apj, 263,
  615

\bibitem[{{Rezzolla} \& {Zanotti}(2013)}]{Rezzolla_book:2013}
{Rezzolla}, L. \& {Zanotti}, O. 2013, Relativistic Hydrodynamics (Oxford, UK:
  Oxford University Press)

\bibitem[{{Ripperda} {et~al.}(2020){Ripperda}, {Bacchini}, \&
  {Philippov}}]{Ripperda2020}
{Ripperda}, B., {Bacchini}, F., \& {Philippov}, A.~A. 2020, \apj, 900, 100

\bibitem[{{Snios} {et~al.}(2019){Snios}, {Nulsen}, {Kraft}, {Cheung}, {Meyer},
  {Forman}, {Jones}, \& {Murray}}]{Snios2019}
{Snios}, B., {Nulsen}, P. E.~J., {Kraft}, R.~P., {et~al.} 2019, Astrophys. J.
  Lett., 879, 8

\bibitem[{{Tchekhovskoy} \& {McKinney}(2012)}]{Tchekhovskoy2012}
{Tchekhovskoy}, A. \& {McKinney}, J.~C. 2012, Mon. Not. R. Astron. Soc., 423,
  L55

\bibitem[{{Tchekhovskoy} {et~al.}(2011){Tchekhovskoy}, {Narayan}, \&
  {McKinney}}]{Tchekhovskoy2011}
{Tchekhovskoy}, A., {Narayan}, R., \& {McKinney}, J.~C. 2011, Mon. Not. R.
  Astron. Soc., 418, L79

\bibitem[{{Vourellis} {et~al.}(2019){Vourellis}, {Fendt}, {Qian}, \&
  {Noble}}]{Vourellis2019}
{Vourellis}, C., {Fendt}, C., {Qian}, Q., \& {Noble}, S.~C. 2019, \apj, 882, 2

\bibitem[{{Walker} {et~al.}(2018){Walker}, {Hardee}, {Davies}, {Ly}, \&
  {Junor}}]{Walker2018}
{Walker}, R.~C., {Hardee}, P.~E., {Davies}, F.~B., {Ly}, C., \& {Junor}, W.
  2018, Astrophys. J., 855, 128

\bibitem[{{Whysong} \& {Antonucci}(2004)}]{Whysong2004}
{Whysong}, D. \& {Antonucci}, R. 2004, Astrophys. J., 602, 116

\bibitem[{{Xiao}(2006)}]{Xiao2006}
{Xiao}, F. 2006, Plasma Physics and Controlled Fusion, 48, 203

\bibitem[{{Yarza} {et~al.}(2020){Yarza}, {Wong}, {Ryan}, \&
  {Gammie}}]{Yarza2020}
{Yarza}, R., {Wong}, G.~N., {Ryan}, B.~R., \& {Gammie}, C.~F. 2020, Astrophys.
  J., 898, 50

\bibitem[{{Younsi} {et~al.}(2020){Younsi}, {Porth}, {Mizuno}, {Fromm}, \&
  {Olivares}}]{Younsi2020}
{Younsi}, Z., {Porth}, O., {Mizuno}, Y., {Fromm}, C.~M., \& {Olivares}, H.
  2020, in Perseus in Sicily: From Black Hole to Cluster Outskirts, ed.
  K.~{Asada}, E.~{de Gouveia Dal Pino}, M.~{Giroletti}, H.~{Nagai}, \&
  R.~{Nemmen}, Vol. 342, 9--12

\bibitem[{{Younsi} {et~al.}(2012){Younsi}, {Wu}, \& {Fuerst}}]{Younsi2012}
{Younsi}, Z., {Wu}, K., \& {Fuerst}, S.~V. 2012, Astron. Astrophys., 545, A13

\bibitem[{{Yuan} \& {Narayan}(2014)}]{Yuan2014}
{Yuan}, F. \& {Narayan}, R. 2014, \araa, 52, 529

\bibitem[{{Zdziarski} {et~al.}(1998){Zdziarski}, {Poutanen}, {Mikolajewska},
  {Gierlinski}, {Ebisawa}, \& {Johnson}}]{Zdziarski1998}
{Zdziarski}, A.~A., {Poutanen}, J., {Mikolajewska}, J., {et~al.} 1998, Mon.
  Not. R. Astron. Soc., 301, 435

\end{thebibliography}

\appendix

\section{Averaged jet morphologies}
\begin{figure*}[h]
\centering
\includegraphics[width=0.99\textwidth]{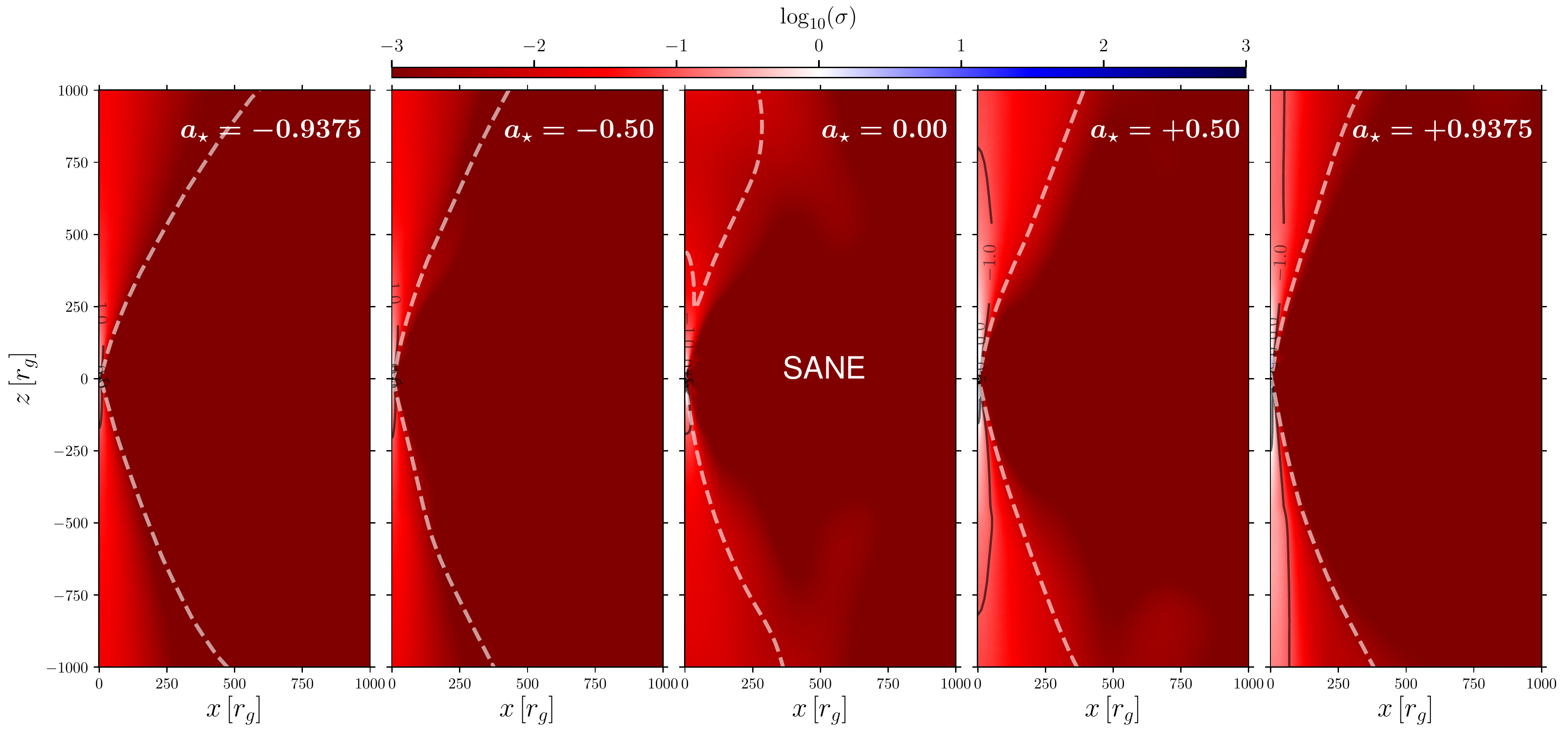}
\vspace{-0.15cm}
\includegraphics[width=0.99\textwidth]{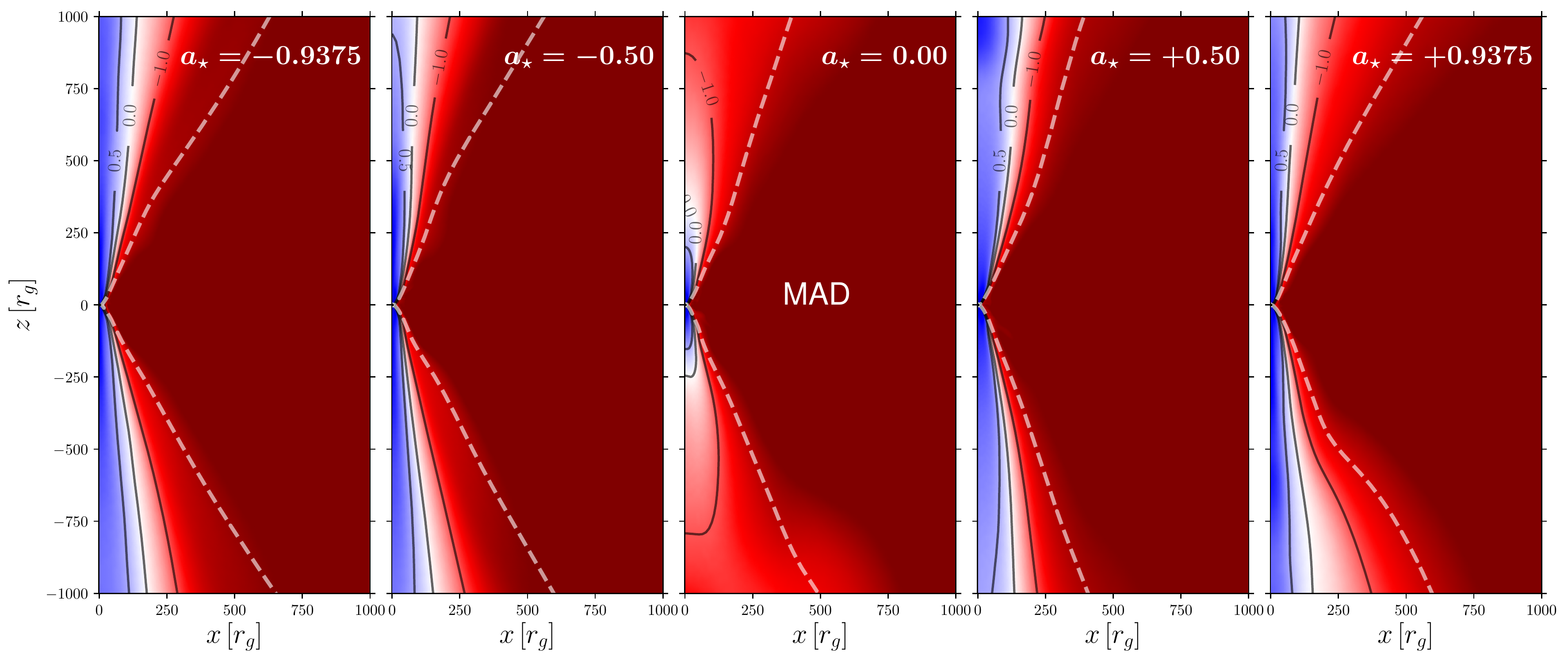}
 \caption{Azimuthal and time averaged distribution of the logarithm of the magnetisation, $
 \sigma$, for different spins (left to right) and for SANE (top) and MAD(bottom) models. The dashed white line corresponds to the Bernoulli parameter, $-hu_t=1.02$ and the black lines show 
$\log_{10}(\sigma)=-1.0,\,0.5,\,\rm{and}\, 
0$ contours.}
\label{fig:spinsigma}
\end{figure*}

\begin{figure*}[h!]
\centering
\includegraphics[width=0.99\textwidth]{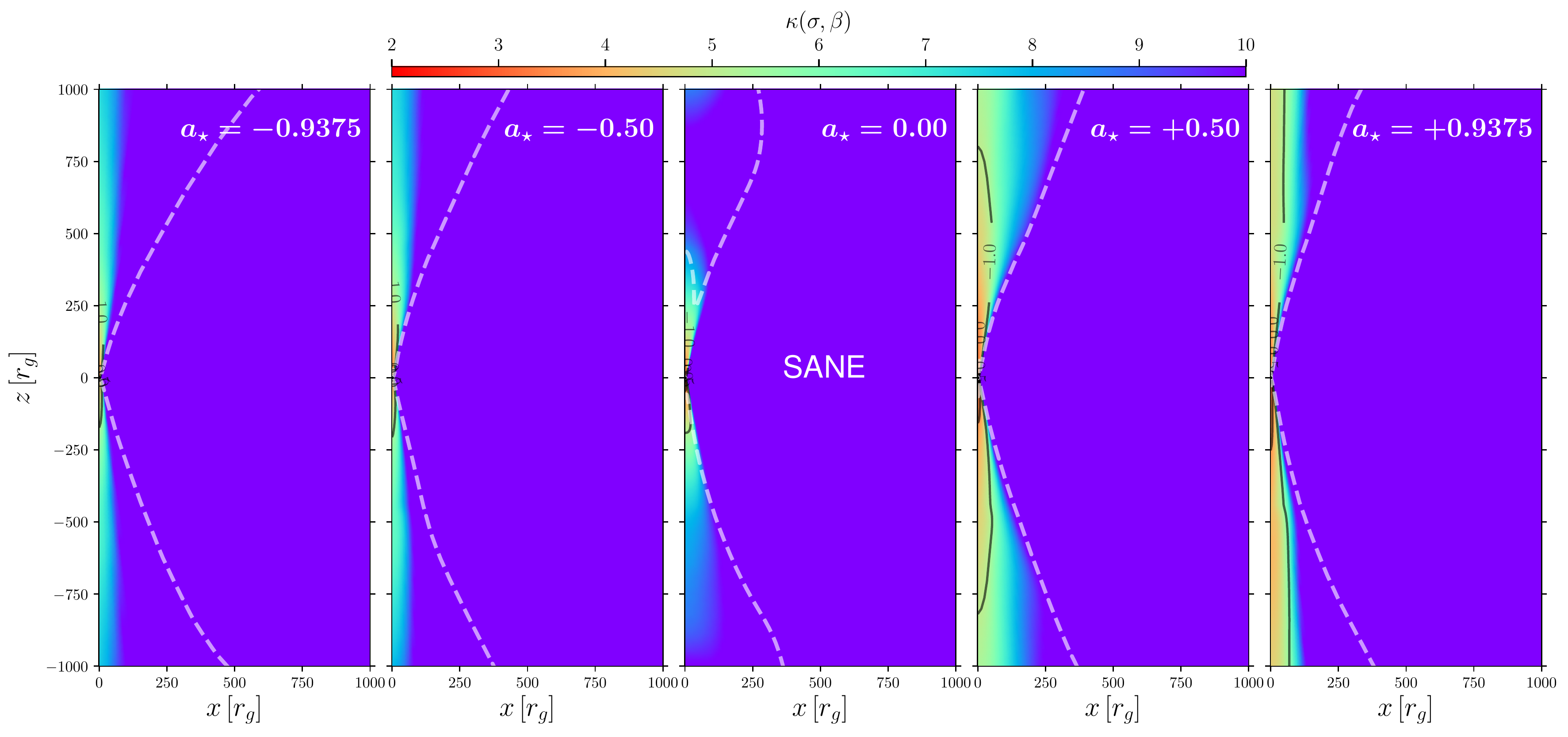}
\vspace{-0.15cm}
\includegraphics[width=0.99\textwidth]{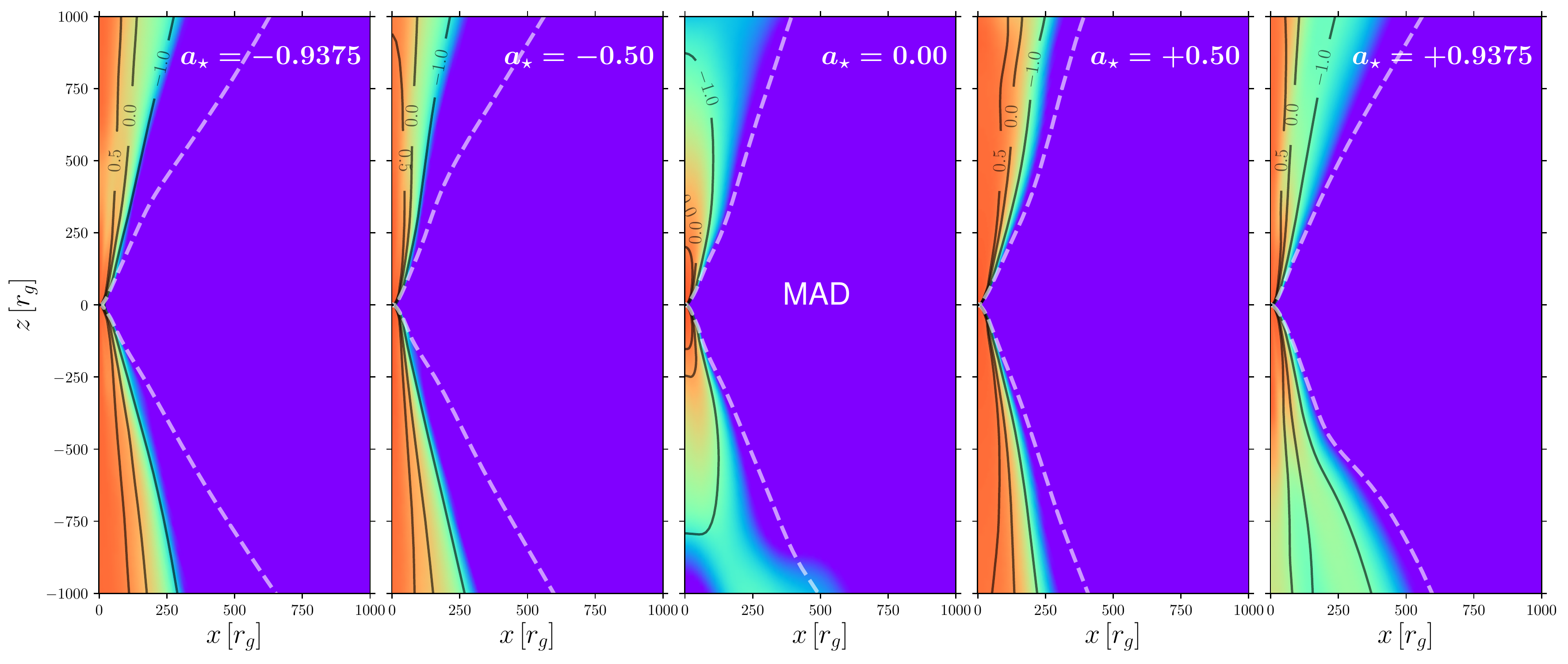}
 \caption{Azimuthal and time averaged distribution of the $\kappa$ value for different spins (left to right) and for SANE (top) and MAD(bottom) models. The dashed white line corresponds to the 
Bernoulli parameter, $-hu_t=1.02$ and the black lines show $\log_{10}(\sigma)=-1.0,\,0.5,\,\rm{and}\, 0$ contours.}
\label{fig:spinkappa}
\end{figure*}

In Fig.~ \ref{fig:spinsigma} we show the azimuthal and time averaged distribution of the logarithm of the magnetisation, $\sigma$, for five different spins (from left to right: -0.9375, -0.50, 0, 0.50, 
and 0.9375) for SANE (top) and MAD (bottom) models. The white dashed line corresponds to $-hu_t=1.02$ and separating between outflowing plasma (region enclosed by the jet axis and the $hu_t$ 
contour line) and bound plasma and the black lines show $\log_{10}(\sigma)=-1.0,\,0.5,\,\rm{and}\, 0$ contours. As mentioned in Sect. \ref{sec:GRMHD} the jet regions in the MAD models exhibit high 
magnetisations in contrast to the SANE models. 
\newline The averaged distribution of the $\kappa$ parameter for different spins and two different accretion models are presented in Fig.~\ref{fig:spinkappa}. Similar to Fig.~\ref{fig:spinsigma} the 
dashed white line corresponds to the Bernoulli parameter, $-hu_t=1.02$ and the black lines show $\log_{10}(\sigma)=-1.0,\,0.5,\,\rm{and}\, 0$ contours. Since the $\kappa$ value depends on both, the 
magnetisation and the plasma-$\beta$ the distribution of the $\kappa$ values roughly follows the distribution of the magnetisation (see Eq.~\ref{eq:kappa}). We find small $\kappa$ values, 
corresponding to flat particle distributions in the MAD models and steep particle distributions in the SANE case (see Eq.~\ref{eq:kappaedf}). Thus, the MAD models have more electrons with large 
Lorentz factors, $\gamma_{e}$ in the jet region than the their SANE counterparts. Notice, that the emission and absorption coefficients derived for by \citet{Pandya2016} for the kappa eDF are only 
valid for $\kappa\leq8$ and we switch to a thermal eDF in these region during the GRRT.  

\section{Multi-frequency observations of M\,87}
In Table ~\ref{Tab:obsSED} we provide the observational data with references used for broad-band spectrum in Figs. \ref{fig:Rhsweep} -\ref{fig:spin}.

\begin{table*}
\centering
\caption{Observed flux densities for M\,87. Columns show the observing frequency in Hz, total flux density and uncertainty both in Jansky, details of the observations and references. The first part of 
the table corresponds to M\,87 in during a quiescent phase where the second part list values during an active phase of M\,87. The third group list values take during 2005 and the last one for 2017.}
\label{Tab:obsSED}
\begin{adjustbox}{max width=0.88\textwidth}
\begin{tabular}{cccll} 
\hline
\hline
   \bf{Frequency (Hz)} & \bf{Flux (Jy)} & \bf{Error\, (Jy)} & \bf{Source / Date} &\bf{Reference}\\
\hline
\hline
$2.06 \times 10^{15}$ & $4.73 \times 10^{-5}$ & $0.47 \times 10^{-5}$  & $1465\, \rm{\AA}$ STIS-F25SRF2 99-5-17 &\citealt{Prieto2016}\\
$1.36 \times 10^{15}$ & $1.33 \times 10^{-4}$ & $0.04 \times 10^{-4}$  & F220W ACS-HRC  03-11-29 &\citealt{Prieto2016}\\
$1.27 \times 10^{15}$ & $1.05 \times 10^{-4}$ & $0.03 \times 10^{-4}$  & 2360A STIS-F25QTZ 03-7-27 &\citealt{Prieto2016}\\ 
$1.11 \times 10^{15}$ & $1.55 \times 10^{-4}$ & $0.03 \times 10^{-4}$  & F250W ACS-HRC 03-05-10 &\citealt{Prieto2016}\\
$8.93 \times 10^{14}$ & $2.16 \times 10^{-4}$ & $0.04 \times 10^{-4}$  & F330W ACS-HRC 03-3-31 &\citealt{Prieto2016}\\
$8.93 \times 10^{14}$ & $2.10 \times 10^{-4}$ & $0.04 \times 10^{-4}$  & F330W ACS-HRC 03-5-10 &\citealt{Prieto2016}\\
$6.32 \times 10^{14}$ & $4.13 \times 10^{-4}$ & $0.12 \times 10^{-4}$  & F475W ACS-HRC 03-11-29 &\citealt{Prieto2016}\\
$4.99 \times 10^{14}$ & $6.33 \times 10^{-4}$ & $0.63 \times 10^{-4}$  & F606W ACS-HRC 03-11-29 &\citealt{Prieto2016}\\
$3.70 \times 10^{14}$ & $9.5 \times 10^{-4}$  & $1.9 \times 10^{-4}$     & F814W ACS-HRC 03-11-29 &\citealt{Prieto2016}\\
$3.32 \times 10^{14}$ & $1.38 \times 10^{-3}$ & $0.1 \times 10^{-4}$    & F850LP ACS-WFC 03-1-19 &\citealt{Prieto2016}\\
$2.47 \times 10^{14}$ & $2.06 \times 10^{-3}$ & $0.18 \times 10^{-3}$  & F110W NIC2 97-11-10 &\citealt{Prieto2016}\\
$1.81 \times 10^{14}$ & $3.1 \times 10^{-3}$  & $0.8 \times 10^{-3}$     & F166N NIC3 99-1-16 &\citealt{Prieto2016}\\ 
$1.37 \times 10^{14}$ & $3.3 \times 10^{-3}$  & $0.6 \times 10^{-3}$     & F222M NIC3 98-1-16 &\citealt{Prieto2016}\\
$2.8 \times 10^{13}$  & $1.67 \times 10^{-2}$ & $9.0 \times 10^{-4}$     & Gemini 10.8\, $\mu m$ 01-05 & \citealt{Perlman2001} \\
$2.6 \times 10^{13}$  & $1.3 \times 10^{-2}$   & $2. \times 10^{-3}$       & Keck 11.7\, $\mu m$ 00-1 &\citealt{Whysong2004} \\
$635.0 \times 10^{9}$ & $0.43$                &	$0.09$ & ALMA 12-6-3     &\citealt{Prieto2016}\\
$350.0 \times 10^{9}$ & $0.96$                &	$0.02$ & ALMA 12-6-3     &\citealt{Prieto2016}\\
$286.0 \times 10^{9}$ & $1.28$                &	$0.02$ & ALMA 12-6-3     &\citealt{Prieto2016}\\
$252.0 \times 10^{9}$ & $1.42$                &	$0.02$ & ALMA 12-6-3     &\citealt{Prieto2016}\\
$221.0 \times 10^{9}$ & $1.63$                &	$0.03$ & ALMA 12-6-3     &\citealt{Prieto2016}\\
$108.0 \times 10^{9}$ & $1.91$                &	$0.05$ & ALMA 12-6-3     &\citealt{Prieto2016}\\
$93.7 \times 10^{9}$ & $1.82$                 &	$0.06$ & ALMA 12-6-3     &\citealt{Prieto2016}\\
$22.0 \times 10^{9}$ & $2.0$                  &	$0.1$  & VLA-A 2003-2006 &\citealt{Prieto2016}\\
$15.0 \times 10^{9}$ & $2.7$                  &	$0.1$  & VLA-A aver. 2003-2006 \& 2003-2008 &\citealt{Prieto2016}\\
\hline
$1.36 \times 10^{15}$  & $2.08 \times 10^{-4}$ & $0.06 \times 10^{-4}$ &   F220W ACS-HRC 05-5-9     &\citealt{Prieto2016}\\ 
$1.10 \times 10^{15}$  & $3.0 \times 10^{-4}$  & $0.2 \times 10^{-4}$  &   F250W ACS-HRC 05-5-9     &\citealt{Prieto2016}\\
$8.93 \times 10^{14}$  & $4.2 \times 10^{-4}$  & $0.2 \times 10^{-4}$  &   F330W ACS-HRC 05-5-9     &\citealt{Prieto2016}\\
$6.32 \times 10^{14}$  & $7.61 \times 10^{-4}$ & $0.15 \times 10^{-4}$ &   F475W ACS-HRC 05-5-9     &\citealt{Prieto2016}\\
$5.0 \times 10^{14}$   & $1.02 \times 10^{-3}$ & $0.13 \times 10^{-3}$ &   F606W ACS-HRC 05-5-9     &\citealt{Prieto2016}\\
$5.0 \times 10^{14}$   & $9.52 \times 10^{-4}$ & $0.95 \times 10^{-4}$ &   F606W ACS-HRC 05-6-22    &\citealt{Prieto2016}\\ 
$3.70 \times 10^{14}$  & $1.61 \times 10^{-3}$ & $0.24 \times 10^{-3}$ &   F814W ACS-HRC 05-5-9     &\citealt{Prieto2016}\\
$2.46 \times 10^{14}$  & $1.6 \times 10^{-3}$  & $0.7 \times 10^{-3}$  &   \textit{J}-band VLT-NACO 05-01.       &\citealt{Prieto2016}\\
$1.37 \times 10^{14}$  & $2.67 \times 10^{-3}$ & $0.55 \times 10^{-3}$ &   \textit{K}-band VLT-NACO 05-01        &\citealt{Prieto2016}\\
$3.7 \times 10^{13}$   & $1.0 \times 10^{-2}$  & $0.5 \times 10^{-2}$  &   8\, $\mu m$ Subaru-spec. 05-04 &\citealt{Prieto2016}\\
$22.0 \times 10^{9}$   & $2.75$                & $0.14$	               &   VLA-A 05-5-3                          &\citealt{Prieto2016}\\
$15.0 \times 10^{9}$   & $3.08$                & $0.15$                &   VLA-B 05-5-3                          &\citealt{Prieto2016}\\
\hline
$5.00  \times 10^{14}$ &  $1.01 \times 10^{-3}$  &   $0.10 \times 10^{-3}$   & ACS-HRC-F606W 05-2-9  &\citealt{Prieto2016}\\
$2.46  \times 10^{14}$ &  $1.00 \times 10^{-3}$  &   $0.19 \times 10^{-3}$  & NACO-J-band, 05-1-20   &\citealt{Prieto2016}\\
$2.06 \times 10^{15}$  &  $4.14 \times 10^{-5}$ &  $4.1\times 10^{-6}$   & $1465\, \rm{\AA}$ STIS-F25SRF2 99-5-17 $r = 0\farcs12$ &\citealt{Prieto2016}\\
$1.36 \times 10^{15}$  &  $1.08 \times 10^{-4}$ &  $0.02 \times 10^{-4}$ & F220W ACS-HRC 03-11-29 $r = 0\farcs13$     &\citealt{Prieto2016} \\
$1.27 \times 10^{15}$  &  $1.81 \times 10^{-4}$ &  $2. \times 10^{-6}$   & 2360A STIS-F25QTZ 01-7-30  $r = 0\farcs12$ &\citealt{Prieto2016}\\
$1.10 \times 10^{15}$  &  $1.28 \times 10^{-4}$ &  $0.01 \times 10^{-4}$ & F250W ACS-HRC 03-05-10 $r = 0\farcs13$ &\citealt{Prieto2016}\\
$8.93 \times 10^{14}$  &  $1.79 \times 10^{-4}$ &  $1.4 \times 10^{-5}$  & F330W ACS-HRC 03-3-31  $r = 0\farcs13$ &\citealt{Prieto2016}\\
$6.32 \times 10^{14}$  &  $4.13 \times 10^{-4}$ &  $0.54 \times 10^{-4}$ & F475W ACS-HRC 03-11-29 $r = 0\farcs13$ &\citealt{Prieto2016}\\
$4.99 \times 10^{14}$  &  $4.24 \times 10^{-4}$ &  $0.24 \times 10^{-4}$ & F606W ACS-HRC 03-11-29 $r = 0\farcs13$ &\citealt{Prieto2016}\\  
$3.70 \times 10^{14}$  &  $6.34 \times 10^{-4}$ &  $1.26 \times 10^{-4}$ & F814W ACS-HRC 03-11-29 $r = 0\farcs15$ &\citealt{Prieto2016}\\  
$3.32 \times 10^{14}$  &  $1.30 \times 10^{-3}$ &  $0.14 \times 10^{-3}$ & F850LP ACS-WF 03-1-19  $r = 0\farcs14$ &\citealt{Prieto2016}\\
$2.47 \times 10^{14}$  &  $1.61 \times 10^{-3}$ &  $0.16 \times 10^{-3}$ & F110W NIC2 97-11-10 $r = 0\farcs15$    &\citealt{Prieto2016}\\
$86.0 \times 10^{9}$   &  $1.39$                &  $0.2$                 & EB, ON, PB, VLBA 09-05-09    &\citealt{Kim2018a}\\
$2.3 \times 10^{11}$   &  $0.98$	            &  $0.04$	             & VLBI $1.3\, \rm{mm}$  2009,  &\citealt{Doeleman2012}\\
\hline
$230.0 \times 10^{9}$   &  $0.98$                &  $0.05$               & VLBI 15-03-2012      &\citealt{Akiyama2015}\\
$22.0 \times 10^{9}$    &  $1.2$                &  $0.1$                & KaVA, VLBA 2013-2014 &\citealt{Hada2017}\\
$15.4 \times 10^{9}$    &  $1.3$                &  $0.1$                & MOJAVE 2001-2011     &\citealt{Lister2018}\\
\hline
$129.0\times 10^{9}$    &  $0.91$               &  $0.27$               & KVN  19-04-2017   &\citealt{An2018}\\
$86.0 \times 10^{9}$    &  $1.12$               &  $0.17$               & KVN  19-04-2017   &\citealt{An2018}\\
$43.0 \times 10^{9}$    &  $1.12$               &  $0.11$               & EAVN 18-04-2017   &\citealt{Kim2018b}\\
$43.0 \times 10^{9}$    &  $1.18$               &  $0.12$               & EAVN 14-04-2017   &\citealt{Kim2018b}\\
$43.0 \times 10^{9}$    &  $1.10$               &  $0.11$               & EAVN 09-04-2017   &\citealt{Kim2018b}\\
$43.0 \times 10^{9}$    &  $1.21$               &  $0.12$               & EAVN 04-04-2017   &\citealt{Kim2018b}\\
$22.0 \times 10^{9}$    &  $1.32$               &  $0.13$               & EAVN 17-04-2017   &\citealt{Kim2018b}\\
$22.0 \times 10^{9}$    &  $1.25$               &  $0.13$               & EAVN 03-04-2017   &\citealt{Kim2018b}\\
\hline
\hline
 \end{tabular}
 \end{adjustbox}
\end{table*}
\end{document}